\documentclass[11pt]{article}
\usepackage{amssymb,cite,amsmath,graphicx,epsf}
\usepackage{epstopdf}

\usepackage[totalwidth=160mm,totalheight=247mm]{geometry}

\newcommand\blank[1]{}

\newcommand{\fract}[2]{{\textstyle\frac{#1}{#2}}}

\newcommand{\ri}{\right}

\newcommand{\eps}{\varepsilon}
\newcommand{\lf}{\left}

\newcommand{\CS}{{\cal S}}

\newcommand\ZZ{{\mathbb Z}}
\newcommand\RR{{\mathbb R}}
\newcommand\NN{{\mathbb N}}

\newcommand{\balpha}{\alpha\kern -6.7pt\alpha}
\newcommand{\bbalpha}{\alpha\kern -4.95pt\alpha}

\newcommand{\CaC}{{\cal C}}
\newcommand{\CH}{{\cal H}}

\newcommand{\PT}{\mathcal{PT}}
\newcommand\eq{\begin{equation}}
\newcommand\en{\end{equation}}
\newcommand\bea{\begin{eqnarray}}
\newcommand\eea{\end{eqnarray}}
\newcommand\nn{\nonumber}
\newcommand\half{{\textstyle\frac{1}{2}}}
\newcommand\hf{\fract{1}{2}}

\newcommand{\One}{{\hbox{{\rm 1{\hbox to 1.5pt{\hss\rm1}}}}}}
\renewcommand{\One}{{\mathbb 1}}
\renewcommand{\One}{{\rm 1\!\!1}}

\newcommand\al{\alpha}

\newcommand\ep{\epsilon}

\newenvironment{tab}{\linespread{1.0} \begin{table}}{\end{table}%
\linespread{1.3}}
\newcommand{\lamt}{\tilde \lambda}
\newcommand{\at}{\tilde \alpha}

\begin{document}

\begin{titlepage}
\vskip 0.5cm
\begin{flushright}
DCPT-08/37  \\
\end{flushright}
\vskip 1.8cm
\begin{center}
{\Large {\bf ${\cal PT}$ symmetry breaking and exceptional points
for a class of
inhomogeneous complex potentials}} \\[5pt]
{\Large {\bf } }
\end{center}
\vskip 0.8cm

\centerline{Patrick Dorey%
\footnote{{\tt p.e.dorey@durham.ac.uk}},
Clare Dunning%
\footnote{{\tt t.c.dunning@kent.ac.uk}},
Anna Lishman%
\footnote{{\tt AnnaLishman@dunelm.org.uk}}
and Roberto Tateo%
\footnote{{\tt tateo@to.infn.it}}}
\vskip 0.9cm
\centerline{${}^{1,3}$\sl\small Dept.~of Mathematical Sciences,
University of Durham, Durham DH1 3LE, UK\,}
\vskip 0.3cm \centerline{${}^{2}$\sl\small IMSAS, University of
Kent, Canterbury, UK CT2 7NF, United Kingdom}
\vskip 0.3cm \centerline{${}^{4}$\sl\small Dip.\ di Fisica Teorica
and INFN, Universit\`a di Torino,} \centerline{\sl\small Via P.\
Giuria 1, 10125 Torino, Italy}
\vskip 1.25cm
\vskip 0.9cm
\begin{abstract}
\vskip0.15cm
\noindent
We study a three-parameter family of $\PT$-symmetric Hamiltonians,
related via the ODE/IM correspondence to the
Perk-Schultz models. We show that real eigenvalues merge and become
complex at quadratic and cubic exceptional points, and explore the
corresponding Jordon block structures by exploiting the quasi-exact
solvability of a subset of the models. The mapping of the phase
diagram is completed using a combination of numerical, analytical
and perturbative approaches. Among other things this reveals some
novel properties of the Bender-Dunne polynomials, and gives a 
new insight into a phase transition to infinitely-many complex
eigenvalues that was first observed by Bender and Boettcher.
 A new exactly-solvable limit, the inhomogeneous complex
square well, is also identified.
\end{abstract}
\end{titlepage}
\setcounter{footnote}{0}
\newcommand{\resection}[1]{\setcounter{equation}{0}\section{#1}}
\newcommand{\appsection}[1]{\addtocounter{section}{1}
\setcounter{equation}{0} \section*{Appendix \Alph{section}~~#1}}
\renewcommand{\theequation}{\thesection.\arabic{equation}}
\def\thefootnote{\fnsymbol{footnote}}
\resection{Introduction}
\label{intro}
In this paper we return to the
spectra of a family of $\PT$-symmetric eigenvalue
problems first studied in detail in~\cite{DDT3,Dorey:2001hi}. 
Consider the following differential operator:
\eq
\CH=
-\frac{d^2}{dx^2}-(i x)^{2M}
-\alpha (i x)^{M-1}+ \frac{\lambda^2-\frac{1}{4}}{x^2} ~ ,
\label{PTg}
\en
where $M$, $\alpha$ and $\lambda$ are real numbers, with $M>0$, and
the powers of $ix$ are rendered single-valued by placing a cut along
the positive imaginary $x$ axis. Then an eigenvalue problem, with a
discrete spectrum, can be defined as
\eq
\CH\,\psi(x)=E\,\psi(x)\, ;\qquad
\psi(x) \in L^2(\CaC)\,,
\label{PTbc}
\en
where ${\cal C}$ is an infinite contour in the complex
plane, which must pass below the origin
whenever $\lambda^2 \neq \frac{1}{4}$ or $M\notin\ZZ$. 
For $M<2$ the ends of this contour can asymptote to the negative and
positive real axes, while for $M\ge 2$ they must be deformed down
into the complex plane so as to continue the $M<2$
spectral problem smoothly~\cite{BB}. This is
illustrated in figure~\ref{contour}.

\[
\begin{array}{c}
\!\!\!\!\!\!\includegraphics[width=0.44\linewidth]{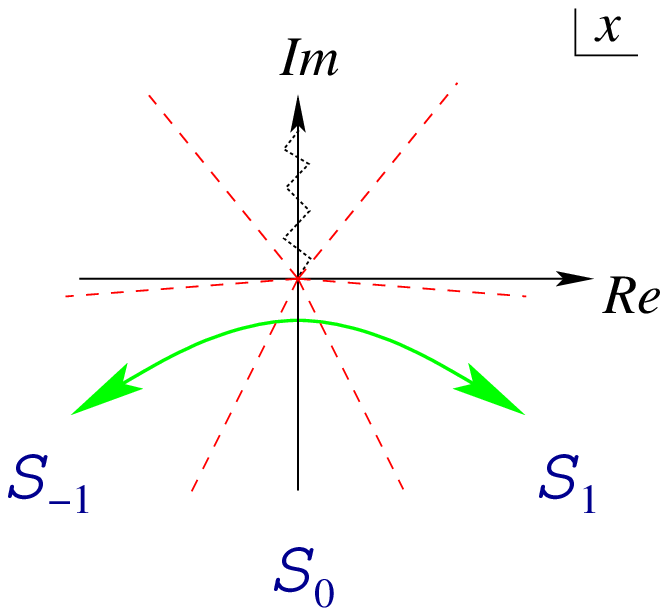}
\\[11pt]
\parbox{0.54\linewidth}{
\small Figure \protect\ref{contour}: A possible quantisation contour 
${\cal C}$ for $M$ just larger than $2$, together with some of the
Stokes sectors.
}
\end{array}
\]
\refstepcounter{figure}
\protect\label{contour}

In~\cite{DDT3,Dorey:2001hi}, the eigenvalue problem was initially
specified in terms of $l\equiv \lambda-\frac{1}{2}$, but with
boundary conditions imposed at infinity, the choice of $\lambda$ is
more natural.

An alternative specification of the boundary conditions,
which holds for all values of $M$, starts from the
Stokes sectors for (\ref{PTg}), which we
denote by 
\eq
\CS_k=\left\{ x\in{\mathbb C}\,:\,
\left|\arg(ix)-\frac{2\pi
k}{2M{+}2}\right|<\frac{\pi}{2M+2}\right\}~,\quad k\in\ZZ\,.
\label{stokesdef}
\en
For all $M$ the requirement (\ref{PTbc}) is 
equivalent to the demand
that $\psi(x)\to 0$ as $x\to\infty$ in the sectors
${\cal S}_{-1}$ and ${\cal S}_1$. This allows for a
convenient rephrasing of the eigenvalue condition in terms of the
vanishing of a certain Wronskian. Following Hsieh and Sibuya
\cite{HS,SibuyaBook,SibuyaArticle}, let
$y^{~}_0(x,E,\alpha,\lambda)$ be the solution to (\ref{PTg}) 
that is (uniquely) defined by the following asymptotic for
$x\to\infty$ on the negative imaginary axis:
\eq
y^{~}_0(x,E,\alpha,\lambda)\sim
\frac{i}{\surd 2}\,
(ix)^{-M/2-\alpha/2}\,\exp\left(-\frac{(ix)^{M+1}}{M{+}1}\right)\,,\quad
x\to -i\,\infty\,,
\label{y0def}
\en
and then set
\eq
\omega=e^{i\pi/(M{+}1)}
\en
and define a sequence of further solutions $y_k$ to (\ref{PTg})
by
\eq
y_k(x,E,\alpha,\lambda)=
\omega^{k/2-(-1)^k k\alpha/2}\,
y^{~}_0(\omega^{-k}x,\omega^{-2Mk}E,(-1)^k\alpha,\lambda)\,.
\label{ykdef}
\en
It is easily checked that $y_k$ decays, or is subdominant, 
in ${\cal S}_k$, and that the `nearest-neighbour'
Wronskians $W[y_k,y_{k+1}]= y_ky'_{k+1}-y'_ky_{k+1}$ 
are all equal to $1$.\footnote{Note, a propagating typo in
\cite{DDT3} and  \cite{Dorey:2007zx} resulted in the factor 
of $(-1)^k$ multiplying $k\alpha/2$ in
the exponent of $\omega$ being omitted from the definition of $y_k$
given in those papers. None of 
the other formulae in \cite{DDT3,Dorey:2007zx} are affected.}
The eigenvalue condition is then that $y_{-1}$ and $y_1$ should be
proportional to each other, in other words that $E$ should be a zero
of the
`next-nearest-neighbour' Wronskian
\eq
T(E,\alpha,\lambda)=W[y_{-1},y_1]\,.
\en
{}From this characterisation, and the analyticity of $T$ as a function
of its arguments, a number of important properties, such as the
discreteness of the spectrum, immediately follow. In addition to being
a spectral determinant, via the ODE/IM correspondence  of
\cite{DTa} (see \cite{Dorey:2007zx} for a review) $T$ encodes 
the properties of the ground state of an integrable quantum field
theory, in this case the Perk-Schultz model 
\cite{Perk:1981nb,Suzuki:2000fc}.
This correspondence is based in part on the fact that $T$
is a Stokes multiplier for (\ref{PTg}), in that the following
equation holds \cite{Dorey:1999uk,DDT3}:
\eq
T(E,\alpha,\lambda)y_0(x,E,\alpha,\lambda)=
y_{-1}(x,E,\alpha,\lambda)+
y_{1}(x,E,\alpha,\lambda)\,.
\label{TQ}
\en

A feature the eigenvalue problem (\ref{PTbc}) 
shares with
many other ${\cal PT}$-symmetric problems is the reality of its
spectrum for many values of the free parameters.
In particular, for real $M>1$, $\alpha$ and
$\lambda$, the spectrum of
(\ref{PTg}) can be proved to be
\bea
\bullet&\! \mbox{~~{\em real}~~~~ if}&\alpha<M+1+2\,|\lambda|~;
\label{rres}\\[3pt]
\bullet&\!\! \mbox{{\em positive}~ if}&\alpha<M+1-2\,|\lambda|~.
\label{pres}
\eea
These results were established in \cite{DDT3} using techniques inspired 
by the ODE/IM correspondence. One of the main aims of this paper is to
refine this picture and to explore in more detail how and where
spectral reality is lost as the region (\ref{rres}) is left.

Along the lines $\alpha=M+1\pm2\lambda$ which form the
frontiers of the region (\ref{rres}) of
guaranteed reality, the model
has an exactly-zero energy level, as in supersymmetric quantum
mechanics. The `protection' of this level
can be seen as the mechanism by which the first levels become complex
\cite{Dorey:2001hi}.
However, numerical investigations at $M=3$, reported in
\cite{Dorey:2001hi},
showed that the region within which
the spectrum of (\ref{PTg}) is complex has considerably more
structure than (\ref{rres}) might suggest.
The curved, cusped line of figure~\ref{scan} indicates
where the first pair of complex eigenvalues is formed as the region of
complete spectral reality is left; it
touches the lines $\alpha=M+1\pm 2\lambda$ at isolated points, where
the protected zero-energy level coincides with another level.

\[
\begin{array}{c}
\!\!\!\!\!\!\includegraphics[width=0.6\linewidth]{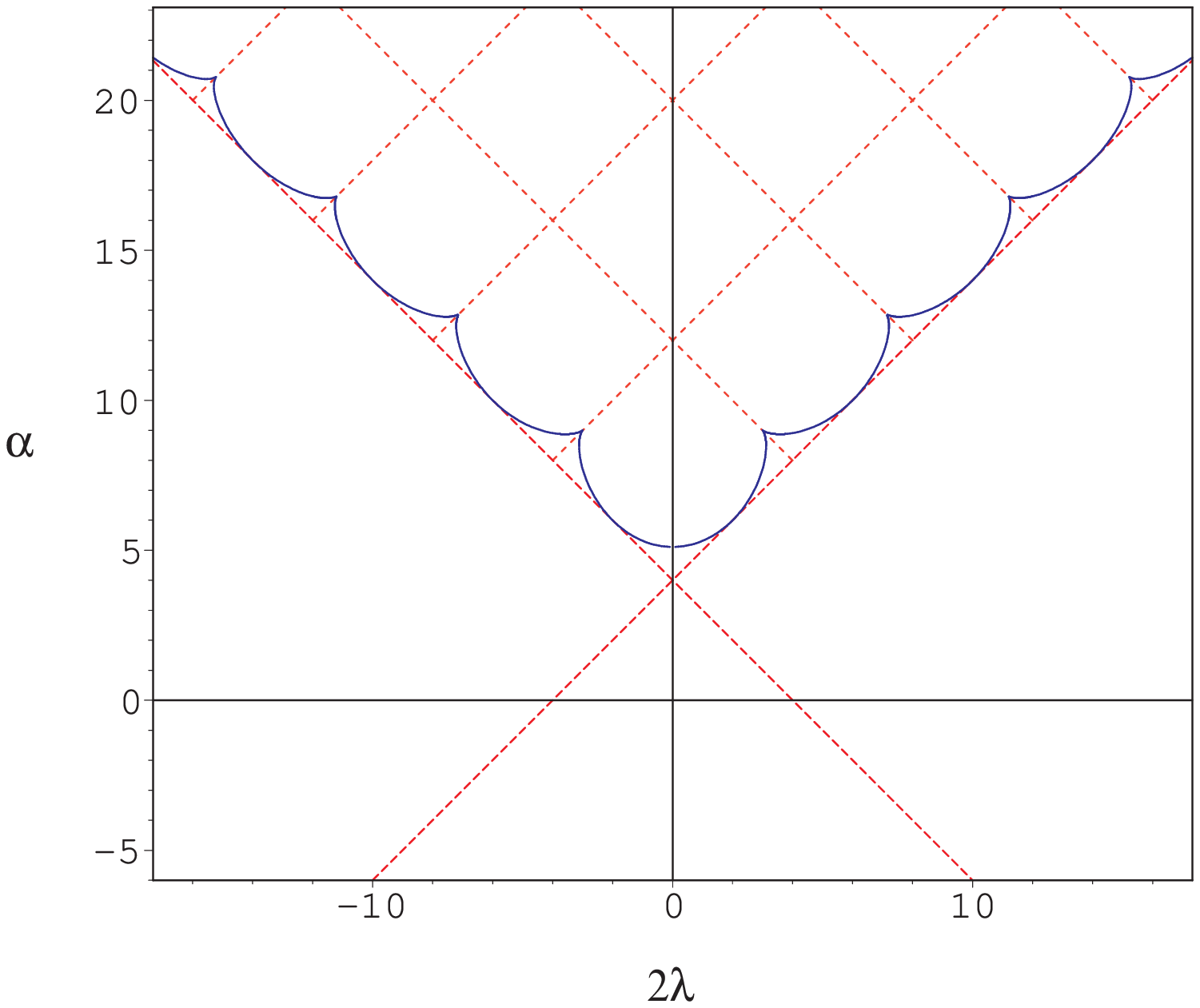}
\\[11pt]
\parbox{0.6\linewidth}{
\small Figure \protect\ref{scan}: The domain of unreality in the
$(2\lambda,\alpha)$ plane for $M=3$, with
portions of lines with a protected zero-energy level also shown. The
horizontal axis is $2\lambda=2l+1$.
}
\end{array}
\]
\refstepcounter{figure}
\protect\label{scan}

\medskip

The additional dotted lines on the figure,
also at angles of $\pm 45^{\circ}$,
 show points {\em within}\/
the region $\alpha>M+1+2\,|\lambda|$ where the model has  an
exactly-zero energy level; exceptionally for $M=3$, the model is also
quasi-exactly solvable along these lines. It is notable that, to
within numerical accuracy, the cusps on the boundary of the region of
unreality appear to lie exactly on these lines.

It is natural to ask where further pairs of complex eigenvalues are
formed. For $M=3$ the answer is shown in figure~\ref{fullscan},
adapted from \cite{Dorey:2007zx}; the same pattern was
found independently by Sorrell \cite{Sorrell:2007tr} via a complex WKB
treatment of the problem. The pattern of cusps is repeated, with
the cusps again appearing to lie
on the lines of protected zero-energy levels.

\[
\begin{array}{c}
\!\!\!\!\!\!\includegraphics[width=0.6\linewidth]{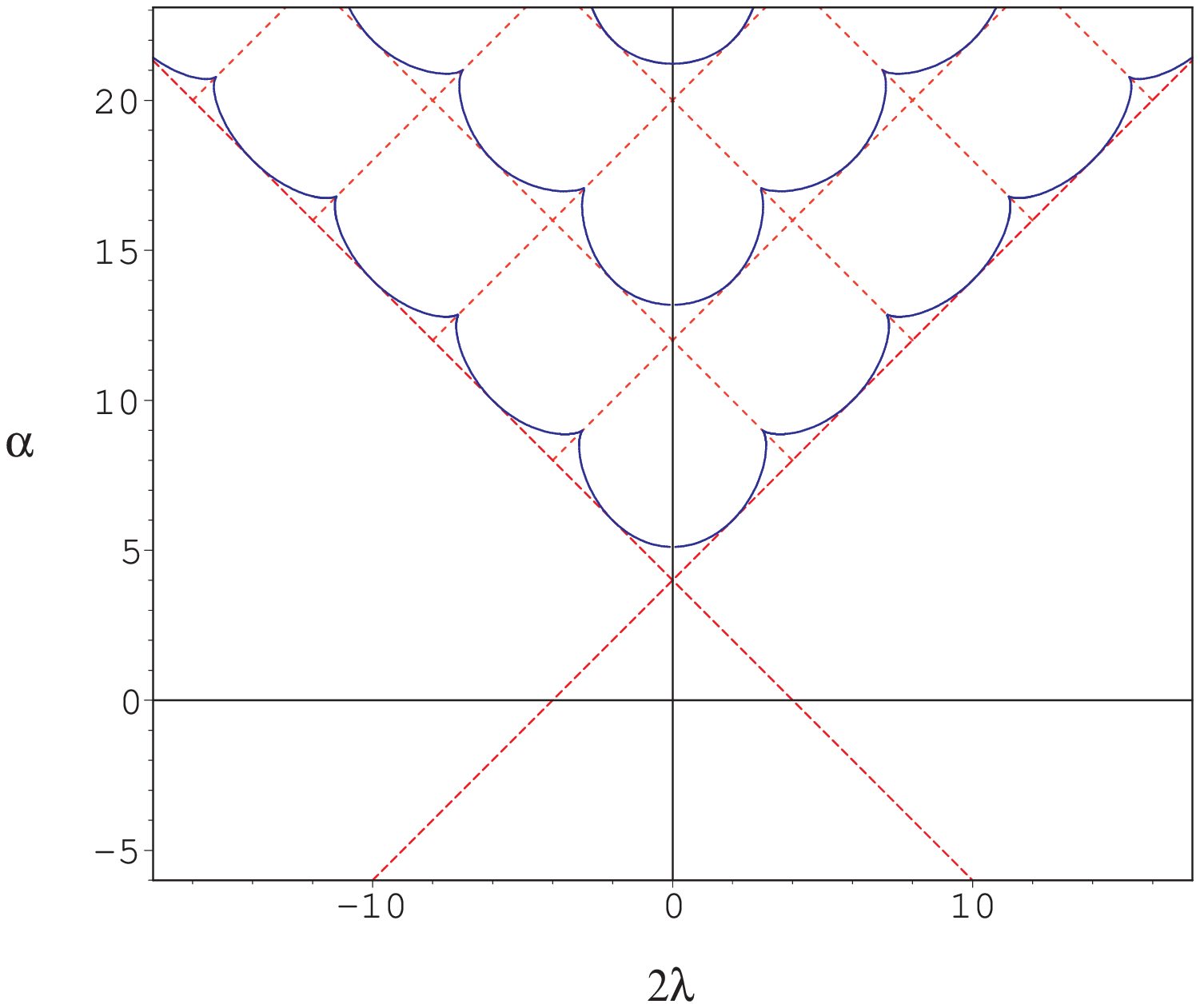}
\\[11pt]
\parbox{0.6\linewidth}{
\small Figure \protect\ref{fullscan}: The $(2\lambda,\alpha)$ plane for
$M=3$, showing lines across which further pairs of complex
eigenvalues are formed.
}
\end{array}
\]
\refstepcounter{figure}
\label{fullscan}

\medskip

The analysis of \cite{Dorey:2001hi} left a number of questions open.
Whilst the merging and subsequent complexification of levels was
suggestive of exceptional points and
a Jordan block structure for the Hamiltonian, this was not demonstrated
explicitly. The apparent siting of the cusps on lines with
simultaneous quasi-exact solvability and
protected zero-energy levels was not proved; in particular, it was not
clear whether this feature should be associated with the
zero-energy level (in which case it should persist for $M\neq 3$)
or with the quasi-exact solvability (in which case it might be lost
for $M\neq 3$). Finally, and connected with this last question, the
general pattern away from $M=3$ was not explored.

In this paper we revisit these issues.
For $M=3$ we investigate the positions
of the cusps, proving that they do indeed lie on QES lines,  and look
at
the  exceptional points in the spectrum and the Jordan form at such
points. We then explore the situation for $M \ne 3$ numerically,
and verify the picture that emerges
with  detailed perturbative studies near $M=1$ and $M=\infty$.
The perturbative treatment near $M=1$ also gives a new insight into the
transition to infinitely-many complex eigenvalues for $M<1$, first
observed by Bender and Boettcher for the $\alpha=0$,
$\lambda^2=\frac{1}{4}$ case of (\ref{PTg}).

\resection{Exact locations of special exceptional points}
\subsection{Generalities and previous results}
Exact formulae for the full curves of exceptional points
are unlikely to exist, even for $M=3$.
However, certain exceptional points {\em can}\/ be
located exactly, and this information turns out to be very useful in
mapping the full phase diagram.
As in \cite{Dorey:2001hi}, we begin by
introducing an alternative set of coordinates on
the $(2\lambda,\alpha)$ plane, defined by
\eq \label{apm}
\al_{\pm}=\frac{1}{2M{+}2}\,[\,\al-M-1\pm2\lambda\,]\,.
\en
For $M=3$ these coordinates are illustrated in figure
\ref{cusplocations}.
The lines $\alpha_+\in\NN$ and $\alpha_-\in\NN$ correspond to the
dotted lines on figures \ref{scan} and \ref{fullscan}, along which
the model (\ref{PTg}) has an
exactly-zero energy level.
For $M=3$ the existence of this level can be understood in terms of
quasi-exact solvability and a hidden ${\cal N}$-fold supersymmetry
\cite{DDT3,nfold}. There are also exactly-zero energy levels along
the lines $\alpha_{\pm}=0$, related for all values of $M$
to standard quantum-mechanical supersymmetry 
\cite{Dorey:2001hi}.

Exceptional points occur in the spectrum of an
eigenvalue problem whenever the coalescence of two or more
eigenvalues is accompanied by a coalescence of the corresponding
eigenvectors; at such points there is a branching of the spectral
surface \cite{kato,heiss03,Sokolov:2006vj,GRS}.
In $\PT$-symmetric systems, eigenvalues are all either real, or
in complex-conjugate pairs. Complex eigenvalues
can therefore be  formed only via the intermediate
coincidence of two (or more) previously-real eigenvalues. For
one-dimensional problems of the sort under discussion here genuine
degeneracies of levels are impossible -- since, for example, the
Wronskian of any two solutions which both decay exponentially
in the same asymptotic
direction must vanish -- and so levels in our problem can only coincide
at exceptional points. Hence the cusped lines on figures
\ref{scan} and \ref{fullscan} are lines of exceptional points. In
fact, we shall see that points on the (codimension one)
smooth segments of the cusped lines are quadratically exceptional,
with two levels coalescing, while the cusps themselves, of codimension
two, are cubic exceptional points. In the following, we 
will often refer to a connected union
of quadratically-exceptional lines and cubically-exceptional points as
an exceptional line.

\[
\begin{array}{c}
\includegraphics[width=0.6\linewidth]{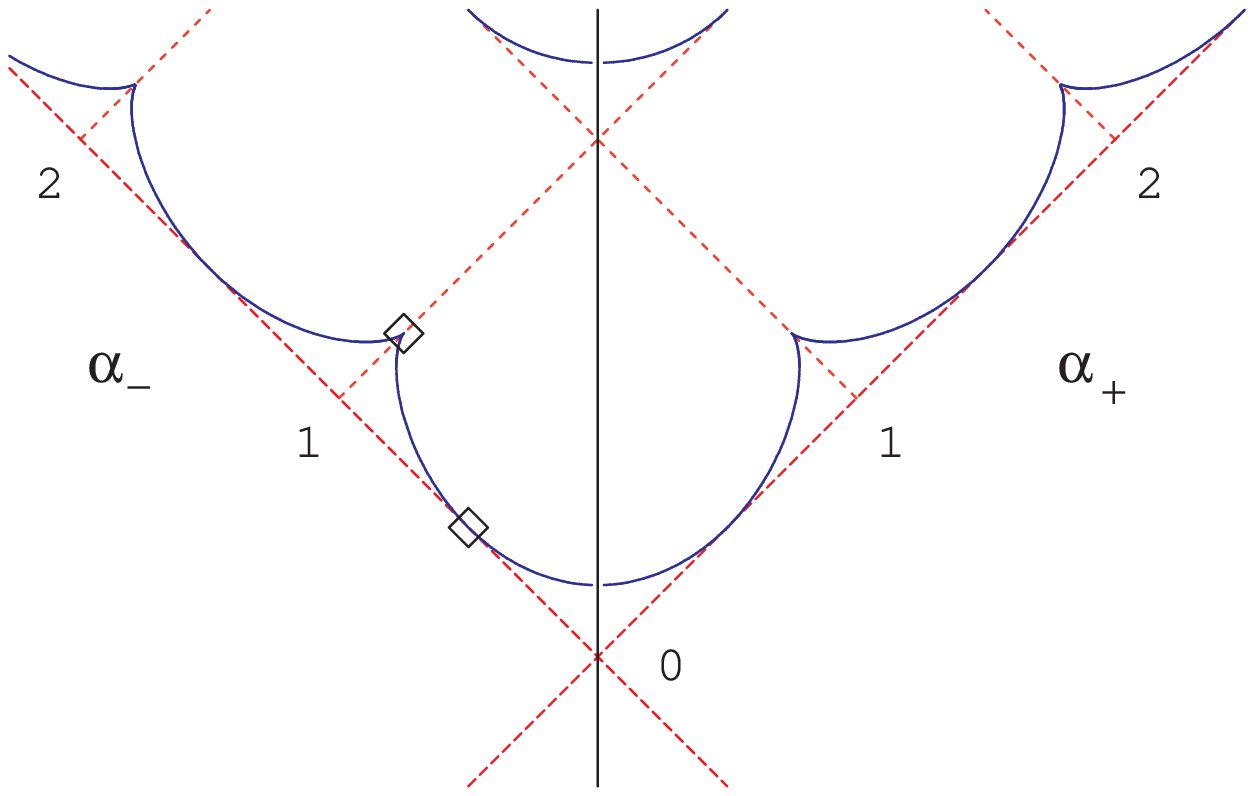}
\\[11pt]
\parbox{0.7\linewidth}{
\small Figure \protect\ref{cusplocations}:
An enlarged view of the $M=3$ phase diagram, showing the
$(\alpha_+,\alpha_-)$ coordinates. The boxes indicate the
locations of the quadratic and cubic exceptional points, at
$(\alpha_+,\alpha_-)=(0,1/2)$ and $(1/4,1)$, which are discussed
later in the main text.
}
\end{array}
\]
\refstepcounter{figure}
\label{cusplocations}

The exactly-zero energy levels can be used to control
the pairing-off of eigenvalues and
the associated formation of exceptional points
\cite{Dorey:2001hi}.
On the `supersymmetric' lines $\alpha_{\pm}=0$
there is always at
least one zero-energy eigenvalue,
for any value of $M$.
The points where this eigenvalue becomes degenerate with a
second one can be found by looking for zero eigenvalues of the
supersymmetric partner potential, the partner for
$(\alpha_+,0)$ being
$(\alpha_+-\fract{M-1}{M+1}\,,\,-1)$ and that for
$(0,\alpha_-)$ being
$(-1\,,\,\alpha_--\fract{M-1}{M+1})$.
This idea was used in \cite{Dorey:2001hi}
to show the existence of quadratic exceptional points
for
\eq
(\alpha_+\,,\,\alpha_-)=(0\,,\,m-\fract{2}{M+1})
\quad\mbox{and}\quad
(\alpha_+\,,\,\alpha_-)=(m-\fract{2}{M+1}\,,\,0)
\label{quadp}
\en
where $m\in\NN\equiv \{1,2,\dots\}$. At $M=3$ these are the
points on figure~\ref{scan} where the cusped curve touches the lines
$\alpha_{\pm}=0$. For $M=3$, a similar argument can be applied on
the other lines $\alpha_{\pm}=n\in\NN$ on which there is an exact
zero-energy level, using a higher-order supersymmetry 
to eliminate this level together with $2n$ others
\cite{DDT3,Dorey:2001hi}.
This establishes the existence of quadratic exceptional points at
\eq
(\alpha_+\,,\,\alpha_-)=(n\,,\,m-\fract{1}{2})
\quad\mbox{and}\quad
(\alpha_+\,,\,\alpha_-)=(m-\fract{1}{2}\,,\,n)\,,\quad
m\in\NN\,,~ n \in \ZZ^+ \quad(M=3)~.
\label{qeps}
\en
These are the points on figure \ref{fullscan}
where cusped curves touch the other lines $\alpha_{\pm}\in\ZZ^+$.
In the next section we will generalise these results to other values
of $M$.

\subsection{Locating exceptional points using self-orthogonality}
\label{locself}
Our alternative argument starts from the idea, discussed in,
for example,
\cite{Sokolov:2006vj}, that at an exceptional point at least one
state will be self-orthogonal, in the sense that its inner product
with itself under a suitable symmetric inner product must vanish.
For the present paper we take this inner product to be
\eq
(f|g) \equiv
\int_{\cal C} f(x)g(x)\,dx
\label{cprod}
\en
where the contour ${\cal C}$ is as in section 1. This
inner product is bilinear rather than sesquilinear, and -- at least in
cases where the contour ${\cal C}$ is the real axis -- it is 
sometimes referred to as the $c$-product\,\cite{moiseyev}.
Correspondingly we will refer to $\sqrt{(f|f)}$ as the $c$-norm of $f$;
note that there is no need for this to be a real number.
The $c$-product is well-defined for any pair
of functions which decay exponentially as $|x|\to\infty$ along ${\cal
C}$, and $\CH$ is symmetric with respect to it:
$(f|\CH g)=(\CH f|g)$.

At an exceptional point, associated with some eigenvalue $E$,
the Hamiltonian acquires a Jordan block form,
and a so-called Jordan chain $\{\psi^{(0)}\dots \psi^{(k-1)}\}$
can be defined which spans the subspace
of the $k$ merging levels, such that
\eq
(\CH-E)\psi^{(j)}=\psi^{(j-1)}\,,~~ j=0\dots k{-}1\,,
\quad \psi^{(-1)}\equiv 0\,.
\en
Then $(\psi^{(0)}|\psi^{(0)})=
(\psi^{(0)}|(\CH{-}E)\psi^{(1)})=
((\CH{-}E)\psi^{(0)}|\psi^{(1)})=0$, and so the state $\psi^{(0)}$ is
indeed self-orthogonal with respect to the $c$-product. Conversely, 
suppose that some eigenstate $\psi$, with eigenvalue $E$,
has vanishing $c$-norm, so that $(\psi|\psi)=0$. We would like to show
that this implies that our system is lying at an exceptional point, and
to this end we recall a useful result, previously exploited in this
context by Trinh~\cite{Trinh:2005zr}.
Suppose that $E_n$ is an eigenvalue, so that $y_{-1}$ and $y_1$ are
proportional to each other and $T(E_n,\alpha,\lambda)=0$. In fact, from
(\ref{TQ}), for such an $E$ we have $y_{-1}(x,E,\alpha,\lambda)=-
y_1(x,E,\alpha,\lambda)$\,. Writing $\psi=y_{-1}=-y_1$, the relevant
result,  converted to the normalisations used in this paper, 
is
\eq
(\psi|\psi)\big|_{E_n}=T'(E_n)
\label{psipsiT}
\en
where the prime denotes differentiation with respect to $E$, and the
dependence of $T$ on $\alpha$ and $\lambda$ has been left implicit.
Now suppose that, as a function of some combination of $\alpha$ and
$\lambda$, $(\psi|\psi)$ has an isolated zero. Then at this point,
$T(E)$ as a function of $E$ has a multiple zero which it does not
possess in the neighbourhood of this point. The zero of
$(\psi|\psi)$ must therefore mark a point where two or 
more eigenvalues of the
eigenproblem have collided. Given the 
impossibility of genuine degeneracies in this problem, this must
be an exceptional point, as claimed.

These results are useful in the present context because along the
lines $\alpha_+=n$ and $\alpha_-=n$, $n\in\ZZ^+$, one eigenfunction
{\em can}\/ be found exactly, namely that with eigenvalue
$E=0$. Consider the line $\alpha_-=n$, along which 
\eq
\alpha-2\lambda=(2n{+}1)(M{+}1)\,.
\label{alphamline}
\en
(Corresponding results for the line $\alpha_+=n$ can be obtained
by negating $\lambda$ throughout in the following.) Then the
zero-energy eigenfunction $\psi=y_{-1}=-y_1$, normalised in line with
(\ref{y0def}) and (\ref{ykdef}), is
\eq
\psi(x)=
\frac{1}{\sqrt{2}}\,
\frac{n!\,(M{+}1)^n\!}{2^n}
\,(ix)^{\frac{1}{2}+\lambda}
\,L_n^{(\frac{2\lambda}{M+1})}\!\Bigl(\frac{-2(ix)^{M+1}\!}{M{+}1}\,\Bigr)
\,e^{(ix)^{M+1}\!/(M{+}1)}_{\phantom{l}}
\en
where $L^{(\gamma)}_n(t)$ is the $n^{\rm th}$ generalised Laguerre
polynomial. (To check that $\psi$ has been normalised with
the correct asymptotic, note
the relation (\ref{alphamline}) between $\lambda$ and $\alpha$
and the fact that the highest term of
$L^{(\gamma)}_n(t)$ is $\frac{(-1)^n}{n!}t^n$.)

To evaluate $(\psi|\psi)$, we distort the
contour ${\cal C}$ to the union of rays
$-\gamma_{-1}+\gamma_1$, where
\eq
\gamma_{\pm 1}=\{x=\fract{1}{i}e^{\pm i\pi/(M{+}1)}t,\,t\in
[0,\infty)\},
\label{gencont}
\en
and then use the integral (\ref{lagint}), analytically continuing in
$\lambda$ if necessary to ensure convergence. The final result is
\eq
(\psi|\psi)=
\frac{\pi}{2}
\left(\fract{M{+}1}{2}\right)^{\!2n{-}1+\frac{2\lambda+2}{M+1}}
\frac{1}{\Gamma(1-\frac{2{+}2\lambda}{M+1})}
\,Q_n(\lambda)
\label{psipsi}
\en
where $Q_n(\lambda)$ is a polynomial of degree $n$ in $\lambda$, which
can be expressed in terms 
of the hypergeometric function ${}_3F_2$ and Pochhammer
symbols $(x)_k\equiv x(x{+}1)\dots(x{+}k{-}1)_k$ as
\bea
Q_n(\lambda)
&=&
(1{-}\fract{2}{M+1})_n
(1{+}\fract{2\lambda}{M+1})_n
\,{}_3F_2(-n,\fract{2\lambda+2}{M+1},\fract{2}{M+1};\,
1{+}\fract{2\lambda}{M+1},-n{+}\fract{2}{M+1};\,1)\nn\\[3pt]
&=&
\sum_{k=0}^n(-1)^k
\left(\begin{matrix}n\\k\end{matrix}\right)
(1{-}\fract{2}{M+1}{-}k)_n
(\fract{2\lambda+2}{M+1})_k
(1{+}\fract{2\lambda}{M+1}{+}k)_{n-k}~.
\label{genQ}
\eea
The zeros of (\ref{psipsi}) locate all those
exceptional points on the line $\alpha_-=n$ which involve the merging
of levels at the eigenvalue $E=0$. 
For $M=3$, we will argue in the next subsection that
this captures {\em all}\/ exceptional
points on this line, with the zeros of $Q_n(\lambda)$ being the cubic
exceptional points associated with cusps on the phase diagram.
For other values of $M$, as will be described in
more detail in section \ref{numres}, the cubic exceptional points
move away from the
lines $\alpha_{\pm}=n$, to be 
replaced on these lines by pairs of quadratic
exceptional points, only one of each pair being at $E=0$ and
corresponding to a zero of $Q_n(\lambda)$.

By contrast, the infinitely-many
zeros of the factor $1/\Gamma(1-\frac{2{+}2\lambda}{M+1})$ 
in (\ref{psipsi})
{\em always}\/ correspond to quadratic exceptional
points. These zeros are at
\eq
2\lambda=(M{+}1)m-2\,,\quad m\in\NN
\label{quadeps}
\en
and using
(\ref{alphamline})
they imply the existence of exceptional points at 
\eq
(\alpha_+,\alpha_-)=(n+m-\fract{2}{M+1},n)\,,\quad
m\in\NN\,,~n\in\ZZ^+.
\label{genform}
\en
This result matches and extends the previously-known cases:
for $n=0$, it yields the points (\ref{quadp}), found in
\cite{Dorey:2001hi} using ideas based on supersymmetry, while for
$M=3$ the result (\ref{qeps}) is reproduced.

\subsection{Locating exceptional points using quasi-exact solvability}
\label{locQES}
Self-orthogonality yields important information about the 
phase diagram at general $M$, but it 
fails to identify the degrees of exceptional
points, and it only sees exceptional points which have eigenvalue
zero.
In this subsection we describe a complementary tactic, special
to $M=3$, which avoids these problems by
exploiting the fact that for $M=3$ the model is
quasi-exactly solvable (QES)
on the lines $\alpha_{\pm}\in\NN$. This will allow us to prove
some general statements about the spectrum of the model on these
lines. A key part of the argument, established
in \cite{Dorey:2001hi}, 
is that any complex levels on
the lines $\alpha_{\pm}\in\NN$
must lie in the QES sector of the model.

For the rest of this section and all of the next
we therefore restrict to $M=3$, and,
to minimise the proliferation of factors of $i$, we
replace $x$ by $z=ix$ and set
$\Phi(z)=\psi(z/i)$.
The quantisation contour is
also rotated by $90^{\circ}$, and
the eigenproblem (\ref{PTg}) becomes
\eq
\Bigl[-\frac{d^2}{dz^2}+ z^{6}
+\alpha  z^{2}+ \frac{\lambda^2-\frac{1}{4}}{z^2}
\Bigr]\Phi(z)=-E\,\Phi(z)\, ,
{}~~~\Phi(z) \in L^2(i\,\CaC)\,.
\label{PT3}
\en
A choice for the contour $i{\cal C}$ which avoids all singularities
in the wavefunctions is given in equation (\ref{iC}) below;
alternatively a rotated version of
(\ref{gencont}) can be used, with suitable analytic
continuations whenever divergent integrals are encountered.

If boundary conditions had been imposed at $z=0$ and $z=+\infty$,
the problem
(\ref{PT3}) would have been quasi-exactly solvable whenever
$\al$ and $\lambda$ were related by $\al=-(4J+2\lambda)$ for some
positive integer $J$, with $J$ energy levels exactly computable
\cite{Turbiner:1987nw}. Bender and Dunne \cite{Bender:1995rh} found
an elegant method to find the corresponding wavefunctions,
square integrable along the positive real axis.
We are instead
interested in solutions defined along
the contour $i{\cal C}$, but with minor modifications the approach of
\cite{Bender:1995rh} can still be used\footnote{For earlier
discussions of quasi-exactly solvable $\PT$-symmetric sextic
potentials, see \cite{DDT3,Dorey:2001hi,Bender:2005kr}.}.
We set $J=\alpha/4-\lambda/2$ and look for solutions of the
form
\eq \label{estate}
\Phi(z) = e^{\frac{z^4}{4}}z^{\lambda +\frac{1}{2}}
\sum_{n=0}^{\infty}a_n(\lambda) p_n(E,\lambda,J)\,z^{2n}
\en
where
\eq
a_n(\lambda)=\left(-\frac{1}{4}\right)^n\frac{1}{n!\,\Gamma(n+\lambda +1)}\,.
\en
The function $\Phi(z)$
will solve (\ref{PT3}) if the
coefficients $p_n(E,\lambda,J)$ satisfy the recursion relation
\eq 
p_n=-Ep_{n-1}+16(J-n+1)(n-1)(n-1+\lambda)p_{n-2}\,,\qquad n\geq 1\,.
\label{Pn}
\en
Setting $p_0=1$ fixes the normalisation, and then
$p_1=-E$ follows from (\ref{Pn}) at $n=1$.
If $J$
is a positive integer,
then the second term on the RHS of (\ref{Pn}) vanishes when $n=J+1$,
and so $p_{J+1}$ is proportional
to $p_J$, as are all subsequent
coefficients $p_{m>J+1}$.
At a zero of
$p_J$ the series therefore terminates.
Owing to the sign of the argument of the
exponential prefactor in (\ref{estate})
(opposite to that in \cite{Bender:1995rh}),
the corresponding $\Phi(z)$ will automatically
satisfy the revised boundary conditions. We
define the $J^{\rm th}$ Bender-Dunne polynomial for this problem
to be
\eq
P_J(E,\lambda)=p_J(E,\lambda,J)~.
\en
This is a polynomial of degree $J$ in $E$, and degree $J-1$ in 
$\lambda$.
By the above reasoning,
its zeros in $E$
give the $J$ quasi-exactly solvable (QES) levels that the
model possesses on the line $\alpha=4J{+}2\lambda$.
Since boundary conditions are not
imposed at the origin, replacing $\lambda$ by $-\lambda$ throughout
also leads to an
acceptable solution, and so for each $J\in\NN$
there are two lines of
quasi-exact solvability in the $(2\lambda,\alpha)$ plane:
$\alpha=4J+2\lambda$ and $\alpha=4J-2\lambda$. In the
$(\alpha_+,\alpha_-)$ coordinates these lines are
$(\alpha_+,\alpha_-)=
(\frac{1}{2}(J{-}1{+}\lambda),\frac{1}{2}(J{-}1))$ and
$(\frac{1}{2}(J{-}1),\frac{1}{2}(J{-}1{-}\lambda))$ respectively.
Figure \ref{proof}, below, shows the QES lines on 
the $(2\lambda,\alpha)$ plane.

These lines are very useful in mapping the exceptional 
points on the whole $(2\lambda,\alpha)$ plane.  The reasoning is 
best explained via a sequence of lemmas, which may be of
independent interest.

\smallskip

\noindent
{\bf Lemma 1:} The Bender-Dunne polynomials satisfy the `reflection
symmetry'
\eq
P_J(E,\lambda)=(-i)^JP_J(iE,-J-\lambda)
\label{BDsymmetry}
\en
{\bf Proof:} Introduce a set of polynomials defined by
$r_n(E,\lambda,J)=(-i)^np_n(iE,-J{-}\lambda,J)$. Direct substitution
into (\ref{Pn}) shows that
the $r_n$ satisfy the recursion
\eq
r_n=-Er_{n-1}+16(J-n+1)(n-1)(J-n+1+\lambda)r_{n-2}\,,\qquad n\geq 1
\label{Qn}
\en
with initial conditions $r_{-1}=0$, $r_0=1$. The claimed symmetry is
equivalent to $r_J(E,\lambda,J)=p_J(E,\lambda,J)$. 
Now consider a more general recursion 
\eq
u_n=-Eu_{n-1}+b_{n-1}u_{n-2}\,,\qquad n\geq 1\,, u_{-1}=0\,,~u_0=1
\label{Un}
\en
with some set of coefficients $\{b_n\}$.
It is straightforward to verify that the general
solution is
\eq
u_n=(-1)^n\sum_{k=0}^{[n/2]}
\Biggl(
\sum_{\genfrac{}{}{0pt}{}{\{0<i_1<\dots <i_k<n\}}{|i_{j+1}-i_j|>1}}
b_{i_1} b_{i_2}\dots b_{i_k} \Biggr)
E^{n-2k}~.
\label{gensol}
\en
In particular, $p_J$ 
is given by (\ref{gensol}) with $n=J$ and
$b_i=16(J{-}i{+}1)(i{-}1)(i{-}1{+}\lambda)$, and $r_J$ by
(\ref{gensol}) with $n=J$ and
$b_i=16(J{-}i{+}1)(i{-}1)(J{-}i{+}1{-}\lambda)$.
Thus the two differ by the substitution $b_i\to b_{J-i}$, and since
(\ref{gensol}) for $n=J$ is itself symmetrical under this mapping, the
lemma is proved.

\smallskip

\noindent
{\bf Lemma 2:} If $\lambda<1-J$, then all zeros of $P_J(E,\lambda)$
are real and distinct.\\[2pt]
{\bf Proof:}
The given values of $\lambda$ correspond to the
points on the QES line $\alpha=4J+2\lambda$ which lie in the region
$\alpha<4+2|\lambda|$. The reality result (\ref{rres}) then
implies that the spectrum of the eigenproblem (\ref{PT3}), which 
includes the QES sector described by the zeros of $P_J(E,\lambda)$,
is real. These zeros must therefore all be real.
To show that the zeros are simple, we use the fifth spectral
equivalence from \cite{DDT3} to map our problem on to one which, for
the given range of $\lambda$, is hermitian. Converted into the current
coordinates, this equivalence states that the spectrum of (\ref{PT3})
is the same as that of the following radial problem for 
functions defined on
$\RR^+$ and decaying at $x\to +\infty$:
\eq
\Bigl[-\frac{d^2}{dx^2}+ x^{6}
+\alpha'  x^{2}+ \frac{\lambda'^{\,2}-\frac{1}{4}}{x^2}
\Bigr]\phi(x)=E\,\phi(x)\, ,
{}~~~\phi(x)|_{x\to 0} \sim x^{\lambda'+1/2}\,,
\label{herm1}
\en
where $\lambda'$ and
$\alpha'$ are related to $\lambda$ and $\alpha$ by
\eq
\left(
\begin{matrix}2\lambda'\\ \alpha'\end{matrix}
\right)
=
\frac{1}{2}
\left(
\begin{matrix} -1&-1\\3&-1\end{matrix}
\right)
\left(
\begin{matrix}2\lambda\\ \alpha\end{matrix}
\right).
\label{dualrel}
\en
This equivalence is illustrated in figure \ref{duals};
the dual problem on the right is hermitian for
$\lambda'+1/2>-1/2$, which translates into $2\lambda+\alpha<4$ on the
left. 
The QES line under discussion
is $\alpha=4J+2\lambda$; it maps onto the diagonally-oriented QES
line $\alpha'=-4J-2\lambda'$ 
on the right-hand diagram,
 and is in the hermitian region of that plane for
$\lambda<1-J$. Since all
eigenvalues of the hermitian spectral problem are distinct,
so must be the zeros in $E$
of $P_J(E,\lambda)$, for all $\lambda<1-J$. 
(It is worth remarking that the 
standard proof of the simplicity of the eigenvalues for the
hermitian problem uses essentially the same steps as 
lead to (\ref{psipsiT}), together with the fact
that for hermitian problems the eigenfunctions can be taken entirely
real, so that the LHS of (\ref{psipsiT}) never vanishes.)

\smallskip

\[
\begin{array}{ccc}
\makebox{\includegraphics[width=0.43\linewidth]{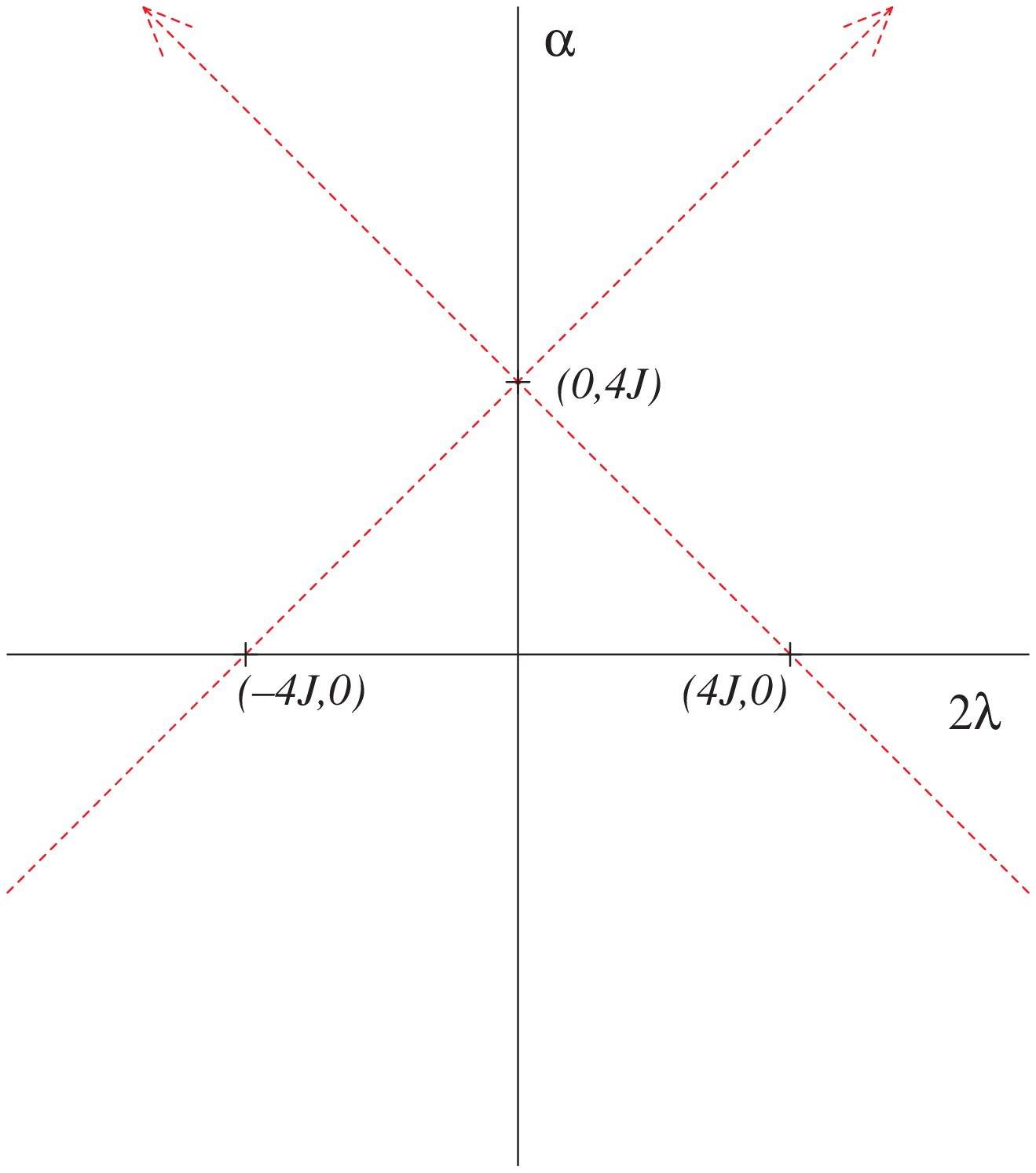}}&
\raisebox{99pt}{~~~$\leftrightarrow$~~~}&
\makebox{\includegraphics[width=0.43\linewidth]{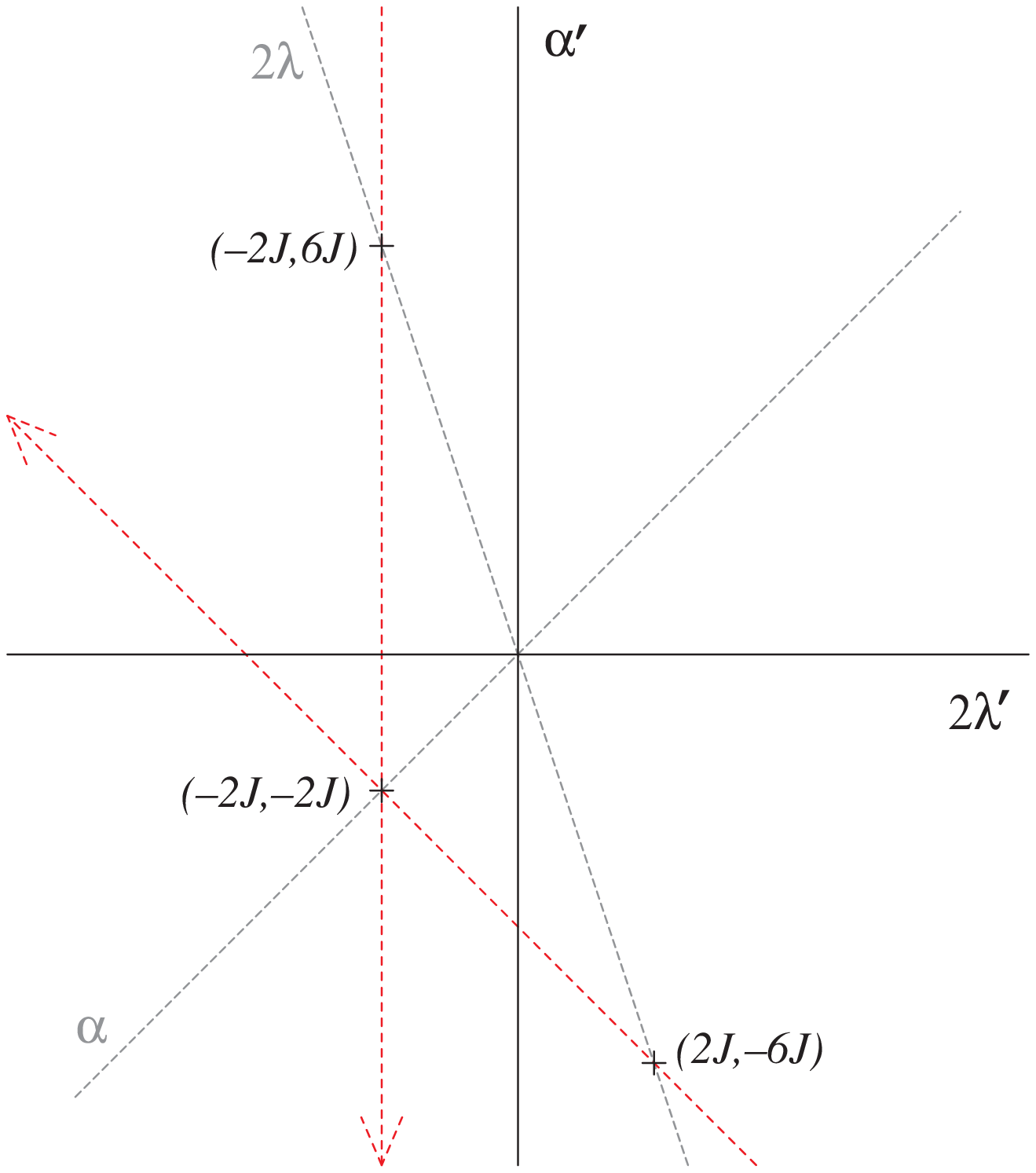}}
\\[11pt]
\multicolumn{3}{c}{
\parbox{0.9\linewidth}{
\small Figure \protect\ref{duals}:
The QES lines $\alpha=4J+2\lambda$ and
$\alpha=4J-2\lambda$ in the $(2\lambda,\alpha)$ plane, and their 
images in the ($2\lambda',\alpha')$
plane under the mapping (\ref{dualrel}). On the right,
the images of the $2\lambda$
and $\alpha$ axes are also shown. The left-hand
diagram corresponds to lateral boundary conditions, while those for
the diagram on the right are radial, and 
hermitian when $2\lambda'>-2$. 
}}
\end{array}
\]
\refstepcounter{figure}
\label{duals}

\noindent
{\em Aside:}\/ To make the proof of lemma 2 self-contained, it is
possible to show directly that $P_J(E,\lambda)$, as defined
for the non-hermitian problem (\ref{PT3}), is equal to the standard
Bender-Dunne polynomial for the
Hermitian problem (\ref{herm1}). QES levels for (\ref{herm1}) occur
when $J'\equiv -(\alpha'+2\lambda')/4$ is a positive integer, or in
other words when $\alpha'=-4J'-2\lambda'$. 
Defining $p'_n(E,\lambda',J')$
to satisfy the recursion
\eq
p'_n=Ep'_{n-1}+16(n-1)(n-J'-1)(n+\lambda'-1)p'_{n-2}\,,
\label{pprec}
\en
with $p'_0=1$, the QES levels are
given by the zeros of the polynomial 
$P'_{J'}(E,\lambda')\equiv p'_{J'}(E,\lambda',J')$
(see \cite{Bender:1995rh} for details). If $\lambda'$ and $\alpha'$
are given in terms of $\lambda$ and $\alpha$ by (\ref{dualrel}), then
(\ref{pprec}) becomes 
\eq
p'_n=Ep'_{n-1}+16(n-1)(n-J-1)(n-J-\lambda-1)p'_{n-2}
\en
where $J=(\alpha-2\lambda)/4=J'$. Since this recursion is, up to a
swap $E\to -E$, the same as (\ref{Qn}), it follows from the proof of
lemma 1 above that $P'_J=P_J$,
as claimed.

\smallskip

\noindent
{\bf Lemma 3:} If $\lambda>-1$, then all zeros of $P_J(E,\lambda)$
are purely imaginary (or zero) and distinct.\\[2pt]
{\bf Proof:} This follows from lemmas 1 and 2.

\smallskip

\noindent
{\bf Lemma 4:} On the infinite segment $\lambda>-1$
of the QES line $\alpha=4J+2\lambda$, 
the eigenproblem (\ref{PT3}) has exactly $J$ distinct
non-real (in fact purely imaginary) eigenvalues for $J$ even, and
$J-1$ non-real (and purely imaginary) eigenvalues and $1$ zero 
eigenvalue for $J$ odd. By the $\lambda\to -\lambda$ symmetry of the
problem, the same holds for the $\lambda<1$ segment of the
$\alpha=4J-2\lambda$ QES line.\\[2pt]
{\bf Proof:} First recall from \cite{Dorey:2001hi} that on QES
lines, the non-QES sector of the spectrum is entirely real. The 
result then follows on combining lemma 3 with the fact that
the degree $J$ polynomial
$P_J(E,\lambda)$ is a function of $E^2$ for $J$ even,
and $E$ times a function of $E^2$ for $J$ odd.

\medskip

\noindent
The results so far show that on the QES lines $\alpha=4J+2\lambda$
the spectrum is entirely real for $\lambda<-J+1$, and has exactly
$2[J/2]$ complex eigenvalues for $\lambda>-1$, where $[x]$ denotes the
largest integer less than or equal to $x$. This is illustrated in
figure~\ref{proof}.

\[
\begin{array}{c}
\includegraphics[width=0.65\linewidth]{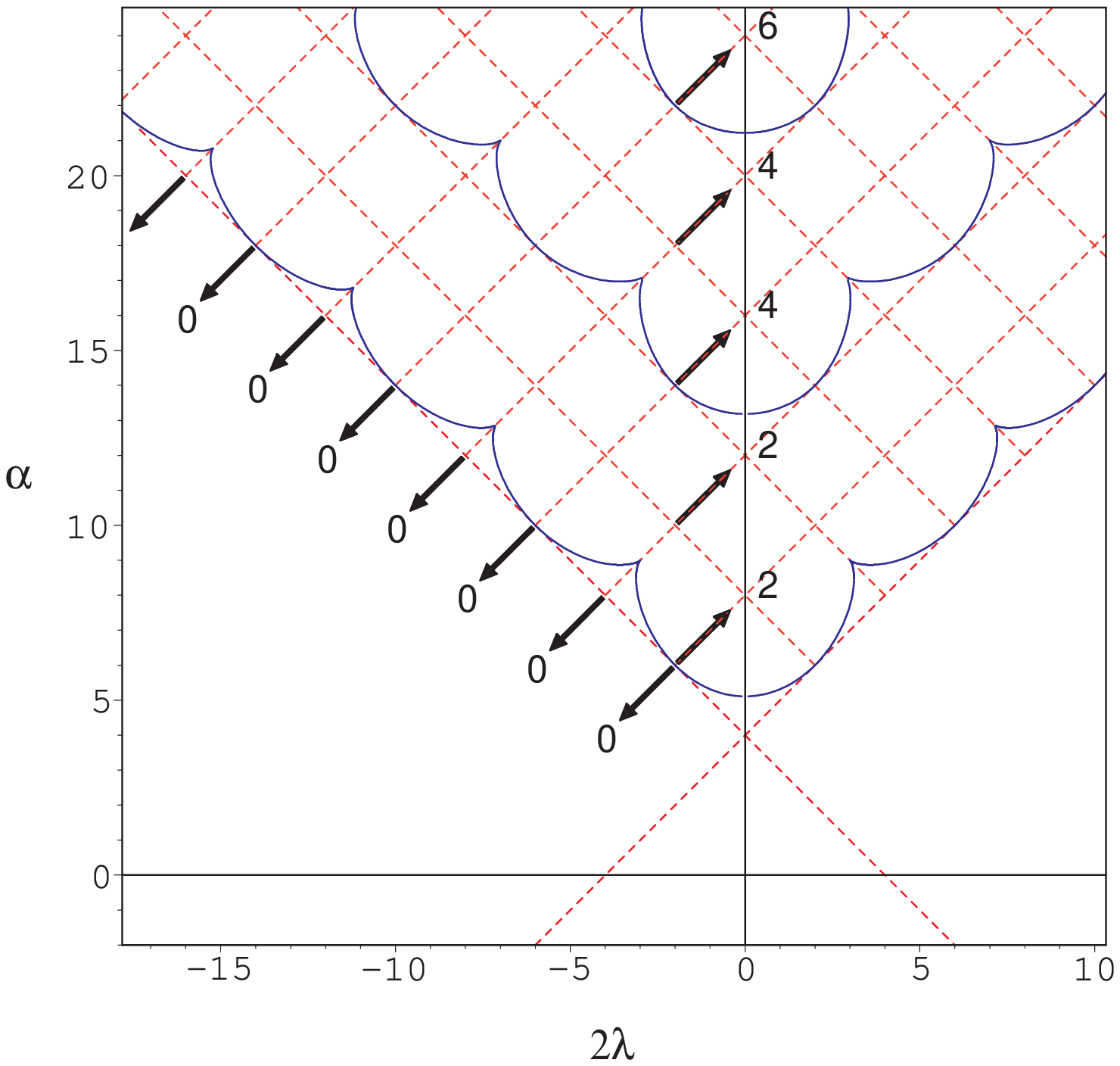}\quad
\\[11pt]
\parbox{0.8\linewidth}{
\small Figure \protect\ref{proof}:
The lines $\alpha=4J\pm 2\lambda$, $J\in \NN$,
of quasi-exact solvability on the $(2\lambda,\alpha)$ plane.
The arrows
indicate the (open) segments of the lines $\alpha=4J+2\lambda$ on 
which the precise numbers of non-real eigenvalues are determined by 
lemmas 2 and~4. Note that these results are consistent with the
locations of the numerically-obtained
curved cusped lines (blue online), across each of which the
number of non-real eigenvalues increases by two.
For $J\neq 1$, only those parts of
the QES lines which lie in the zone of possible unreality 
$\alpha > 4+2|\lambda|$ are shown. The lines with
$J$ odd coincide with the 
protected zero-energy level lines  shown in figures
\ref{scan}, \ref{fullscan} and \ref{cusplocations}.
}
\end{array}
\]
\refstepcounter{figure}
\label{proof}

The situation in the remaining intervals
$-J+1\le \lambda\le -1$ is clarified by lemmas
5 and 6.

\medskip

\noindent
{\bf Lemma 5:} At the points $\lambda=-J+n$, $n=2\dots J{-}1$ on 
the QES line $\alpha=4J+2\lambda$, the problem has exactly 
$2[n/2]$ distinct non-real eigenvalues.\\[2pt]
{\bf Proof:} In addition to being on the line $\alpha=4J+2\lambda$,
the given points lie on the lines $\alpha=4n-2\lambda$ in a region
where lemma 4 applies.

\medskip

\noindent
{\bf Lemma 6:} On each segment $-J+2m-1\le\lambda\le -J+2m$ of the QES
line $\alpha=4J+2\lambda$, where $J\ge 2$ and
$m=1,2 \dots [(J{+}1)/2]-1$, there is 
at least one point where the eigenproblem  has an exceptional point 
with eigenvalue zero. The same is true of the segments
$J-2m\le\lambda\le J-2m+1$,
$m=1,2 \dots [(J{+}1)/2]-1$ of the QES line $\alpha=4J-2\lambda$.\\[2pt]
{\bf Proof:} By lemma 5, when $\lambda=-J+2m-1$ the number of
non-real eigenvalues is $2m-2$, while when $\lambda=-J+2m$ it is $2m$.
The number of non-real eigenvalues thus changes by 
two as $\lambda$ moves from $-J+2m-1$ to $-J+2m$. 
By the $\PT$ symmetry of the problem,  non-real eigenvalues always
occur in complex-conjugate
pairs; combining this with the $E\to -E$ symmetry of 
the QES
sector, any non-real eigenvalues created or destroyed
away from $E=0$ must appear 
in quartets. So to change the number of non-real eigenvalues by 
two, at least one pair must be created or destroyed
at zero, and
hence there must be at least one exceptional point with eigenvalue
zero in each interval $-J+2m-1\le\lambda\le -J+2m$. The final
statement of the lemma then follows from the $\lambda\to -\lambda$
symmetry of the problem. 

\medskip

\noindent
{\bf Lemma 7:} For $M=3$, the zeros of the polynomials $Q_n(\lambda)$,
defined in equation (\ref{genQ}) above, are all real and simple, with
one in each interval $\lambda\in [2m{-}1,2m]$, $m=1\dots n$.\\[2pt]
{\bf Proof:} 
Specialising the discussion of subsection \ref{locself} to $M=3$,
$Q_n(\lambda)$ has  a real zero at every point on the line
$\alpha=4(2n{+}1)+2\lambda$, $\lambda\in \RR$ where the eigenproblem
has an exceptional point with eigenvalue zero.
Combining this with lemma 6 taken
at $J=2n+1$, $Q_n(\lambda)$ has at least one real zero in each
interval $[2m{-}1,2m]$, $m=1\dots n$\,; but since
$Q_n(\lambda)$ is a polynomial of degree $n$ this must exhaust all of
its zeros, which must therefore also all occur singly.

\medskip

Lemmas 6 and 7 show that on the lines $\alpha=4J\pm 2\lambda$ with
$J=2n+1$, there are $n$ odd-order exceptional points with eigenvalue
zero.  If all of these have the lowest possible degree, that is three,
then at each a
pair of complex eigenvalues is created, and would account precisely
for the total number of complex QES levels which must appear as the
QES line is traversed. In principle, one could imagine more
complicated scenarios where further pairs of complex levels appear at
these exceptional points and then annihilate with each other later,
but our numerical results 
show no evidence of such behaviour and we shall proceed
on the assumption that it does not occur. If we further assume that
the triply-exceptional points are isolated in the sense that there are
no other triply-exceptional points in their immediate neighbourhoods in
the $(2\lambda,\alpha)$ plane, then these points must be occurring
where two lines of double degeneracy meet at a cusp, as
illustrated in figure~\ref{surf} below. Thus, subject to
the two assumptions just mentioned, we have proved that for $M=3$
the cusps in the exceptional lines
do indeed lie on the lines of protected zero-energy levels,
as conjectured earlier.

\medskip

\[ 
\begin{array}{c}
\!\!\!\includegraphics[width=0.42\linewidth]{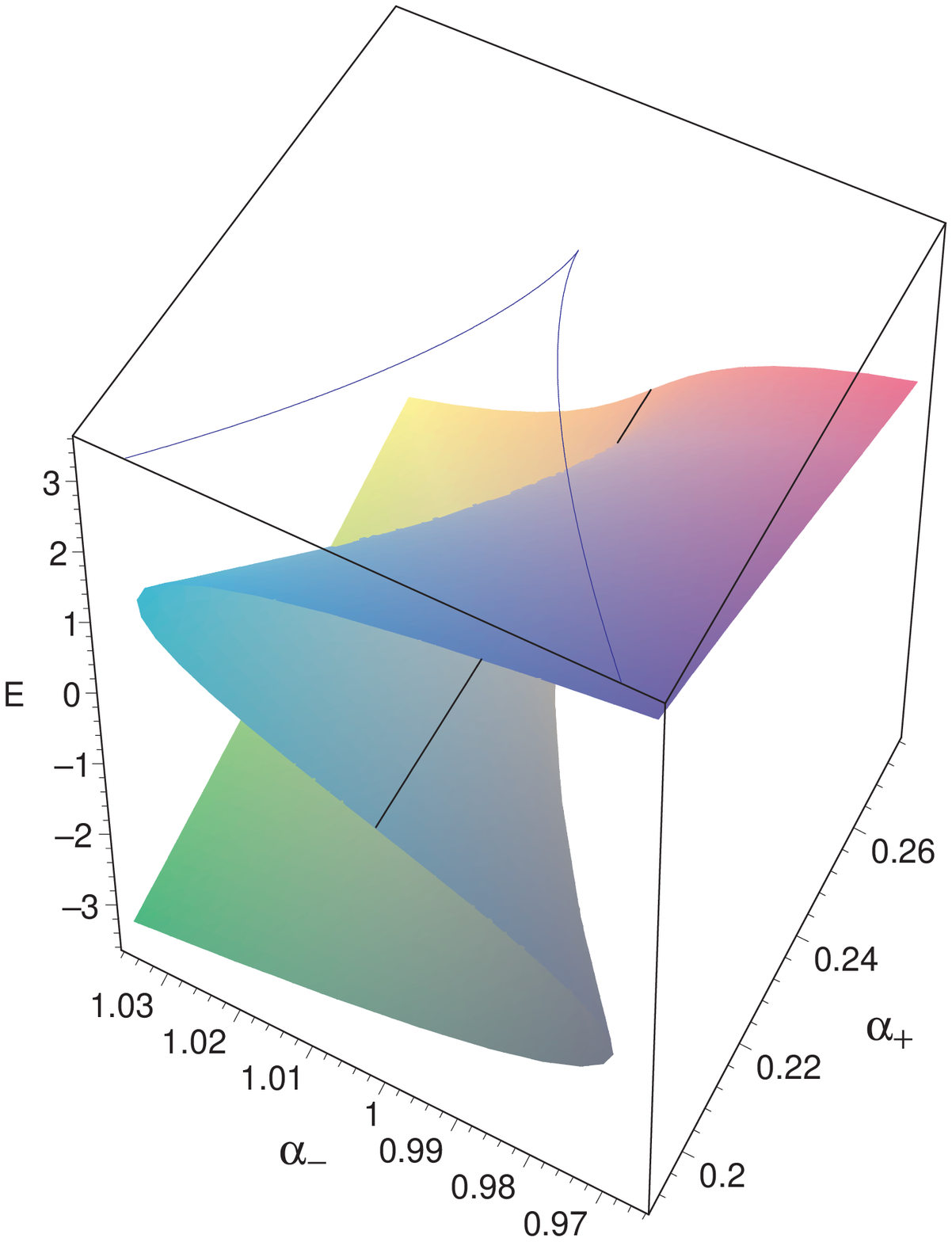}
\\[11pt]
\parbox{0.6\linewidth}{
{\small Figure \ref{surf}: 
The behaviour of the
energy level surface  $E(\alpha_+,\alpha_-)$ in the vicinity of a cubic
exceptional point.
}}
\end{array}
\]
\refstepcounter{figure}
\label{surf}

The above results capture the $[J/2]$
transitions to complex levels that occur as
the QES line $\alpha=4J+2\lambda$ is traversed, at each of which a pair
of complex levels is created.
However this does not necessarily exhaust all of the
exceptional points on the corresponding QES line -- indeed, for $J$
odd (\ref{quadeps}) shows that there are infinitely-many more
exceptional points on the line $\alpha=4J+2\lambda$, 
at $(2\lambda,\alpha)=(4m{-}2,4J{+}4m{-}2)$, $m\in\NN$. 
Examination of figure \ref{proof} reveals
that at each of these points an exceptional line touches, but does not
cross, the
QES line, so that the exceptional point does not cause the creation 
of further complex levels while motion is restricted to the QES line.
Once the QES line is left, the $J$ QES levels start to mix with the 
non-QES sector, and further complex levels can be formed. 
(In fact, if the point $(4m{-}2,4J{+}4m{-}2)$
on the QES line $\alpha=4J+2\lambda$ is left in a direction
perpendicular to that line, a pair of complex levels is created
in a {\em different}\/ QES sector, that
for the QES line $\alpha=4(J{+}2m{-}1)-2\lambda$.) This general
picture, and
also the claimed isolation of the triply-exceptional points as
illustrated in figure \ref{surf}, will be supported by some 
perturbative calculations in the next section.

\medskip

One would also like to be able to rule out the more exotic
scenarios for the behaviour of 
levels in the QES sector, mentioned in the
discussion following lemma 7. We do not have a rigorous argument for
this, but we do have extensive numerical, and some analytical,
evidence for the following conjecture which, if true, would
eliminate such possibilities:

\medskip

\noindent
{\bf Conjecture:} For all $\lambda$, the squared zeros (in $E$)
of the
Bender-Dunne polynomials $P_J(E,\lambda)$ are real, and, apart 
from the zero at $E=0$ when $J$ is odd, they
are monotonically-decreasing functions of $\lambda$.

\medskip

\noindent
Immediate consequences of a proof of
the conjecture would be that QES levels, once
complex, remain so, and that the only way that QES levels can become
complex is via $E=0$. In turn this would prove that the zeros of
the polynomials $Q_n(\lambda)$ do indeed
correspond to triply-degenerate points
in the spectrum, since one can easily rule out the presence of zero
eigenvalues in the non-QES sector at the relevant points.

The conjecture is similar in spirit to the Feynman-Hellman theorem,
but the eigenproblem here is not hermitian, and this invalidates any
variant of the
standard proof. Note also that the conjecture is certainly false for
the non-QES part of the spectrum, the (un-squared) levels of which can
pass through zero while remaining real.
As a sample of our numerical checks,
figures \ref{Q20} and \ref{Q21}
 show the squared QES levels for
$J=20$ and $J=21$. Apart from the $E^2=0$ line on figure \ref{Q21}, 
all the
curves are monotonic; given this, the fact that they all pass through 
zero between $\lambda=1-J$ and
$\lambda=-1$ follows from lemmas 2 and 3 above.

\[ 
\begin{array}{c}\textsc{}
\hskip -30pt
\includegraphics[height=0.46\linewidth]{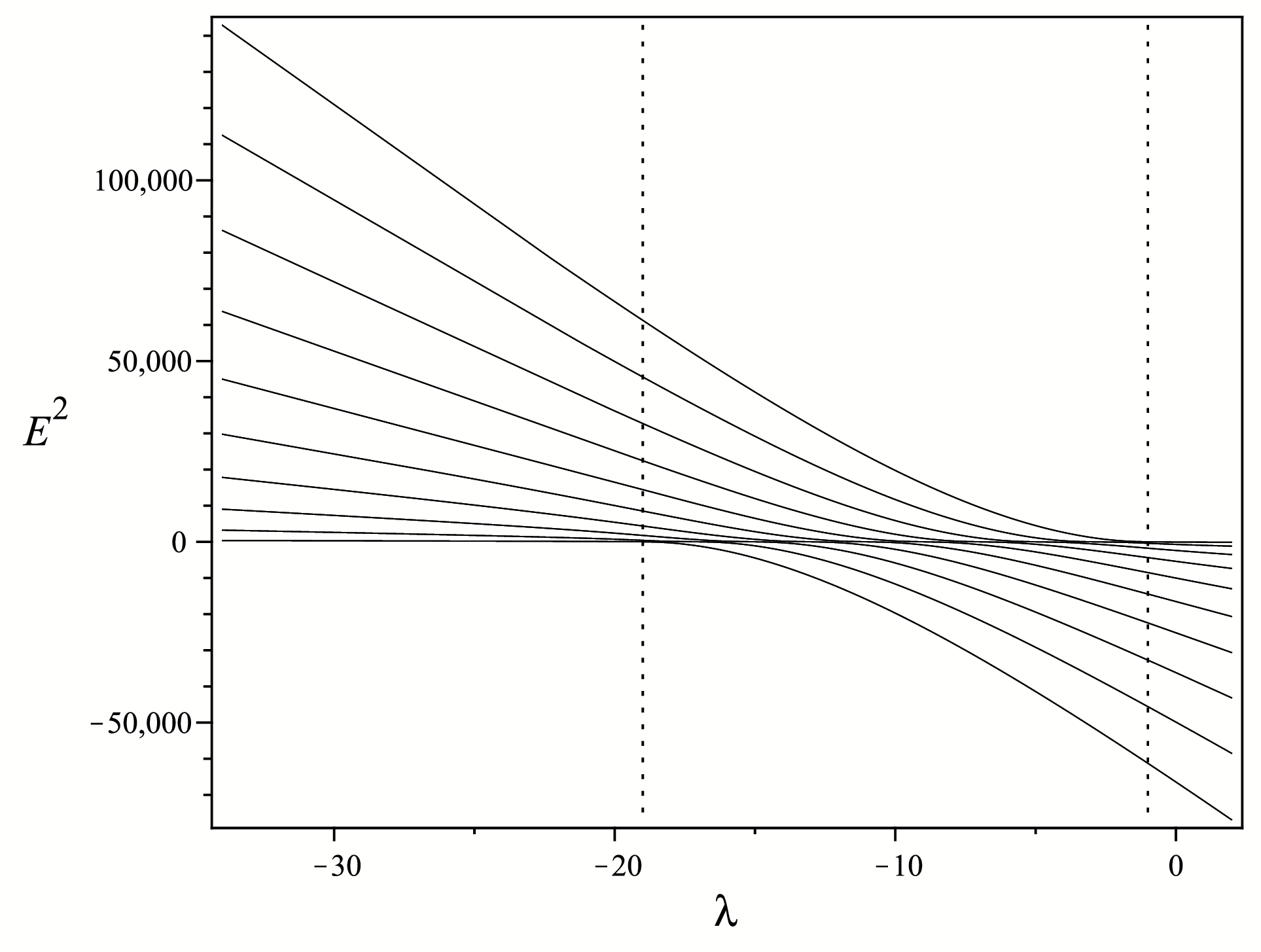}
\\[7pt]
\parbox{0.85\linewidth}{
{\small Figure \ref{Q20}: Squared QES
levels for $J=20$. The dotted vertical lines
are at $\lambda=1-J$ and $\lambda=-1$\,; all transitions from real to
imaginary eigenvalues occur for $1-J\le\lambda\le -1$.
}}
\end{array}
\]
\refstepcounter{figure}
\label{Q20}
\[ 
\begin{array}{c}
\hskip -30pt
\includegraphics[height=0.46\linewidth]{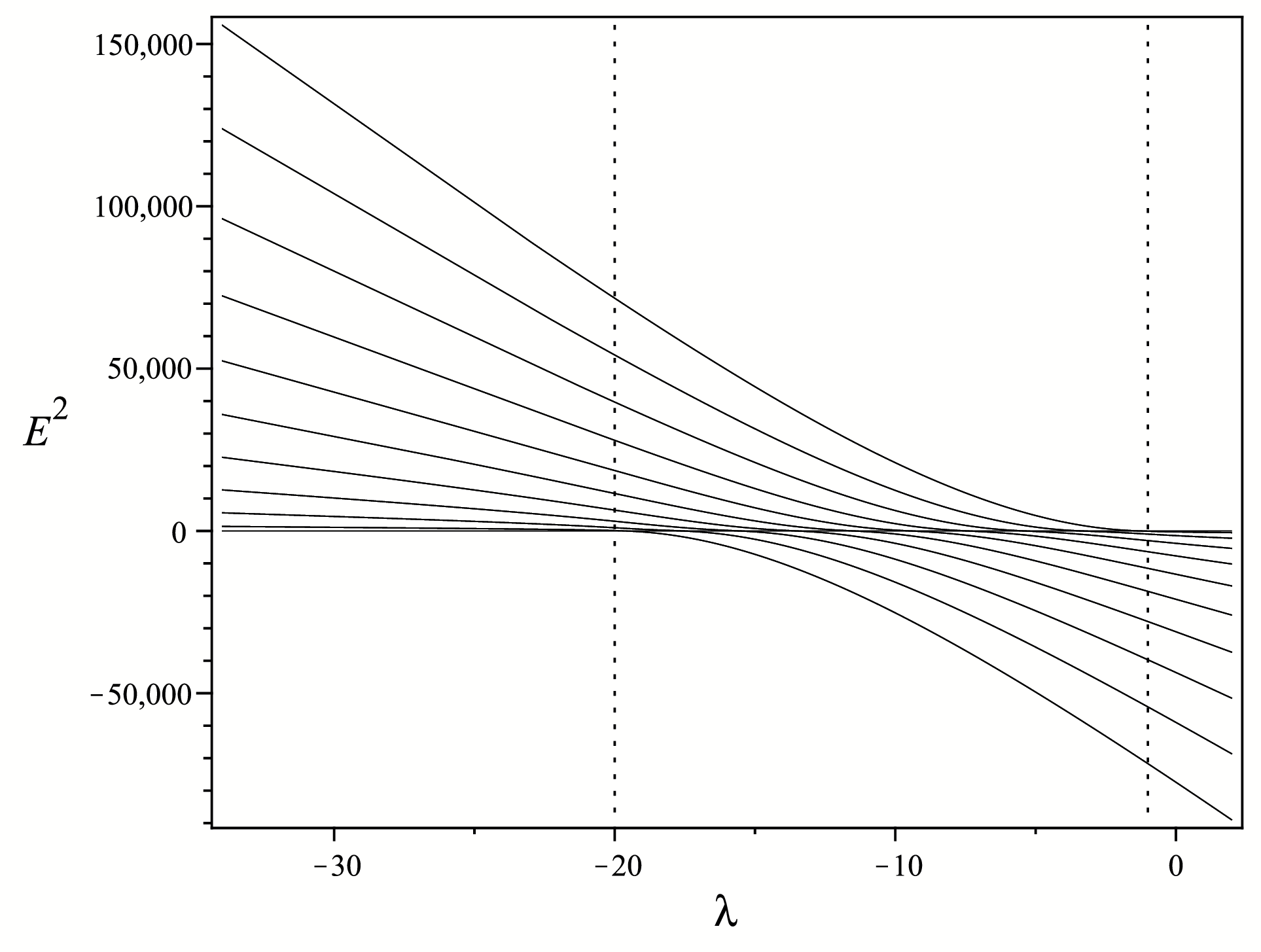}
\\[7pt]
\parbox{0.85\linewidth}{
{\small Figure \ref{Q21}: As figure \ref{Q20}, but for $J=21$.
Again, all transitions to imaginary eigenvalues occur for
$1-J\le\lambda\le -1$.
}}
\end{array}
\]
\refstepcounter{figure}
\label{Q21}

\medskip

\noindent
The curves shown on figures \ref{Q20} and \ref{Q21}
appear to asymptote to linear
functions of $\lambda$ as $\lambda\to\pm\infty$. This turns out to be
the case, as will be shown below, where the slopes of these functions
will also be found exactly.
Figures \ref{dQ20} and \ref{dQ21} illustrate these results, and further
support the monotonicity conjecture, by plotting the derivatives of
the squared QES levels, again for $J=20$ and $J=21$. 

\medskip

\[ 
\begin{array}{c}
\hskip -40pt
\includegraphics[height=0.46\linewidth]{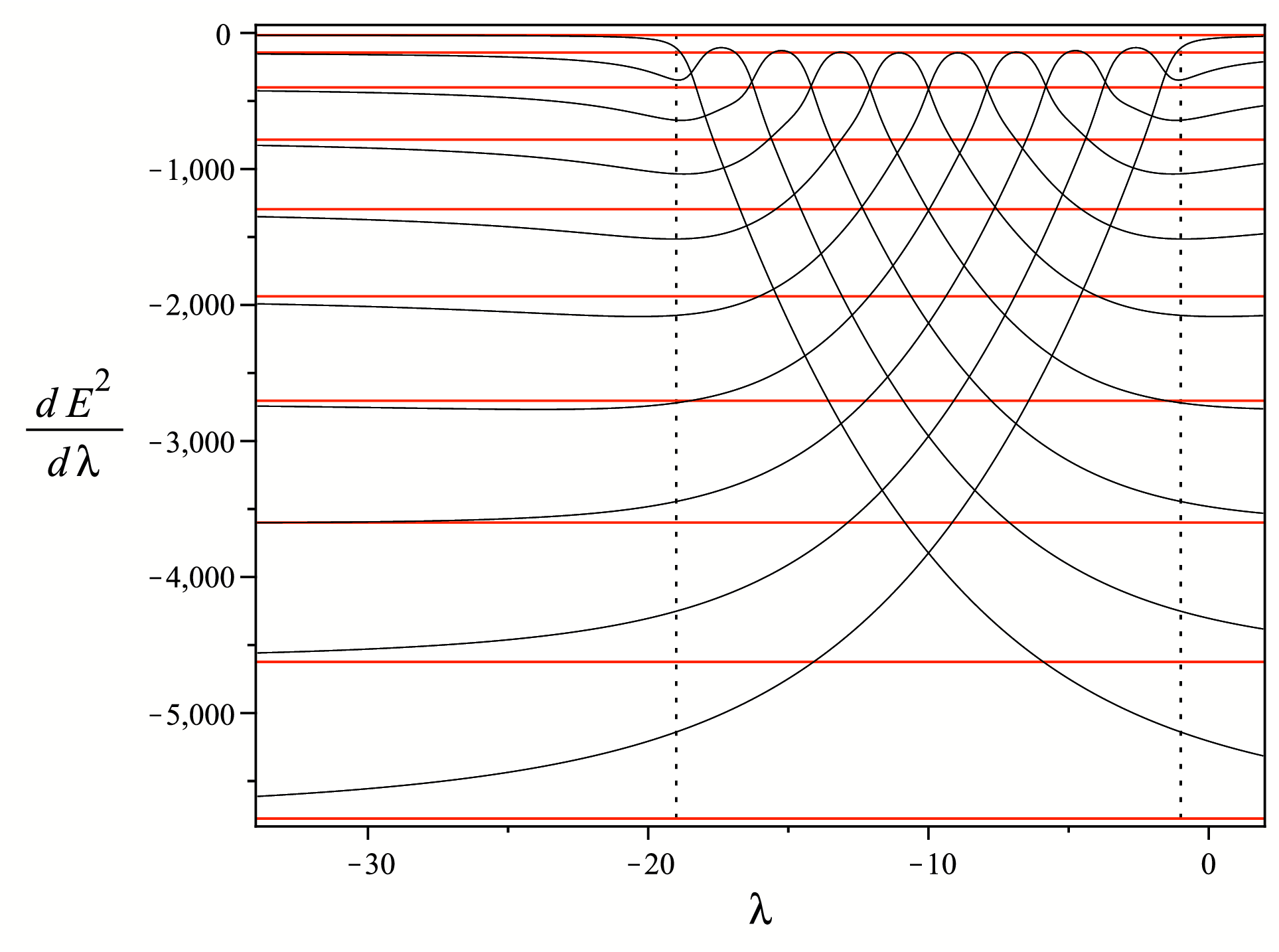}
\\[7pt]
\parbox{0.85\linewidth}{
{\small Figure \ref{dQ20}: Derivatives of squared QES
levels for $J=20$. The straight horizontal lines (red online) 
show the predicted asymptotic values for these derivatives, 
which are everywhere negative.
The symmetry of the plot about $\lambda=J/2$ follows from the
reflection symmetry proved in lemma 1.
}}
\end{array}
\]
\refstepcounter{figure}
\label{dQ20}
\[ 
\begin{array}{c}
\hskip -40pt
\includegraphics[height=0.46\linewidth]{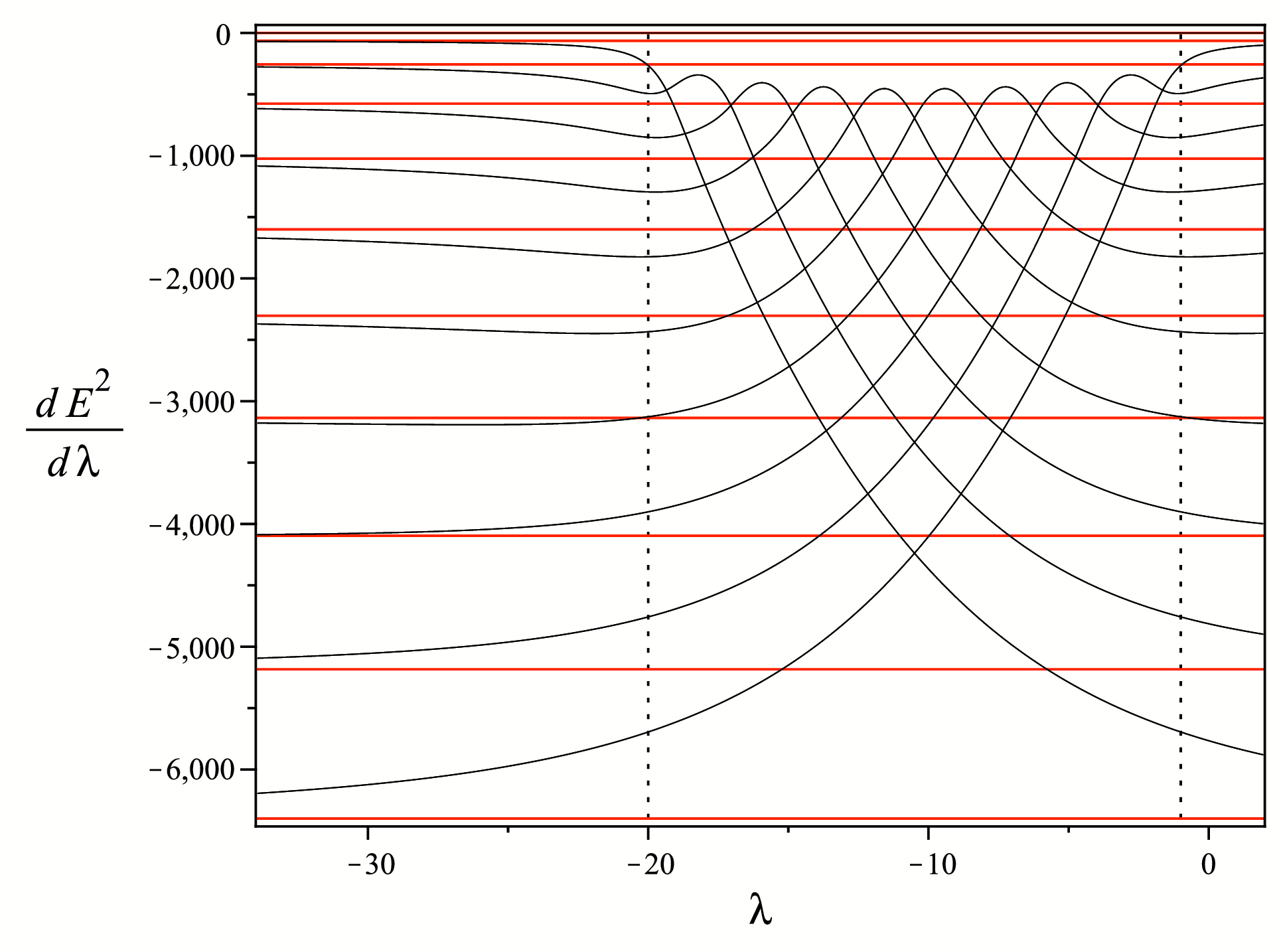}
\\[7pt]
\parbox{0.85\linewidth}{
{\small Figure \ref{dQ21}: As figure \ref{dQ20}, but for $J=21$. All
but one of the plotted functions are negative, the exception
corresponding to the level at $E=0$.
}}
\end{array}
\]
\refstepcounter{figure}
\label{dQ21}

\medskip

To treat the large-$|\lambda|$ behaviour of the QES levels
analytically, a first approach is to return to the Bender-Dunne
recursion relation (\ref{Pn}). In the limit $|\lambda|\gg J$, and
keeping $n\le J$, this simplifies to
\eq 
p_n=-Ep_{n-1}+16\lambda (J-n+1)(n-1)p_{n-2}\,,\quad p_0=1\,.
\label{Pna}
\en
Solving for low-lying values of $J$, a remarkable 
simplification occurs precisely at $n=J$, where the asymptotic
Bender-Dunne polynomials $P^{\rm asymp}_J$ are found. For $J$ even,
\eq
P_J^{\rm asymp}(E)=\prod_{k=1}^{J/2}\left(E^2+16(2k{-}1)^2\lambda\right)\,,
\label{oddasympt}
\en
while for $J$ odd,
\eq
P_J^{\rm asymp}(E)=-E\prod_{k=1}^{(J{-}1)/2}\left(E^2+16(2k)^2\lambda\right)\,.
\label{evenasympt}
\en
Hence the squared QES levels are indeed linear functions of $\lambda$
in this limit, with slopes which are negative, and
proportional to the squares of
odd or even integers.
Rather than prove these formulae directly from the asymptotic
Bender-Dunne recursion relation, we will use the same spectral
equivalence as employed in the proof of lemma 2 above. The polynomial
$P_J(E,\lambda)$ encodes in its zeros the QES levels along the 
line $\alpha=4J+2\lambda$; by the
reflection symmetry (\ref{BDsymmetry}) it will suffice to 
consider the asymptotic behaviour of the levels in just one direction,
which we choose to be $\lambda\to -\infty$. Then, instead of applying
the mapping (\ref{dualrel}) immediately, we precede it by the trivial
symmetry $(2\lambda,\alpha)\to (-2\lambda,\alpha)$ of the $\PT$-symmetric
problems, which flips between the two QES lines shown on the left-hand
diagram of figure~\ref{duals}. The
line $(2\lambda,4J+2\lambda)$, $\lambda\in\RR$, is now
mapped to $(-2J,-4\lambda)$ on the $(2\lambda',\alpha')$ plane, the
vertical line on the right-hand diagram of figure~\ref{duals}. The
limit $\lambda\to -\infty$ 
is thus mapped to $\alpha'\to +\infty$, $2\lambda'=-2J$ in the 
spectrally-equivalent lateral problem (\ref{herm1}). For the QES
sector, we are only interested in the $J$ lowest-lying levels, where
$J$ remains fixed as $\alpha'=-4\lambda\to +\infty$. In this limit the
quadratic term in (\ref{herm1}) comes to dominate and the problem
reduces to the (scaled) radial simple harmonic oscillator
\eq
\Bigl[-\frac{d^2}{dx^2}
-4\lambda  x^{2}+ \frac{J^{\,2}-\frac{1}{4}}{x^2}
\Bigr]\phi(x)=E\,\phi(x)\, ,
{}~~~\phi(x)|_{x\to 0} \sim x^{-J+1/2}\,.
\label{herm2}
\en
Since $J\ge 1$, the boundary conditions at the
origin are irregular. Nevertheless, the problem can be solved exactly for
any value of $J$, the first
$J$ levels being 
\eq
E=\sqrt{-4\lambda}(-2J+4n-2)\,,\qquad n=1,2,\dots J\,.
\en
(Strictly speaking, a resonance when $J$ is an integer means it is
safest to shift $J$ slightly away from integer values for the
calculation, but the final result is unaffected.) It is
straightforward to see that this confirms equations (\ref{oddasympt}) and
(\ref{evenasympt}) above, as desired. (The overall normalisation of
$P^{\rm asympt}_J(E)$ can be fixed by considering the coefficient of
$E^J$.) Note that these results
imply the truth of the monotonicity conjecture in the limits
$|\lambda|\to\infty$ and thus give some supporting evidence for its
general validity. Unfortunately, the regions $|\lambda|\to\infty$ are
not of interest from the point of view of reality properties, since
they are already covered by lemmas 2 and 3 above, and so a full proof
of monotonicity of the squared eigenvalues would still be worthwhile. 

\bigskip

Finally, an alternative way to locate
the cusps corresponding to the collision of levels at $E=0$
is to examine the odd Bender-Dunne polynomials $P_{2m+1}(E,\lambda)$
directly.  These
factorise as $E$ times a polynomial in $E^2$,
and there will be a multiply-degenerate zero-energy level whenever
this polynomial vanishes at $E=0$, or equivalently whenever
\eq
\frac{d}{dE}P_{2m+1}(E,\lambda)\Bigr |_{E=0} =0~.
\label{pe}
\en
For fixed $m$ and $J\equiv 2m+1$, consider the sequence of polynomials 
$p_{2n+1}(E,\lambda,J)$, $n=0\dots m$, and define 
\eq
q_n(\lambda,m)=\frac{-1~}{2^{7n}n!}\,\frac{d}{dE}p_{2n+1}(E,\lambda,2m{+}1)\bigr|_{E=0}~. 
\en
A consideration of the Bender-Dunne recurrence (\ref{Pn}) and its
derivative at $E=0$ shows that $q_n(\lambda,m)$ satisfies the first order
recurrence
\eq
q_n=
(m-n+\fract{1}{2})(n+\fract{1}{2}\lambda)q_{n-1} +
\left(\begin{matrix}m\\ n\end{matrix}\right)
\prod^n_{k=1}(k-\fract{1}{2})(k-\fract{1}{2}+\fract{1}{2}\lambda)\,,
\label{per}
\en
with initial condition $q_0=1$. Anticipating the final result in our
notation, we set
\eq
Q_m(\lambda)=q_m(\lambda,m)
\label{Qq}
\en
so that $Q_m(\lambda)$
is a polynomial in $\lambda$ of degree $m$, and its zeros are
the points identified by (\ref{pe}). For example: 
\bea
Q_1(\lambda)&=&\frac{1}{4}(3+ 2\lambda)\\
Q_2(\lambda)&=& \frac{1}{16}(41+ 40\lambda+8\lambda^2)\\
Q_3(\lambda)&=& \frac{3}{64}(7+ 2\lambda)(63+ 56\lambda+8\lambda^2)\,.
\eea
It turns out that the general solution to (\ref{per}) can be expressed
using the hypergeometric function ${}_3F_2$ and Pochhammer
symbols $(x)_k\equiv x(x{+}1)\dots(x{+}k{-}1)$,
$(x)_k=(-1)^k(1{-}k{-}x)_k$\,. For $n=m$ this solution is
\bea
Q_m(\lambda)=
&\!=\!&(1{+}\fract{1}{2}\lambda)_m
(\fract{1}{2})_m\,
 {}_3F_2(-m,\fract{1}{2},\fract{1}{2}{+}\fract{1}{2}\lambda\,;\,
 1{+}\fract{1}{2}\lambda,\fract{1}{2}{-}m\,;\,1)\qquad\nn\\[3pt]
&\!=\!&\sum_{k=0}^m(-1)^k
\left(\begin{matrix}m\\k\end{matrix}\right)
(\fract{1}{2}{-}k)_m(\fract{1}{2}{+}\fract{1}{2}\lambda)_k
(1{+}\fract{1}{2}\lambda+k)_{m-k}~.
\label{recsol}
\eea
This is the $M=3$ case of the general formula (\ref{genQ}), 
here derived by a completely different route.
Since terms being summed in (\ref{recsol}) are invariant
up to a factor of $(-1)^m$
under the simultaneous exchange $k\to m{-}k$, $\lambda\to
-2m{-}1{-}\lambda$, the $M=3$ 
polynomial $Q_m(\lambda)$ is invariant up to a sign under $\lambda\to
-2m{-}1{-}\lambda$ and its zeros are symmetrically distributed about
$\lambda=-m-1/2$. This reflects the more general result (\ref{BDsymmetry}).
For $m$ odd, this means that
$Q_m(-m-1/2)= 0$.

Returning to the $\alpha_{\pm}$ coordinates, if $J=2m{+}1$ and $m\in\NN$
then the QES line $\alpha=4J+2\lambda$ corresponds to $(\alpha_+,\alpha_-)
=(m{+}\lambda/2,m)$, and the $m$ cusps on this line occur at the
zeros of $Q_m(\lambda)$, while for $\alpha=4J-2\lambda$
the cusps lie on the line $(\alpha_+,\alpha_-)=(m,m{-}\lambda/2)$ with
$\lambda$ a zero of $Q_m(-\lambda)$. The first few cases from this
second set are given in table~\ref{tabcusp}; to within our numerical
accuracy, they match the cusp positions shown in
figure~\ref{fullscan}. Notice that the symmetrical distribution of
the zeros of $Q_m(\lambda)$ mentioned at the end of the last
paragraph implies a relationship between the locations of pairs of a
priori unrelated cusps and  gives  a simple  formula for the
remaining `unpaired' cusps: $(\alpha_+,\alpha_-)=(m,m/2{-}1/4)$ and
$(\alpha_+,\alpha_-)=(m/2{-}1/4,m)$ for all odd $m\in\NN$.

\begin{tab}[tb]
\begin{center}
\begin{tabular}{c|c}
$\alpha_{+}$ & $\alpha_{-}$  \\[2pt]  \hline
&\\[-9pt]
$1$ & $\frac{1}{4}$ \\[3pt]
$2$ & $ \frac{3}{4}\pm\frac{3\sqrt{2}}{8}$ \\[3pt]
$3$ &  $\frac{5}{4} , \frac{5}{4} \pm \frac{\sqrt{70}}{8} $ \\[3pt]
$4$ & $ \frac{7}{4} \pm \frac{\sqrt{86+5\sqrt{190}}   } {8},
\frac{7}{4} \pm \frac{\sqrt{86-5\sqrt{190}}   } {8}  $  \\[4pt]
$5$ & $ \frac{9}{4}, \frac{9}{4} \pm \frac{\sqrt{170+7\sqrt{214}}   }
{8},
\frac{9}{4} \pm \frac{\sqrt{170-7\sqrt{214}}   } {8}  $
\end{tabular}
\end{center}
\caption{\footnotesize Location of some of the cusps in the
   $(\alpha_{+} ,
     \alpha_{-})$-plane  for $M=3$.}
\label{tabcusp}
\end{tab}

\resection{Jordan blocks for $M=3$}
\subsection{The Jordan block at a quadratic exceptional point}
\label{secquad}
We now investigate the exceptional points and their neighbourhoods in
more detail, beginning
with a quadratic example.
The first step is to find the
Hamiltonian $H_0$ at the
exceptional point, restricted to the two-dimensional
space of states which
merge at that point. This will have a Jordan block form.
We then perturb about this point by writing
the full Hamiltonian, $H$, as $H=H_0+V$, and expand
$H$ using the wavefunctions of $H_0$ as a basis.
Thus we will need to calculate
\eq
H_{mn}=\langle\tilde{\phi}_m |H| \phi_n\rangle
=\langle\tilde{\phi}_m |H_0| \phi_n\rangle +
\langle\tilde{\phi}_m |V| \phi_n\rangle\,,
\en
where
$\{\phi_m,\phi_n\}$
is a basis for the Jordan block form of $H_0$,
and
$\{\tilde\phi_m,\tilde\phi_n\}$
is an appropriate dual basis, so that the functions together form
a part of a biorthogonal system, as discussed in
\cite{Curtright:2005zk} for generic cases and
\cite{Sokolov:2006vj} in the presence of exceptional points.
In the current setting, wavefunctions decay as $|z|\to\infty$
along $i{\cal C}$ and a suitable pairing between functions
$g(z)$ and `dual functions' \cite{Curtright:2005zk} $\tilde f(z)$ is
a rotated version of the $c$-product (\ref{cprod}):
\eq
\langle \tilde f|g\rangle=\int_{i{\cal C}} \tilde f(z) g(z)\,dz~.
\en
Here and below a convenient choice for $i{\cal C}$,
beginning and ending in the (rotated) Stokes sectors $i\CS_{-1}$ and
$i\CS_1$,
will be
\eq
i{\cal C}=-\gamma_{-1}+\gamma_0+\gamma_1
\label{iC}
\en
where $\gamma_{\pm 1}=\{t e^{\pm \pi i/4}, t\in [\varepsilon,\infty)\}$,
$\gamma_0=\{\varepsilon e^{it},t\in [-\pi/4,\pi/4]\}$, and
the small positive number
$\varepsilon$ ensures that any
singularities at $z=0$ are avoided. For later use we note the
following basic integral along the contour $i{\cal C}$,
which holds for arbitrary $a\in\RR$:
\eq
\int_{i{\cal C}}z^a\,e^{z^4/2}dz=
\frac{2^{(a-3)/4}\pi i}{\Gamma(\frac{1}{4}(3{-}a))}\,.
\label{intformula}
\en
This is easily checked, either by analytic continuation from $a>-1$ or
 using the real integral
$\int_{\varepsilon}^{\infty}t^ae^{-t^4/2}dt=
2^{(a-7)/4}\Gamma(\frac{1}{4}(a{+}1),\frac{1}{2}\varepsilon^4)$, where
$\Gamma(a,z)$ is the incomplete gamma function.

As our example we take the exceptional point at
$(2\lambda,\alpha)=(2,6)$, $(\alpha_+,\alpha_-)=(1/2,0)$, shown with
a box on figure~\ref{cusplocations}.
This lies on two lines of quasi-exact solvability,
$(\alpha=4J+2\lambda) |_{J=1}$ and
$(\alpha=4J-2\lambda) |_{J=2}$\,, and we shall focus on
the second of these.
Along this line $\alpha=8-2\lambda$ and the eigenvalue
problem is
\eq
\left(-\frac{d^2}{dz^2}+z^6+(8-2\lambda)z^2+
\frac{\lambda^2-\frac{1}{4}}{z^2}+E\right)\Phi=0~.
\en
Setting $\lambda=1-2\ep$, 
this corresponds to $(\al_+,\al_-)=(1/2,\ep)$ and
the exceptional point, where
two levels merge and the Hamiltonian can be
written in a Jordan block form, occurs at $\ep=0$.
The recursion relation
(\ref{Pn})  for $p_n(E,-\lambda,J)$ becomes
\eq
p_n=-Ep_{n-1}+16(3-n)(n-1)(n+2\ep-2)p_{n-2}
\en
and, as expected, the second term on the RHS vanishes when $n=J+1=3$.
The energy eigenvalues of the two QES levels are
given by the roots of $p_2(E,-\lambda,2)$:
$E_{0,1}=\pm 4i\sqrt{2\ep}=\pm 4\sqrt{-2\ep}$. The
corresponding (unnormalized) eigenvectors are, from (\ref{estate}),
\eq
\Psi_{0,1} =
z^{2\ep-\frac{1}{2}}
\sqrt{2\ep}\,\Bigl(1\mp \frac{iz^2}{\sqrt{2\ep}}\Bigr)
e^{\frac{z^4}{4}}\,,
\en
where 
$\Psi_{0,1}=\frac{\sqrt{2\ep}}{\Gamma(2\ep)}\Phi^{\rm eq.(\ref{estate})}|_{E_{0,1}}$\,.    
At $\ep=0$ these two eigenvectors coincide, and to
see the Jordan form of the Hamiltonian we proceed as in
appendix \ref{jordan}
and construct
\bea
\phi_0 &=& \left.\Psi_0\right|_{\ep=0} = -iz^{3/2}e^{z^4/4} \nn\\
\phi_1 &=& \left.2a\sqrt{\ep}\,\frac{d\Psi_0}{d\ep}\right|_{\ep=0}
= a\sqrt{2}\,z^{-1/2}e^{z^4/4}
\eea
where $a$ is a constant. The Hamiltonian at $\ep=0$ is
\eq
H_0=-\frac{d^2}{dz^2}+z^6+6z^2+\frac{3}{4z^2}
\en
and requiring $\phi_0$ and $\phi_1$ to satisfy the `Jordan
chain' relations $H_0\,\phi_0 = 0$ and $H_0\,\phi_1=\phi_0$
fixes $a=-\frac{i}{4\sqrt{2}}$\,, and shows that the
Hamiltonian has the desired Jordan block form.
However, this basis is not unique: replacing
$\{\phi_0,\phi_1\}$ by
$\{\mu\phi_0,\mu\phi_1+\nu\phi_0\}$
preserves the Jordan chain for any constants $\mu$ and $\nu$.
This freedom can be used to make a convenient choice for our
biorthogonal system. Dropping a common factor of $-i$,
the general Jordan basis is
\bea
\phi_0 &=& \mu z^{3/2}e^{z^4/4} \\
\phi_1 &=&
\left(\fract{1}{4}\mu z^{-1/2}+\nu z^{3/2}\right)
e^{z^4/4}~.
\eea
The integrals $\int_{i{\cal C}}\phi_m\phi_n\,dz$ with
$m,n\in\{0,1\}$ can be evaluated using (\ref{intformula}) and are
\bea
\int_{i{\cal C}}\phi_0\phi_0\,dz
 &=& \mu^2\!\int_{i{\cal C}}z^3\,e^{z^4/2}\,dz = 0 \\
\int_{i{\cal C}}\phi_0\phi_1\,dz
 &=& \int_{i{\cal C}}
\left(\fract{1}{4}\mu^2 z+\mu\nu z^3\right) e^{z^4/2}\,dz
= \frac{i\mu^2}{8}\sqrt{2\pi} \\
\int_{i{\cal C}}\phi_1\phi_1\,dz
 &=& \int_{i{\cal C}}
\left(\fract{1}{16}\mu^2z^{-1}-\fract{1}{2}\mu\nu z+
\nu^2 z^3\right)
e^{z^4/2}\,dz
= \frac{i\mu}{4}\left(
\fract{1}{8}\pi\mu+ \sqrt{2\pi}\nu \right).
\eea
Therefore, if
\eq
i\mu^2=\frac{4\sqrt{2}}{\sqrt{\pi}}~,\quad
\nu=-\frac{\sqrt{\pi}}{8\sqrt{2}}\,\alpha
\en
then $\int_{i{\cal C}}\phi_1\phi_1\,dz=0$
and $\int_{i{\cal C}}\phi_0\phi_1\,dz=1$, allowing us to take the dual
basis to be
$\tilde{\phi}_0=\phi_1$ and $\tilde{\phi}_1=\phi_0$.

The spectrum in the neighbourhood of $H_0$ can now be investigated.
For this a two-parameter family of perturbations is required, and
we take one of these parameters to be $\ep$, and the other, $\eta$,
to be orthogonal to this in the $(\alpha_+,\alpha_-)$ coordinates so
that $(\alpha_+,\alpha_-)=(1/2+\eta,\ep)$ and
\eq
H=H_0+4(\eta+\ep)z^2+4(1+\eta-\ep)(\eta-\ep)z^{-2}\,.
\en
The matrix elements of interest are, in an obvious notation,
\bea
\langle\tilde{\phi}_{0,1}|z^2|\phi_{0,1}\rangle &=&
\left(\begin{matrix}
\frac{\sqrt{2\pi}}{4}&\frac{1}{4}-\frac{\pi}{32}\\[3pt]
-4&\frac{\sqrt{2\pi}}{4}
\end{matrix}\right)
\\
\langle\tilde{\phi}_{0,1}|z^{-2}|\phi_{0,1}\rangle &=&
\left(\begin{matrix}
\frac{\sqrt{2\pi}}{4}&\frac{1}{4}-\frac{3\pi}{32}\\[3pt]
4&\frac{\sqrt{2\pi}}{4}
\end{matrix}\right)
\blank{
\\
\langle\tilde{\phi}_0|z^2|\phi_0\rangle &=&
\langle\tilde{\phi}_1|z^2|\phi_1\rangle
= \frac{\sqrt{2\pi}}{4} \\
\langle\tilde{\phi}_0|z^{-2}|\phi_0\rangle &=&
\langle\tilde{\phi}_1|z^{-2}|\phi_1\rangle
= \frac{\sqrt{2\pi}}{4} \\
\langle\tilde{\phi}_0|z^2|\phi_1\rangle &=&
\frac{1}{4}\left(1-\frac{\pi}{8}\right) \\
\langle\tilde{\phi}_0|z^{-2}|\phi_1\rangle &=&
-\frac{1}{4}\left(1+\frac{3\pi}{2}\right) \\
\langle\tilde{\phi}_1|z^2|\phi_0\rangle &=& -4 \\
\langle\tilde{\phi}_1|z^{-2}|\phi_0\rangle &=& 4\,.
}
\eea
and so the truncated Hamiltonian, to leading order, is
\eq
H_{pert}\approx
\left(\begin{array}{cc}
2\sqrt{2\pi}\eta & 1 \\
16\eta^2-32\ep & 2\sqrt{2\pi}\eta \\
\end{array}\right)\,.
\label{approxH}
\en
The approximate energy levels are thus
the roots of the characteristic
polynomial of this matrix:
\eq
E=2\sqrt{2\pi}\,\eta \pm 4\sqrt{\eta^2-2\ep}\,.
\en
Two special cases deserve comment.
For $\eta=0$ the QES levels $E=\pm 4\sqrt{-2\ep}$ are recovered, as
expected given that these QES levels were used to set up the
approximation scheme in the first place.
More interesting is the fact that for $\ep=0$ the energy levels remain
real as the exceptional point is traversed. (The same phenomenon was
remarked in a finite-dimensional setting in \cite{heiss03}\,.)
This corresponds to the fact that the approximation correctly
predicts the {\em direction}\/
of the line of exceptional points away from the
QES point $(\alpha_+,\alpha_-)=(1/2,0)$. However one should be wary of
trusting the approximation any further --
one might expect that
the {\em curvature}\/ of the line of exceptional points could be
recovered from the line of points where
the discriminant of the characteristic polynomial of
(\ref{approxH}) vanishes, which is $\eta^2-2\ep=0$\,,
or $\al_-=\fract{1}{2}\left(\fract{1}{2}-\al_+\right)^2$.
However, a fit to the numerical eigenvalues of the full equation shows
that the shape of the curve of exceptional points near to $(1/2,0)$ is
rather given by
$\al_-\approx \kappa\left(\fract{1}{2}-\al_+\right)^2$ with
$\kappa\approx 0.78$. Given that this curvature is controlled by
sub-leading effects, this failure should not be too surprising, but it
does highlight the delicacy of perturbation theory about exceptional
points. A more systematic investigation of this issue would be
valuable, but for now we will pass on to an examination of a typical
cubic exceptional point.

\subsection{The Jordan block at a cubic exceptional point}
{}From table~\ref{tabcusp}, the first cubic exceptional points occur
at $(\alpha_+,\alpha_-)= (1,1/4)$ and $(1/4,1)$, on the $J=3$ QES
lines $\al=12-2\lambda$ and $\al=12+2\lambda$.
We focus on the line $\al=12+2\lambda$  and set 
$\lambda=2\ep-3/2$ so that $(\alpha_+,\alpha_-)= (\ep+1/4,1)$ and the
exceptional point occurs at $\ep=0$.  
The
eigenvalue problem 
in terms of $\ep$ is
\eq
\left(-\frac{d^2}{dz^2}+z^6 +(4\ep+9)z^2+
\frac{(2\ep-1)(2\ep-2)}{z^2}+E\right)
\Psi=0
\en
and the recursion relation for $p_n(E,2\ep-3/2,3)$ is
\eq
p_n=-Ep_{n-1}+16(4-n)(n-1)(n+2\ep-5/2)p_{n-2}~.
\en
The roots of $p_3$ give the energy eigenvalues of the
three QES levels: 
$E_0=0$, $E_\pm=\pm 8\sqrt{-2\ep}$. The corresponding eigenstates are
\bea
\Psi_0 &=& e^{z^4/4}z^{2\ep-1}a\left(1+\frac{2z^4}{4\ep+1}\right) \\
\Psi_{\pm} &=& e^{z^4/4}z^{2\ep-1}a\left(1\mp\frac{4\sqrt{-2\ep}z^2}{4\ep-1}
-\frac{2z^4}{4\ep-1}\right) \nn
\eea
where $a$ is some normalisation to be fixed later.
Note that when $\ep=0$ these three QES eigenstates  merge and we have
only one known eigenstate 
at this point, namely
\eq
\left.\Psi_0\right|_{\ep=0}= a\,e^{z^4/4}\left(\frac{1}{z}+2z^3\right).
\en

\subsubsection{The Jordan basis}
If we perturb away from the cubic exceptional point along the QES
line, the Hamiltonian will  correspond
to a toy model matrix of the form
\eq
L(\ep)=
\left(\begin{array}{ccc}
0 & 1 & 0  \\
\ep/2 & 0 & 1\\
0 & \ep/2 & 0 \\
\end{array}\right).
\label{ll}
\en
The method that we used to calculate the Jordan basis for the
quadratic exceptional 
point in section \ref{secquad}, explained for $n \times n$ Jordan blocks 
in appendix~\ref{jordan}, does not apply here.   This is because the
matrix  considered in appendix~\ref{jordan} would correspond to a
perturbation of   the Hamiltonian 
along  a line perpendicular to the QES line, along which  
we do not know the relevant eigenfunctions  analytically.  Instead,
we will have to find the basis functions for (\ref{ll}) by solving the 
Jordan chain constraints directly, 
to find  wavefunctions $\phi_0$, $\phi_1$ and $\phi_2$ that satisfy 
\bea
H_0\phi_0 &=& 0 \nn \\
H_0\phi_1 &=& \phi_0 \label{con1}\\
H_0\phi_2 &=& \phi_1 \nn
\eea
where $H_0$ is the Hamiltonian at the cubic exceptional point:
\eq
H_0 = -\frac{d^2}{dz^2}+z^6+9z^2+\frac{2}{z^2}~.
\label{h0}
\en
Note that $H_0\left.\Psi_0\right|_{\ep=0}=0$ so we can take
$\phi_0=\left.\Psi_0\right|_{\ep=0}$. Then solving (\ref{con1}) for
$\phi_1$
and $\phi_2$, we find
\bea
\phi_1 &=&
e^{z^4/4}\left(\frac{az}{2}+b\left(\frac{1}{z}+2z^3\right)\right) \\
\phi_2 &=&
e^{z^4/4}\left(\frac{a}{16z}+\frac{bz}{2}+c\left(\frac{1}{z}+2z^3\right)
\right) \nn
\eea
with $a$, $b$ and $c$ constants, arbitrary at this stage.
These are  the most general solutions to (\ref{con1}) that also
satisfy the relevant boundary condition, that is square integrability
along $i{\cal C}$. 

Now that we have a basis, we must find the dual basis
$\tilde{\phi}_0$, $\tilde{\phi}_1$
and $\tilde{\phi}_2$ which satisfies
\bea
\int_{i{\cal C}}\phi_i\tilde{\phi}_i\,dz  &=& 1 \,,\,\textrm{for
  $i=0,1,2$} \nn \\ 
\int_{i {\cal C}}\phi_i\tilde{\phi}_j\,dz &=& 0 \,,\,\textrm{for
  $i\neq j$} . 
\label{con2}
\eea
{}From \cite{Sokolov:2006vj} we expect the
dual basis to be $\tilde{\phi}_0=\phi_2$, $\tilde{\phi}_1=\phi_1$
and $\tilde{\phi}_2=\phi_0$ and this is supported by the
fact that $\int_{i{\cal C}}\phi_0\phi_0\,dz=\int_{i{\cal C}}\phi_0\phi_1\,dz=0$ and
$\int_{i{\cal C}}\phi_1\phi_1\,dz\propto a^2$.
Fixing $\int_{i{\cal C}}\phi_2\phi_2\,dz=0$ and  $\int_{i{\cal C}}\phi_1\phi_2\,dz=0$
constrains the coefficients $b$ and $c$ to be
\bea
b &=& -\frac{a\pi}{16\Gamma(3/4)^2} \nn \\
c &=& \frac{a(3\pi^2-8\Gamma(3/4)^4)}{512\Gamma(3/4)^4}.
\eea
Then requiring  $\int_{i{\cal C}}\phi_1\phi_1\,dz=\int_{i{\cal C}}\phi_0\phi_2\,dz=1$ fixes $a^2$:
\eq
a^2=-\frac{2^{11/4}i}{\Gamma(3/4)}.
\en
Choosing the root with positive real part for $a$ we have fixed the
basis to be 
\bea
\phi_0 &=&
\frac{(1-i)2^{7/8}}{\sqrt{\Gamma\left(\frac{3}{4}\right)}}e^{z^4/4}\left(
\frac{1}{z}+2z^3\right) \nn \\
\phi_1 &=&
\frac{(i-1)2^{7/8}}{16\Gamma\left(\frac{3}{4}\right)^{5/2}}e^{z^4/4}\left(
\frac{\pi}{z}-8z\Gamma\left(\frac{3}{4}\right)^2+2\pi z^3\right)
\label{basis} \\
\phi_2 &=&
\frac{(1-i)2^{7/8}}{512\Gamma\left(\frac{3}{4}\right)^{9/2}}e^{z^4/4}\left(
\frac{24\Gamma\left(\frac{3}{4}\right)^4+3\pi^2}{z}
-16\pi\Gamma\left(\frac{3}{4}\right)^2z+6\pi^2z^3
-16\Gamma\left(\frac{3}{4}\right)^4z^3\right) \nn
\eea
with the dual basis $\tilde{\phi}_0=\phi_2$, $\tilde{\phi}_1=\phi_1$
and $\tilde{\phi}_2=\phi_0$.

\subsubsection{Matrix elements and the cusp singularity}

We first perturb away from the cusp along the QES line
$(\alpha_+,\alpha_-)= (\ep+1/4,1)$ and write 
\eq
H=H_0+V~,
\en
where   $V=\frac{4\ep^2-6\ep}{z^2}+4\ep z^2$ is considered as a
perturbation of $H_0$ (\ref{h0}). The required matrix elements are 
\eq
\langle\tilde{\phi}_2|V|\phi_0\rangle =\frac{128\pi\ep^2}{3\Gamma\left(
\frac{3}{4}\right)^2}\;,
\en
and
\bea
\langle\tilde{\phi}_1|V|\phi_0\rangle &=&
\langle\tilde{\phi}_2|V|\phi_1\rangle \nn \\
&=&
-\frac{8\ep}{3\Gamma\left(\frac{3}{4}\right)^4}\left(24\Gamma\left(
\frac{3}{4}\right)^4-12\Gamma\left(\frac{3}{4}\right)^4\ep+\pi^2\ep\right)
\nn \\
&\approx& -64\ep  
\eea
to leading order in $\ep$. 
To investigate the shape of the cusp we also need to perturb away
from the exceptional point in the direction perpendicular to the QES
line, i.e. along $\eta$ where $\al=-4\eta+9$ and
$\lambda=2\eta-3/2$, or $\al_+=1/4$ and $\al_-=1-\eta$. The
Hamiltonian is now $H=H_0+V+V'$  with
\eq
V'=-4\eta z^2+\frac{2\eta(2\eta-3)}{z^2}~,
\en
and to first order in $\eta$ we find 
\eq
\langle\tilde{\phi}_2|V'|\phi_0\rangle =\frac{128\eta \pi(\eta-3)}{3\Gamma\left(
\frac{3}{4}\right)^2}
\approx -\frac{128\pi\eta}{\Gamma\left(
\frac{3}{4}\right)^2}\;.
\en
The remaining matrix elements effect the energy levels only at
subleading order in $\eta$ and so  they can consistently be ignored. The
resulting truncated Hamiltonian is
\eq
H_{pert} \approx
\left(\begin{array}{ccc}
0 & 1 & 0 \\
-64\ep & 0 & 1 \\
-\frac{128\pi\eta}{\Gamma(3/4)^2} & -64\ep & 0 \\
\end{array}\right)\;.
\label{Hper}
\en
The matrix  (\ref{Hper}) has the characteristic polynomial
$X^3+128\ep X+\frac{128\pi\eta}{\Gamma(3/4)^2}=0$.
 Now the curve of
exceptional points occurs when $dX/d\ep\rightarrow\infty$ (or
equivalently $dX/d\eta\rightarrow\infty$). Since
\eq
\frac{dX}{d\ep}=\frac{-128X}{3X^2+128\ep}
\en
the requirement $dX/d\ep\rightarrow\infty$ fixes
\eq\label{Xe}
X=\pm\sqrt{-\frac{128\ep}{3}}.
\en
Substituting this into the characteristic polynomial above and
restricting to $\ep\leq 0$ gives the following relation between
$\eta$ and $\ep$:
\eq
\eta=\pm\frac{2}{3}\sqrt{\frac{128}{3}}\frac{\Gamma(3/4)^2}{\pi}|\ep|^{3/2}.
\en
For $\ep>0$ the relation (\ref{Xe}) is not real indicating that
there are no exceptional points in this region, which matches 
our numerical results. In terms of the $\al_{\pm}$ notation,
$\al_+=\ep+1/4$ and $\al_-=1-\eta$ so this relation becomes:
\eq
\al_-=
1\pm\frac{2}{3}\sqrt{\frac{128}{3}}\frac{\Gamma(3/4)^2}{\pi}(1/4-\al_+)^{3/2}
\label{pred}
\en
which is valid for $\al_-$ close to 1 and $0<<\al_+\leq 1/4$.

A comparison between the prediction (\ref{pred}) for the line of exceptional
points in the vicinity of the cusp at $(\alpha_{+},\alpha_{-})=(1/4,1)$
and numerical data 
obtained from a direct solution of the eigenvalue problem  is shown
in figure \ref{figpred}. The shape of the curve is accurately
reproduced. In principle the same calculations could be performed for
other cusps, though the relevant wavefunctions become more
complicated.

\medskip

\[
\begin{array}{c}
\!\!\!\!
\!\!\!\!
\!\!\includegraphics[width=0.5\linewidth]{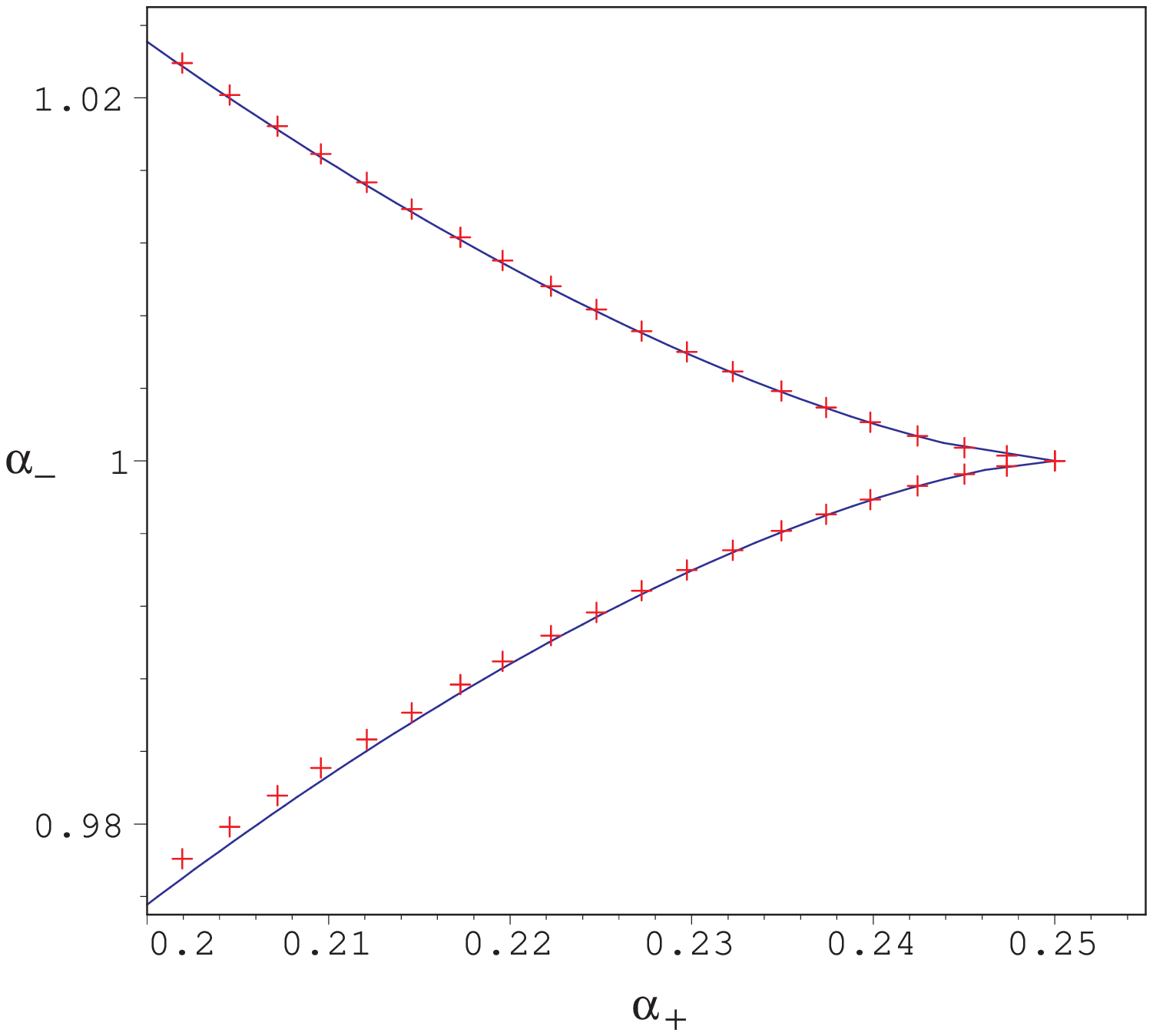}
\\[11pt]
\parbox{0.7\linewidth}{
{\small Figure \ref{figpred}: The first cusp for $M=3$: the crosses show the prediction
(\ref{pred}) while the solid line was found by solving the full problem.}}
\end{array}
\]
\refstepcounter{figure}
\label{figpred}
\resection{Numerical results for $M\neq 3$}
\label{numres}
Having established the existence of quadratic and  cubic exceptional
points at $M=3$, we now explore the
situation at other values of $M$.
Whitney's theorem for mappings from the plane to the plane
\cite{whitney} implies that the fold and cusp singularities
(corresponding to the doubly-exceptional lines and triply-exceptional
cusp points seen at $M=3$) are stable, and so the pattern
of cusped lines must persist, at least while $M$ remains sufficiently
close to $3$.
Recall also that protected zero-energy levels lie on the lines
$\alpha_{\pm}=n$  for all values of $M$.  However, away from $M=3$
quasi-exact solvability is lost,
and so one of the properties which confined the cusps at
$M=3$ to the lines $\alpha_{\pm}=n$, namely the symmetry of
the set of merging levels under $E\to -E$, may no
longer hold.

Figures~\ref{fig2}, \ref{fig1p5} and \ref{fig1p3}
show the exceptional lines for
$M=2$, $1.5$ and $1.3$. The plots were obtained by a direct
numerical solution of the second dual form of the
eigenvalue problem, as described in appendix \ref{dualapp}.

\medskip

\[
\begin{array}{c}
\!\!\!\!\!\!\includegraphics[width=0.6\linewidth]{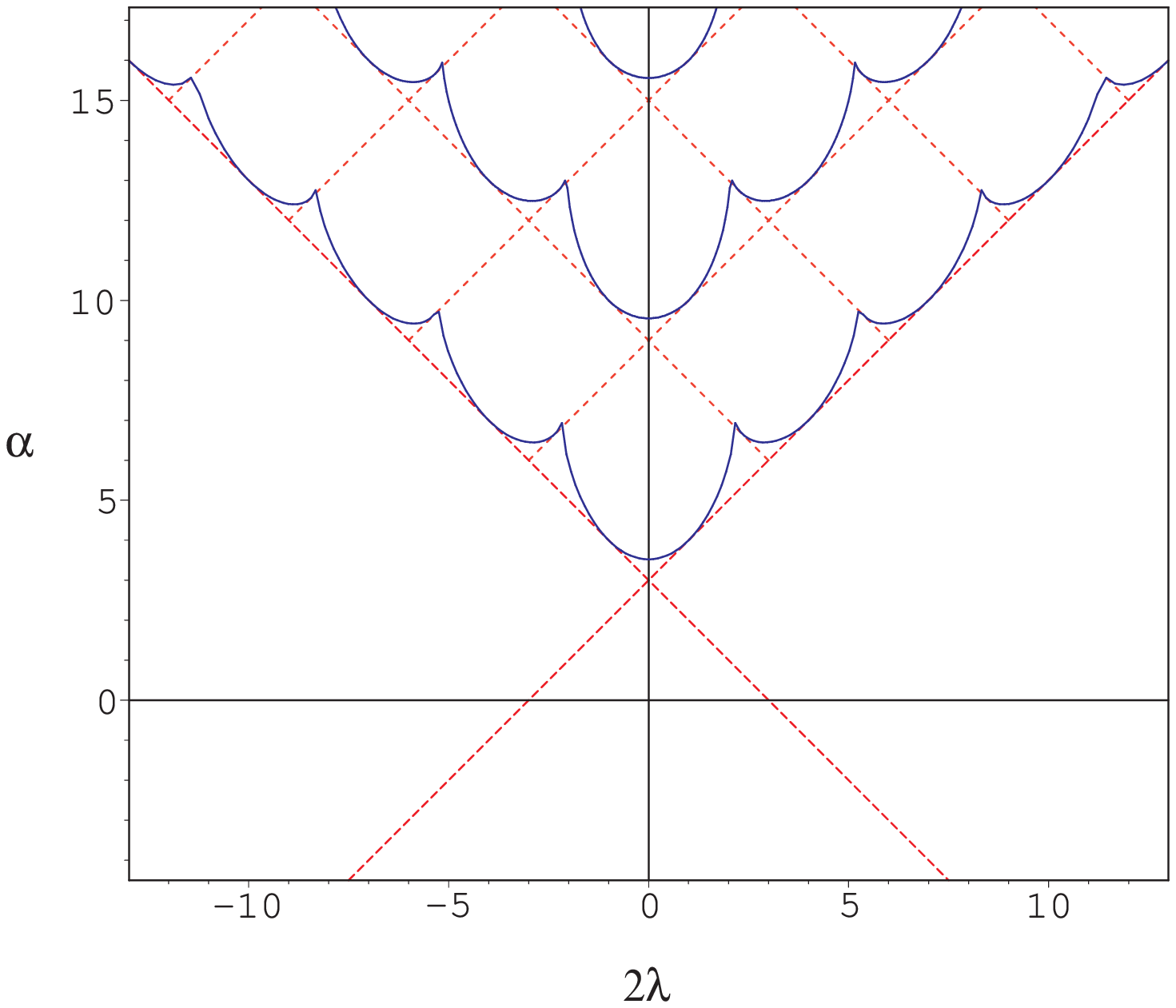}
\\[11pt]
\parbox{0.4\linewidth}{
{\small Figure \ref{fig2}: Exceptional lines at $M=2$.
}}
\end{array}
\]
\refstepcounter{figure}
\label{fig2}

\medskip

As predicted, the overall pattern
remains the same, but the cusps
move away from the protected zero-energy lines.
The points where the outermost cusped line
touches the supersymmetric zero-energy lines
$\alpha_{\pm}=0$ are
known exactly, from (\ref{quadp}). As $M$ decreases from $3$,
they move down from the midpoints
between $\alpha_{\mp}=n$ and $\alpha_{\mp}=n+1$ along the lines
$\alpha_{\pm}=0$, as predicted by the formula 
(\ref{genform}).
At the same time, the numerical data shows that
the cusps move upwards, on the rescaled coordinates of the
plots which keep the lines of protected zero-energy levels at constant
locations. As $M\to 1^+$ the pattern shows signs of simplifying, with
the cusps heading away towards $\alpha=+\infty$ and the regions of
unreality shrinking towards the lines $2\lambda\in 2\ZZ$. 
This behaviour will be discussed further in section \ref{pertsec}.

\[
\begin{array}{c}
\!\!\!\!\!\!\includegraphics[width=0.6\linewidth]{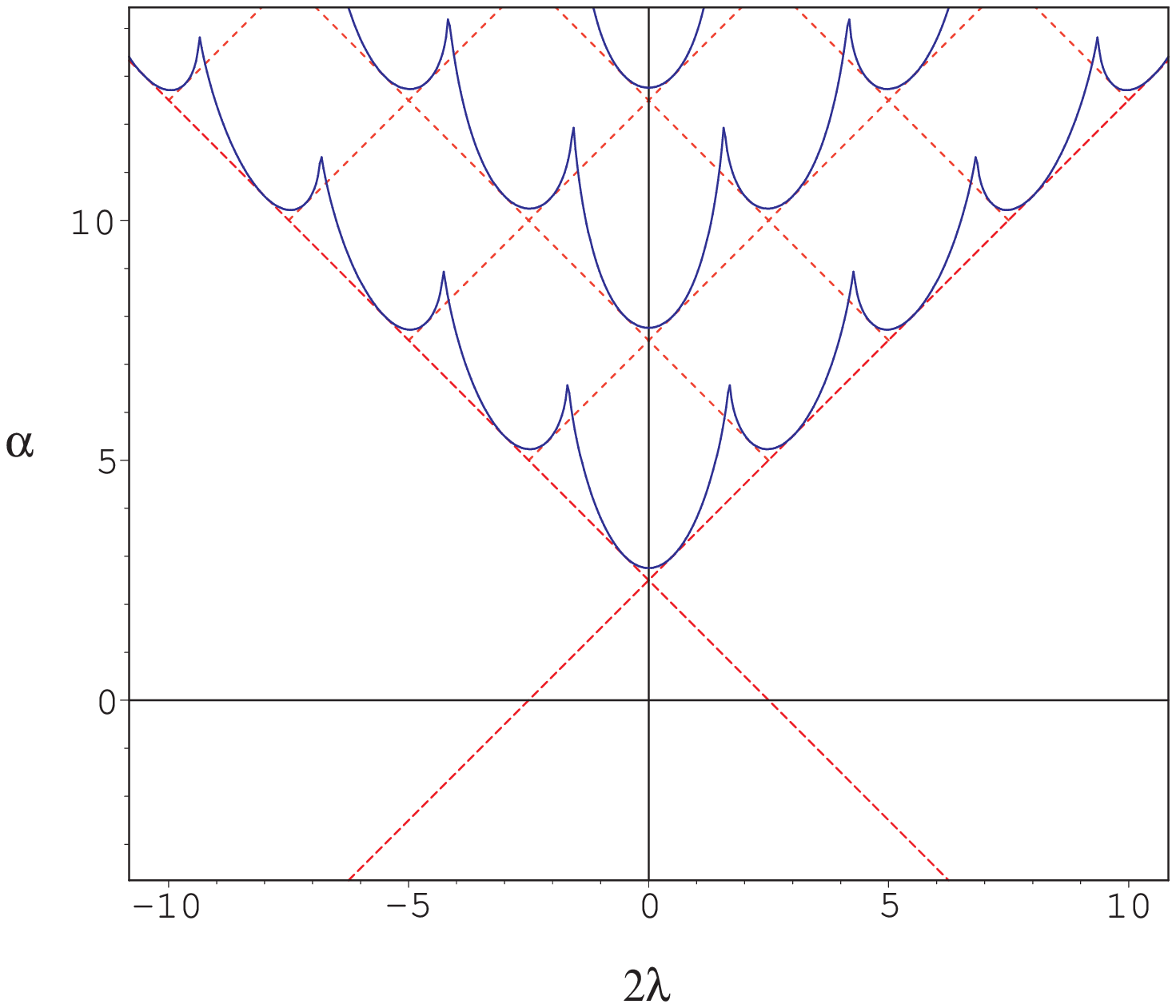}
\\[11pt]
\parbox{0.4\linewidth}{
{\small Figure \ref{fig1p5}: Exceptional lines at $M=1.5$.
}}
\end{array}
\]
\refstepcounter{figure}
\label{fig1p5}

\medskip

\[
\begin{array}{c}
\!\!\!\!\!\!\includegraphics[width=0.6\linewidth]{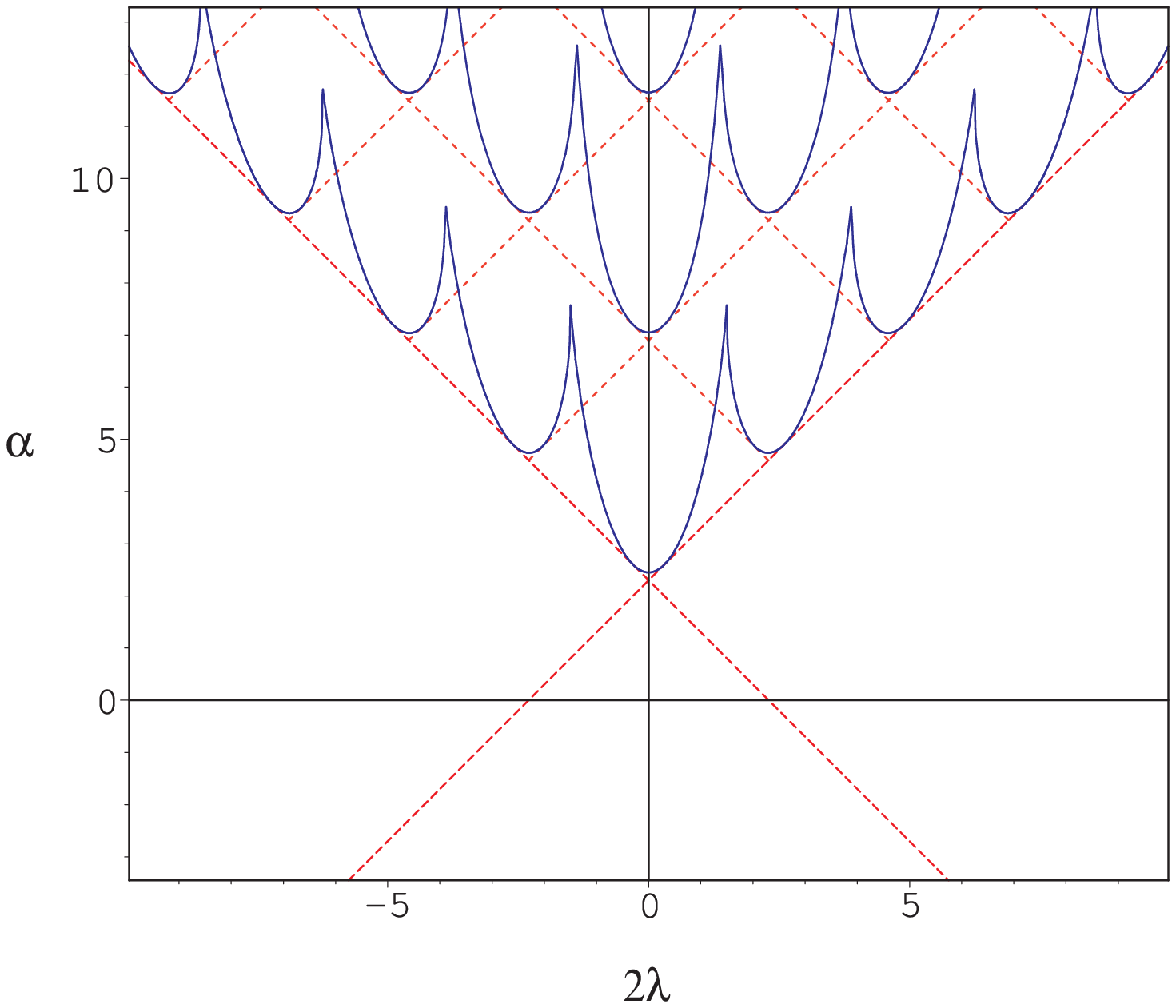}
\\[11pt]
\parbox{0.4\linewidth}{
{\small Figure \ref{fig1p3}: Exceptional lines at $M=1.3$.
}}
\end{array}
\]
\refstepcounter{figure}
\label{fig1p3}

It is interesting to see the fate of the exceptional points
corresponding to the zeros of the polynomials $Q_n(\lambda)$, which at
$M=3$ are triply-exceptional
cusps. For $M\neq 3$ the cusps move away from the
lines $\alpha_{\pm}\in\NN$, and so the zeros of the $Q_n(\lambda)$ are 
no longer cusps, but are
instead only doubly exceptional. Furthermore, the presence of
an exactly-zero level on 
the lines $\alpha_{\pm}\in\NN$
forces the smooth parts of the
exceptional lines to be tangent to 
these lines immediately $M$ moves away from $3$, 
and this leads to a complicated change in the shape of
these curves, illustrated in 
figure~\ref{cuspmovement}.

\[
\!\!
\begin{array}{ccc}
\includegraphics[width=0.315\linewidth]{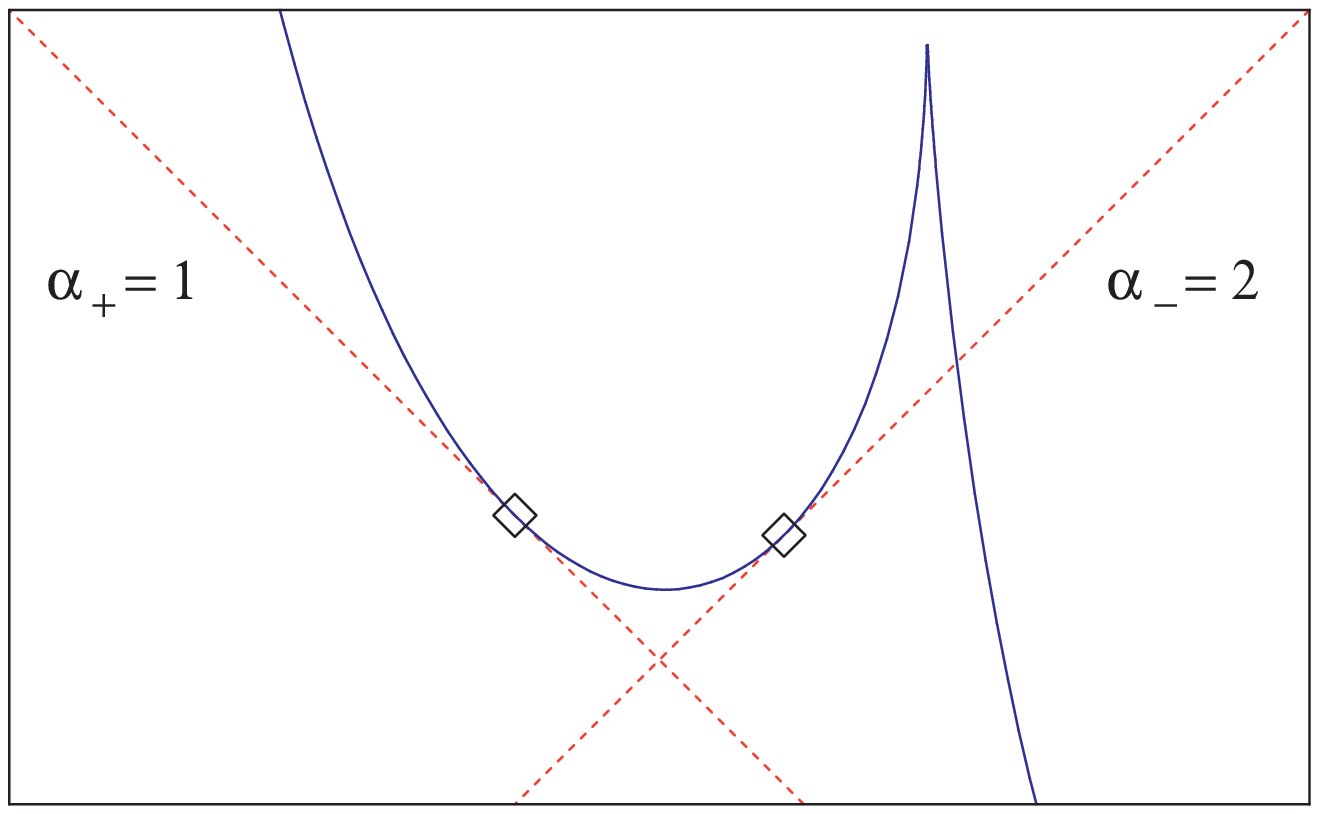}
& \includegraphics[width=0.315\linewidth]{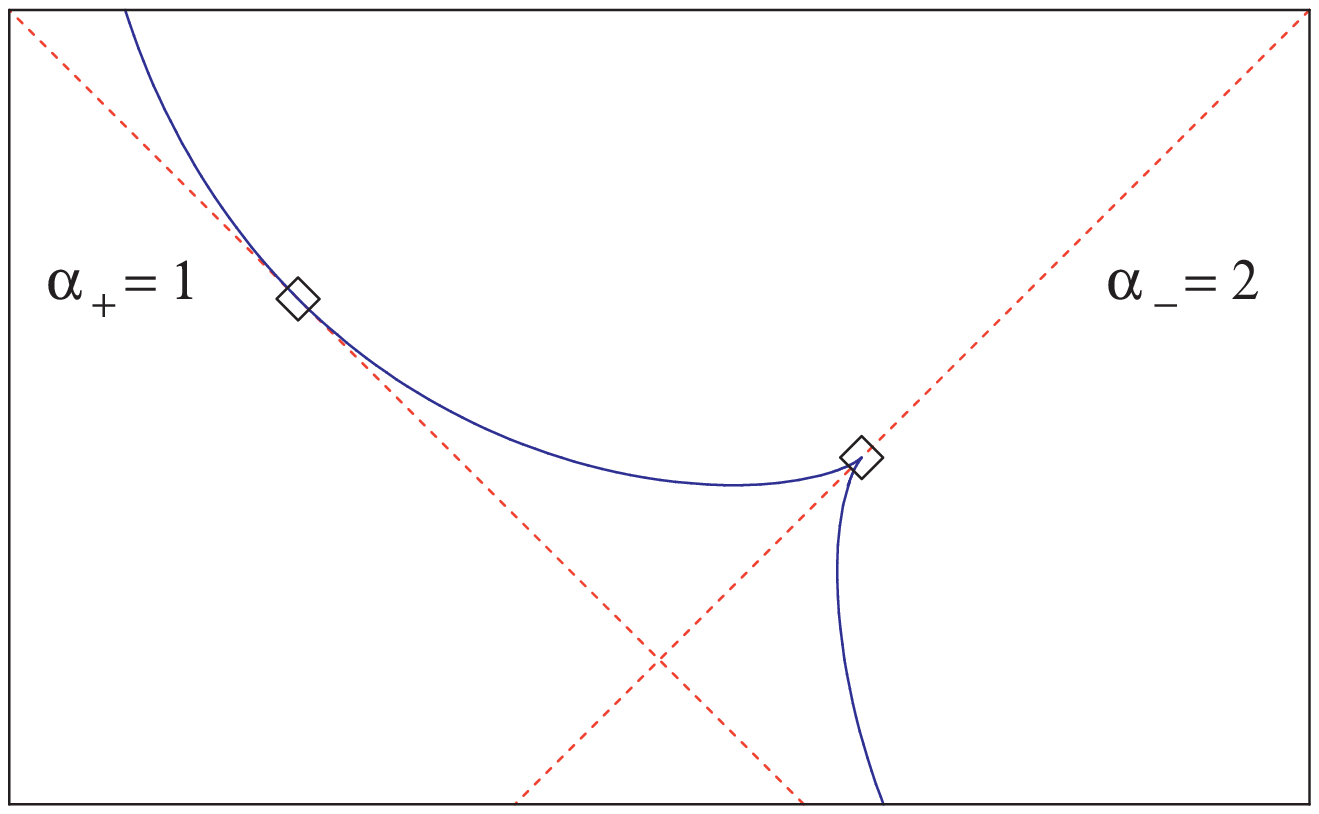}
& \includegraphics[width=0.315\linewidth]{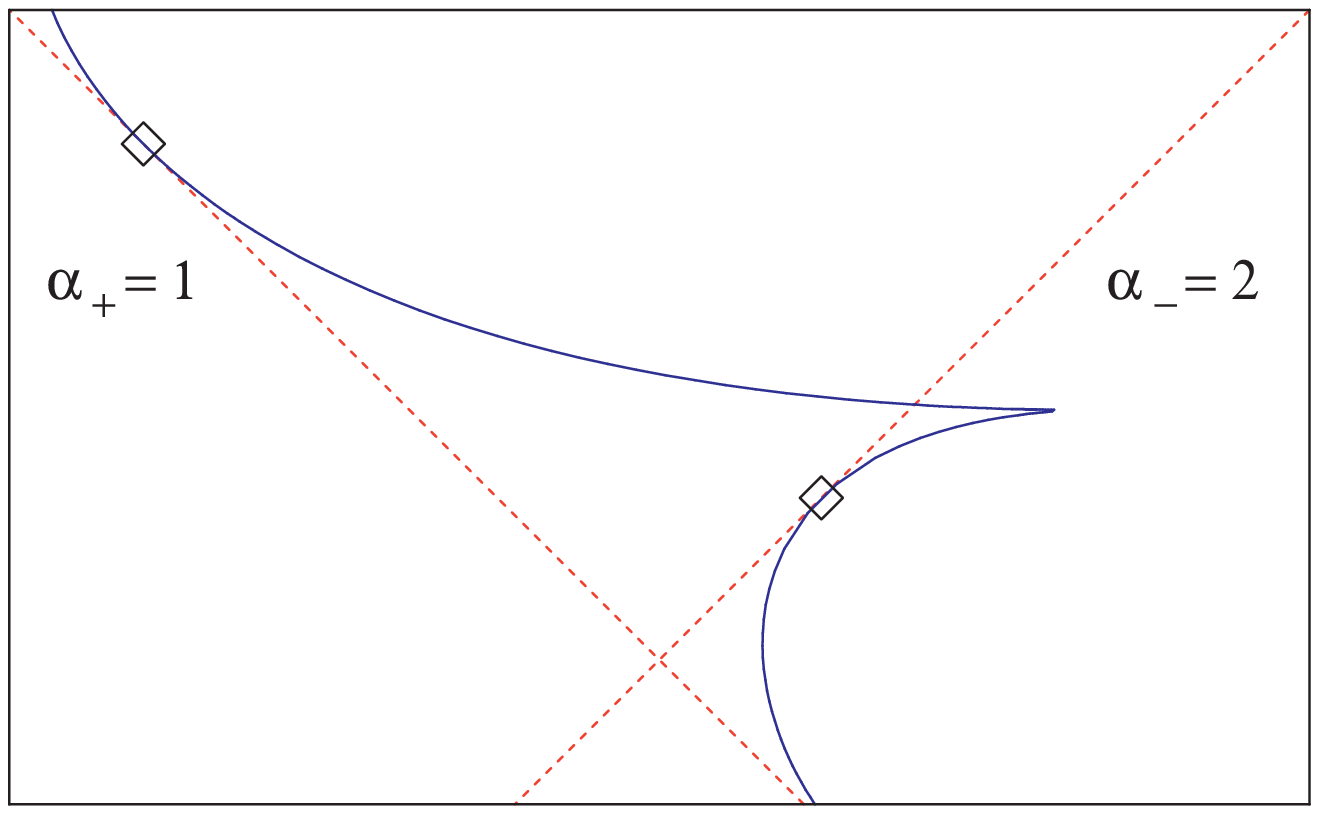}
\\[11pt]
M=1.5
& M=3
& M=6
\\[15pt]
\multicolumn{3}{c}{\parbox{0.5\linewidth}{
{\small Figure \ref{cuspmovement}: The movement of a cusp for $M\neq 3$.
}}}
\end{array}
\]
\refstepcounter{figure}
\protect\label{cuspmovement}

The plots of figure~\ref{cuspmovement} indicate that for $M>3$
the movement of the cusps away from the lines $\alpha_{\pm}\in\NN$
is opposite to that for $M<3$, and this
can also be seen in
figures \ref{fig10p0} and \ref{fig30p0}, which show the exceptional
lines for $M=10$ and $M=30$.
Again, the locations of the zero-energy
exceptional points on the lines $\alpha_{\pm}$ confirm the formula
(\ref{genform}), and there are no hints of any further exceptional
points beyond those predicted by our general considerations.
As the cusps move towards the $\alpha$ axis, they start to merge to 
leave
isolated `islands' of unreality in the phase diagram.
In the theory of singularities, this merging of two cusps is
sometimes called
the `beaks' transition (see, for example, \cite{beaks} and references
therein).
As for $M\to 1^+$, the structure simplifies
as $M\to\infty$.

\medskip

It turns out that the simplifications near to $M=1$ and $M=\infty$
can be understood
analytically, using the fact that
the limiting points $M=1$ and $M=\infty$ are 
exactly solvable,  and allow perturbative
treatments to be set up in their vicinities. 
In the next two sections this will be developed in
detail, starting with the region near to $M=1$ where we will see that
it leads to
a novel insight into the original `Bender-Boettcher' phase transition
to infinitely-many complex levels, which occurs when $M$ becomes
smaller than $1$.

\bigskip

\[
\begin{array}{c}
\!\!\!\!\!\!\includegraphics[width=0.6\linewidth]{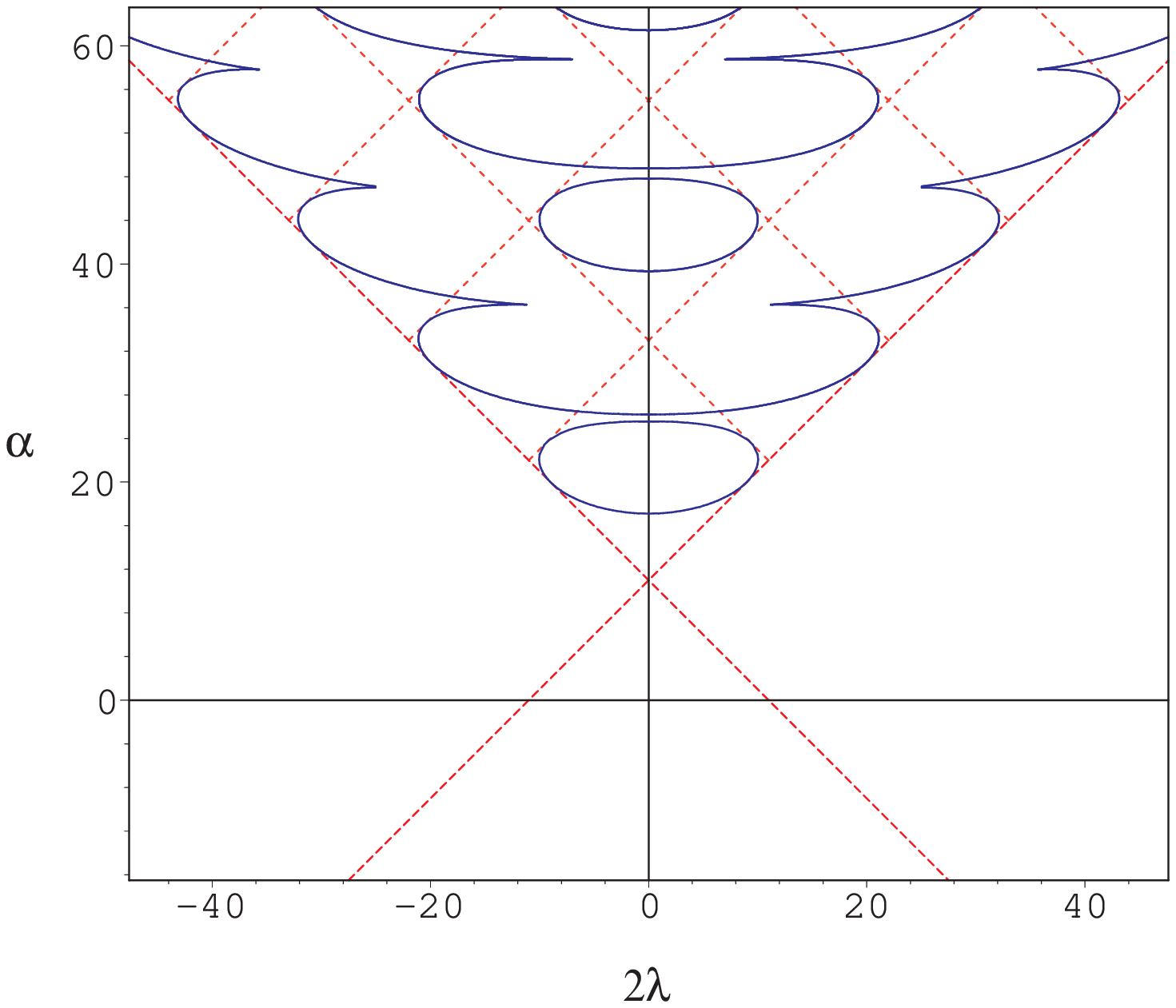}
\\[11pt]
\parbox{0.4\linewidth}{
{\small Figure \ref{fig10p0}: Exceptional lines at $M=10$.
}}
\end{array}
\]
\refstepcounter{figure}
\label{fig10p0}
\[
\begin{array}{c}
\!\!\!\!\!\!\includegraphics[width=0.6\linewidth]{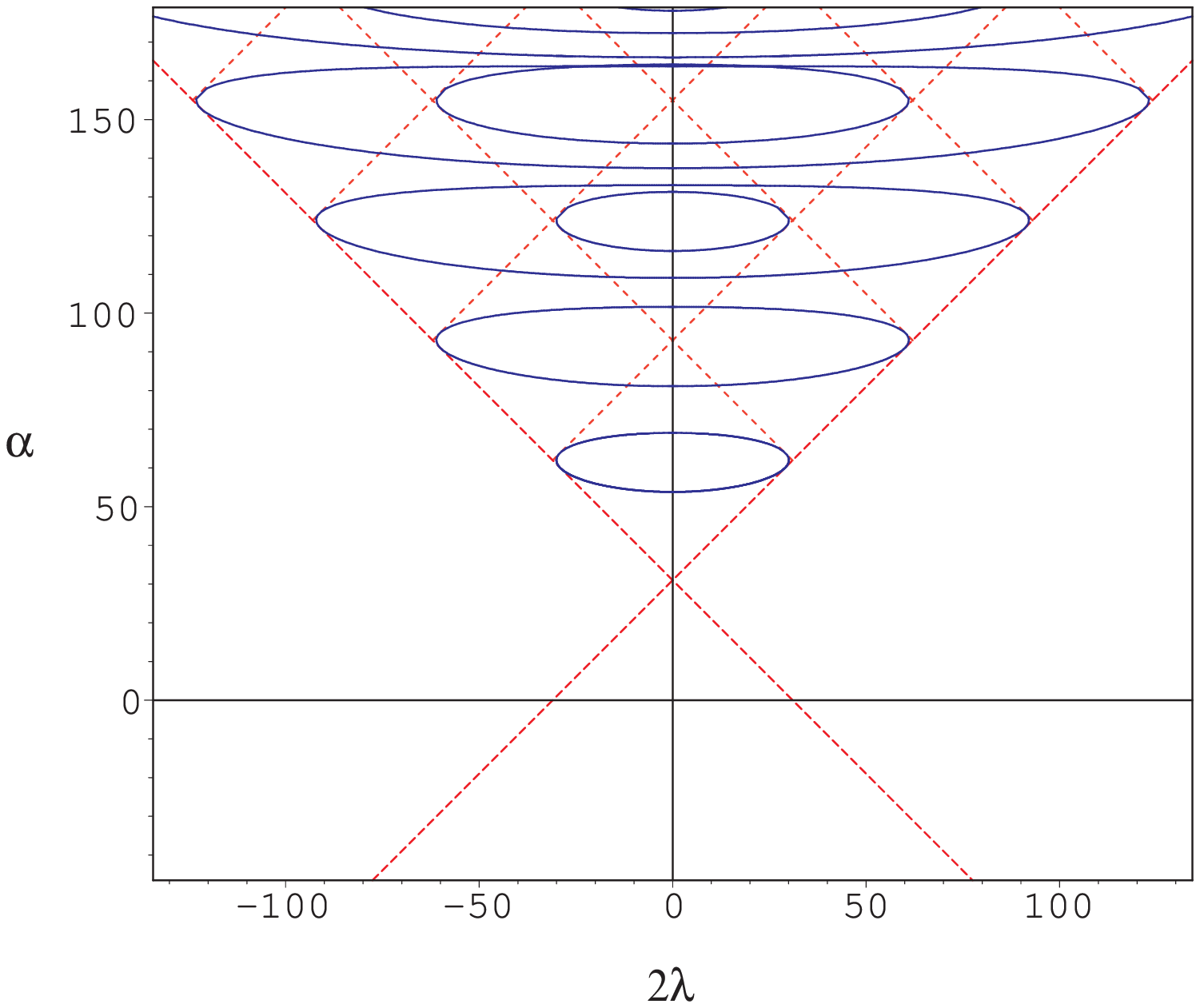}
\\[11pt]
\parbox{0.4\linewidth}{
{\small Figure \ref{fig1p3}: Exceptional lines at $M=30$.
}}
\end{array}
\]
\refstepcounter{figure}
\label{fig30p0}

\resection {Perturbation theory about $M=1$}
\label{pertsec}
\subsection{Exceptional points via near-degenerate
perturbation theory}
In this section we revert to the original formulation of the
eigenvalue problem, namely
\eq
H_M\psi(x) =E\,\psi(x)\, ,
{}~~~\psi(x) \in L^2(\CaC)
\en
where
\eq \label{H_M}
H_M=-\frac{d^2}{dx^2}-(ix)^{2M}-\al(ix)^{M-1}+
\frac{\lambda^2-\frac{1}{4}}{x^2}~.
\en
 For $M=1$ this problem can be solved exactly -- it is
the $\PT$-symmetric simple
harmonic oscillator \cite{Znojil:1999qt,Dorey:1999uk}, and its spectrum is
entirely real. (Note, for $\lambda^2-\frac{1}{4}\neq 0$ the
wavefunctions themselves can be complex,
owing to the singularity of the potential at the origin and
the departure of the quantisation contour from the real axis there.)
As $M$ moves away from $1$, pairs of eigenvalues can become
complex; as discussed earlier, this is always preceded by the
coincidence of two real eigenvalues and so the first complex
eigenvalues will emerge from points in the $(2\lambda,\alpha)$ plane
at which the spectrum has degeneracies for $M=1$.
 We aim to investigate exactly how this occurs.

In \cite{Bender:1998gh}, Bender {\it et al.}\ used a perturbative
approach to study the spectrum for $M$ near $1$ with
$\alpha=0$ and $\lambda^2=\frac{1}{4}$.
The full Hilbert space was truncated to the
subspace spanned by $M=1$ eigenfunctions
$|2n{-}1\rangle$ and $|2n\rangle$, where
$H_1|m\rangle=(2m{+}1)|m\rangle$, $m\in\ZZ^+$\,,
and $H_M$ expanded within that two-dimensional subspace
about $H_{M=1}$. Diagonalising the resulting $2\times 2$ matrix
yielded an approximation to the eigenvalues of
$H_M$.
However, as
shown in \cite{Dorey:2004fk}, this approximation predicts level-merging
for both signs of $M{-}1$ rather than the one sign actually
observed, and when applied to the pair of levels $|2n\rangle$ and
$|2n+1\rangle$, it predicts that they too will merge,
contrary to the actual behaviour of the model. These problems can be
traced to the fact that the $M=1$ eigenvalues at
$\alpha=\lambda^2-\frac{1}{4}=0$
are equally spaced, making the truncation to the subspace spanned by
$|2n{-}1\rangle$ and $|2n\rangle$ unjustified.

For the more general Hamiltonian (\ref{H_M}) the situation can be
improved, as $\alpha$ and
$\lambda$ can be tuned so as to make some pairs of levels close
to each other relative to all of the others. Truncation to these
levels will then be reliable, and as we show below it gives
a good approximation to their behaviour for
$M$ close to $1$.

To see how a consistent
prediction of exceptional points can emerge from this approach,
it is worth examining a simple
$2\times 2$ example which illustrates the main features.
Consider the `unperturbed' Hamiltonian
\eq
H_1(\eta)=
\left(
\begin{matrix}
2\eta&0\\
0&-2\eta
\end{matrix}
\right)
\label{toy1}
\en
where $\eta$ will be considered small but fixed, with the eigenvalues
$\pm 2\eta$ corresponding to the nearby pair of energies in the full
problem.
Add to it a perturbation with both
diagonal and off-diagonal parts:
\eq
V_{\epsilon}(\eta)=
-\frac{\epsilon}{\eta}
\left(
\begin{matrix}
\alpha&i\\[3pt]
i&-\alpha
\end{matrix}
\right)
\label{toy2}
\en
where $\alpha$ is fixed and 
$\epsilon$ is the perturbing parameter (corresponding to $M-1$ in
the full problem). The factor of $1/\eta$ will reflect the fact that
nearby levels in the unperturbed problem interact more strongly
as they approach each other.
 Then $H_{1+\epsilon}=H_1+V_{\epsilon}$
has eigenvalues
\eq
E_{\pm}=\pm \sqrt{(2\eta-\alpha\epsilon/\eta)^2-\epsilon^2/\eta^2}
\label{toyvalues}
\en
and exceptional points at $\epsilon=\pm\frac{2}{1\pm\alpha}\,\eta^2$.
For fixed $\alpha\neq\pm1$ the two
exceptional points are at $\epsilon=O(\eta^2)$,
so, even with the $1/\eta$ factor in its specification,
$V_{\epsilon}(\eta)$
is still small at their locations.
For $\alpha=\pm 1$ one exceptional point
is pushed away to infinity, but the other remains in a region where
the perturbation is still small.

\subsection{Perturbative locations of the exceptional points}
Returning to the original problem, the Hamiltonian at $M=1$ is
\eq \label{H1}
H_1=-\frac{d^2}{dx^2}+x^2+\frac{\lambda^2-\frac{1}{4}}{x^2}-\alpha\,.
\en
With the given boundary conditions, $H_1$ has $c$-normalised
eigenfunctions  \cite{Adam}
\eq
\phi_n^{\pm}(x) =
\frac{\sqrt{2}\sqrt{n!}}{\sqrt{(1-e^{\mp 2\pi i\lambda})
\Gamma(\pm\lambda+n+1)}}\,
x^{1/2\pm\lambda}e^{-\frac{x^2}{2}}
L_n^{\pm\lambda}(x^2)
\,,\quad n=0,1,\dots\,
\label{efns}
\en
where the $L_n^\beta$ are Laguerre polynomials.
The corresponding eigenvalues are
\eq
E_n^{\pm} =-\alpha+4n+2\pm 2\lambda \,.
\label{evals}
\en
A degenerate eigenvalue occurs when $E_n^+=E_m^-$ for some $n$ and $m$,
which requires
\eq
\lambda=
m-n\,.
\en
Thus, on the vertical lines
$2\lambda\in 2\ZZ$ in the $(2\lambda,\alpha)$ plane,
infinitely-many pairs of the eigenfunctions (\ref{efns}) are
proportional to each other.
Indeed, if $\lambda=q$ is a non-negative integer, 
then
for all non-negative integers $p$,
$\phi_{p+q}^-=i(-1)^q\phi_p^+$.
Since $\phi_n^+\rightarrow\phi_n^-$ when $\lambda\rightarrow -\lambda$,
it also follows that
$\phi_{p+q}^+\propto \phi_p^-$ when $\lambda=-q$.

In order to find the eigenvalues of $H_M$ for
$M= 1+\epsilon$,
we treat $H_M=H_{1+\epsilon}$ in a basis of near-degenerate
eigenfunctions of $H_1$ by writing it as
\eq
H_{1+\epsilon}=H_1+V_{\epsilon}
\en
where $H_1$ is given by (\ref{H1}) and
\eq
V_{\epsilon}=-x^2-(ix)^{2+2\epsilon}-\al(ix)^{\epsilon}.
\en
The exact matrix elements of $V_{\epsilon}$ in the truncated
basis of $H_1$ eigenfunctions were found by Millican-Slater
\cite{Adam}, and are reproduced in
appendix \ref{matrixelements}, while those of
$H_1$ are given by (\ref{evals}). Rediagonalising the
resulting $2\times 2$ matrix
gives the approximate energy levels.

To find the exceptional points reliably, we require both that the
perturbation is small, and that the two levels in the truncated
subspace are close. With
$M=1+\ep$ and $\lambda=q+\eta$, this means that $\ep$ and $\eta$ must
be small. In fact, we shall see that the exceptional points occur when 
$\epsilon$ is of order $\eta^2$, and our approximations will be
good in this region.
We shall also assume that
$q\ge 0$, as results for negative $q$ are easily restored using
the $\lambda\to -\lambda$ symmetry of the problem.  For
small values of $\eta$, the pairs of levels
$\{\phi^+_p,\phi^-_{p+q}\}$, $p\ge 0$,
are  almost degenerate;
to lighten the notation, we fix the integer $p\ge 0$ and denote the
corresponding basis by
$\{\phi^+,\phi^-\}\equiv \{\phi^+_p,\phi^-_{p+q}\}$.
The matrix elements of  $H_1$ are
\bea
\langle\phi^+|H_1|\phi^+\rangle &=& 4p+2q+2\eta+2 \\
\langle\phi^-|H_1|\phi^-\rangle &=& 4p+2q-2\eta+2 \\
\langle\phi^+|H_1|\phi^-\rangle &=& \langle\phi^-|H_1|\phi^+\rangle
\, =\, 0~,
\eea
while those of $V_{\ep}$
follow from
(\ref{Phipp}), (\ref{Phipp1})
and (\ref{Phipm}).

Expanding in $\ep$ and $\eta$ and retaining terms proportional to
$\eta$, $\epsilon/\eta$, $\epsilon$ and $\epsilon^2/\eta$
the matrix
elements $H_{ab}\equiv\langle{\phi^a}|H_M|\phi^b\rangle$  are
\bea
H_{++} &\approx&
4p+2q+2-\al+\left(2p+q+1-\frac{\al}{2}\right)\frac{\ep}{\eta}+2\eta
\nn \\
&&~+\left(\left(2p+q+1-\frac{\al}{2}\right)\psi(p+q+1)+2p+2\right)\ep \\
&&~+\left(\left(2p+q+1-\frac{\al}{4}\right)\psi(p+q+1)+2p+1\right)
\frac{\ep^2}{\eta}\,;\nn\\[4pt]
H_{--} &\approx&
4p+2q+2-\al-\left(2p+q+1-\frac{\al}{2}\right)\frac{\ep}{\eta}-2\eta
\nn \\
&&~+\left(\left(2p+q+1-\frac{\al}{2}\right)\psi(p+1)+2p+2q+2\right)\ep \\
&&~-\left(\left(2p+q+1-\frac{\al}{4}\right)\psi(p+q+1)+2p+1\right)
\frac{\ep^2}{\eta}\,;  \nn\\[4pt]
H_{+-} &\approx& \frac{i}{|\eta|}
\left[\left(2p+q+1-\frac{\al}{2}\right)\ep\right. \nn\\
&&~+\left(\frac{1}{2}\left(2p+q+1-\frac{\al}{2}\right)
(\psi(p+q+1)-\psi(p+1))-q\right)\ep\eta \\
&&~+\left.\left(\left(2p+q+1-\frac{\al}{4}\right)
\psi(p+q+1)+2p+1\right)\ep^2\right]\nn
\eea
where $\psi(z)=\Gamma'(z)/\Gamma(z)$.
Diagonalising $H_{ab}$, the approximate
eigenvalues $E_{\pm}$ at $M=1+\ep$, $\lambda=q+\eta$, $\alpha$ are:
\bea
E_{\pm} &=& 4p+2q+2-\al+\bigg(2p+q+2
+ \frac{1}{4}(4p+2q+2-\al)(\psi(p+q+1)+\psi(p+1))\bigg)\ep \nn \\
&&~\pm \bigg[
(8p+4q+4-2\al)\ep+ 4\eta^2
\nn\\& &\quad ~~~
+ \Big((4p+2q+2-\al)(\psi(p+q+1)-\psi(p+1))-4q\Big)\ep\eta 
\nn \\
& &\quad ~~~~
+\Big((8p+4q+4-\al)\psi(p+q+1)+8p+4\Big)\ep^2 \,
\bigg]^{1/2}~.
\label{epa}
\eea
Within this approximation, exceptional points occur on the
curves on the $(2\lambda,\alpha)$ plane where the argument
of the square root in (\ref{epa}) vanishes.
These curves, and their images under $\lambda\to -\lambda$,
are plotted in figures~\ref{figa}
and \ref{figc}
for $\epsilon=0.005$ and
$\ep=0.02$
respectively.
Each shows the exceptional lines corresponding to
$p$ and $q$ equal to
$0,1$ and 2. (The exceptional lines for other values
of $p$ and $q$ are outside the regions shown on the plots.)
The dotted lines indicate $\al_{\pm}\in\ZZ^+$, as previously.\\

\[
\begin{array}{c}
\!\!\!\!\!\!
\!\!\!\!\!\!
\includegraphics[width=0.6\linewidth]{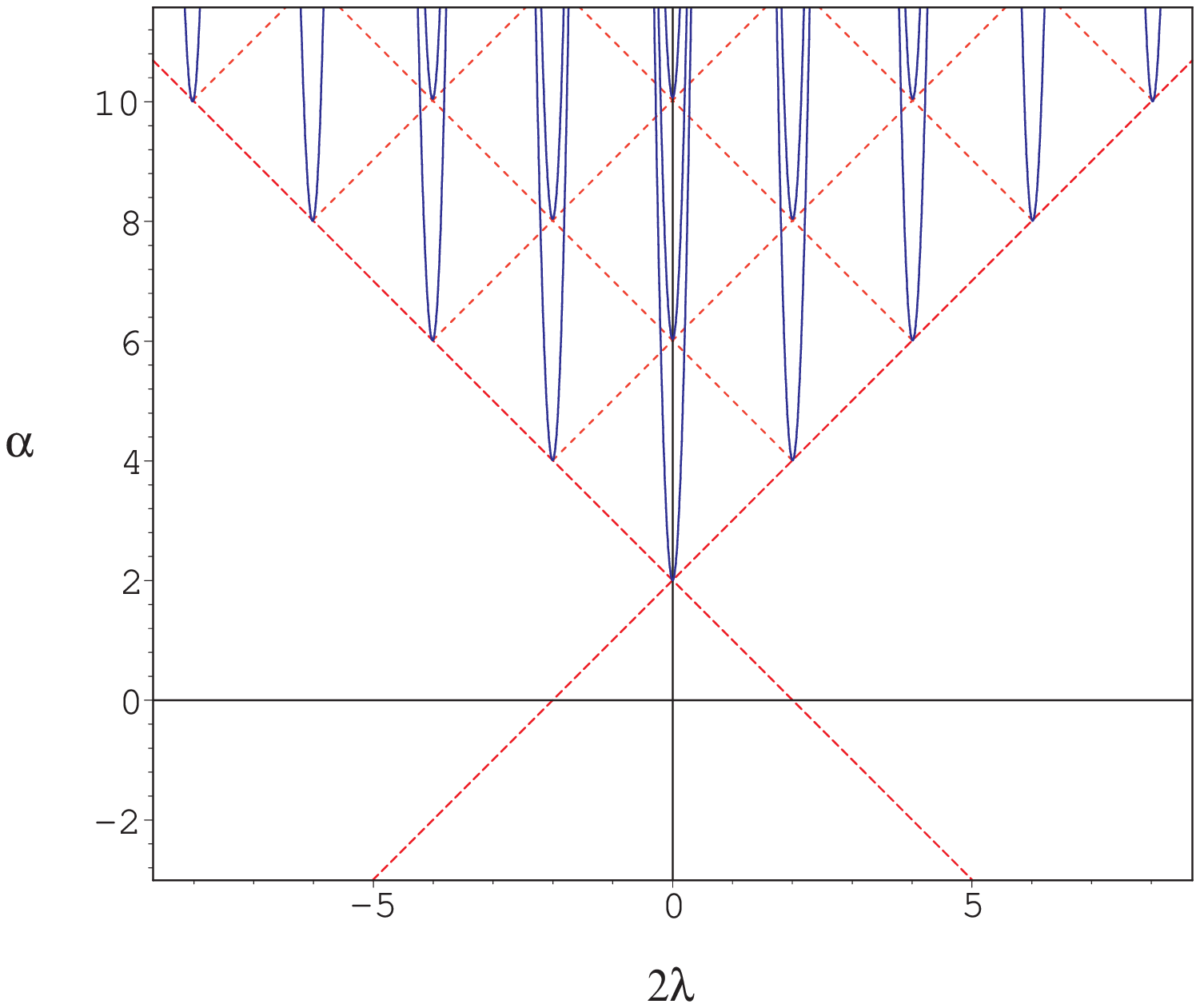}
\\[11pt]
\parbox{0.65\linewidth}{
{\small Figure \ref{figa}: Perturbative lines of exceptional
points for $M=1.005$.
}}
\end{array}
\]
\refstepcounter{figure}
\label{figa}
\medskip

\[
\begin{array}{c}
\!\!\!\!\!\!
\!\!\!\!\!\!\includegraphics[width=0.6\linewidth]{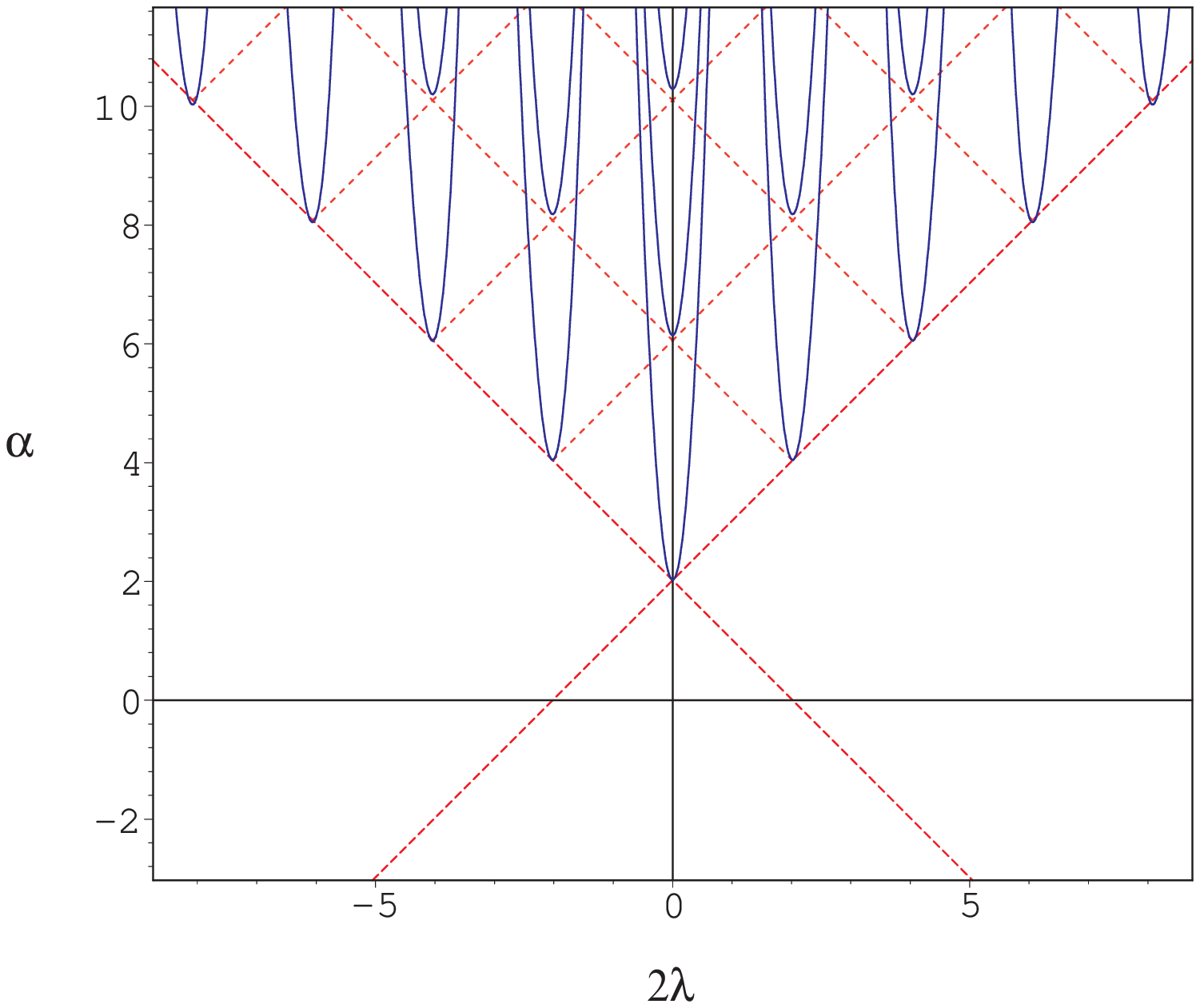}
\\[11pt]
\parbox{0.65\linewidth}{
{\small Figure \ref{figc}: Perturbative lines of exceptional
points for $M=1.02$.
}}
\end{array}
\]
\refstepcounter{figure}
\label{figc}

As $M$ increases, regions of
complex eigenvalues open up from the lines
$\lambda\in\ZZ$, starting near the bottom of the spectrum.
While the mergings of these regions and the joinings of
their exceptional lines to form
cusps cannot be seen within this
approximation (since the truncation is to just two levels), the
pictures are consistent with the numerical evidence in
the last section that the cusps move down from
$\alpha=+\infty$ as $M$ increases from $1$ towards $3$.

The clearest insight into the transitions near $M=1$ comes on
retaining only the leading terms of the matrix elements
for small $\eta$ and $\epsilon$,
namely those proportional to $\eta$ and $\epsilon/\eta$.
For $\lambda=q+\eta$ and $M=1+\epsilon$ as before,
the matrix elements
in the basis
$\{\phi^+,\phi^-\}=
\{\phi_p^+, \phi_{p+q}^-\}$
simplify to
\eq
\left(
\begin{matrix}
H_{++}&H_{+-}\\[3pt]
H_{-+}&H_{--}
\end{matrix}
\right)
\approx
\left(
\begin{matrix}
-2\kappa&0\\[3pt]
0&-2\kappa
\end{matrix}
\right)
+
\left(
\begin{matrix}
2\eta&0\\[3pt]
0&-2\eta
\end{matrix}
\right)
+
\left(
\begin{matrix}
-\kappa\epsilon/\eta&-i\kappa\epsilon/\eta\\[3pt]
-i\kappa\epsilon/\eta&\kappa\epsilon/\eta
\end{matrix}
\right)
\label{mapprox}
\en
where
\eq
\kappa=\fract{1}{2}\al-2p-q-1\,.
\en
The approximate eigenvalues are then
\eq
E_{approx}= -2\kappa   \pm
2\sqrt{\eta^2-\kappa\ep\,}~.
\label{Eapprox}
\en
Apart from the overall shift by $-2\kappa$ and
the replacement of $\epsilon$ by $\kappa\epsilon$, (\ref{mapprox})
and (\ref{Eapprox}) have
exactly the same form as the toy example (\ref{toyvalues}) at
$\alpha=1$, one of the two values for which an exceptional point is
found for only one sign of $\epsilon$. Thus our approximation captures
an important feature of the full problem which was missed by the simpler
approach used in \cite{Bender:1998gh}.
Exceptional points occur when the argument of the square root in
(\ref{Eapprox})
vanishes. At fixed $\epsilon$, and using the $\lambda\to -\lambda$
symmetry, this happens on the parabolas
\eq
\alpha=4p+2q+2+\frac{1}{2\epsilon}(2\lambda\pm 2q)^2
\en
on the $(2\lambda,\alpha)$ plane, where $p$ and $q$ are non-negative
integers. Thus there is a parabola rooted at every intersection of the lines
$\alpha^+\in\ZZ^+$, $\alpha^-\in\ZZ^+$.
However, there is a significant difference between the situations for
$\epsilon>0$ ($M>1$) and for $\epsilon<0$ ($M<1$). For
$\epsilon>0$,
the parabolas are upwards convex, as in figures \ref{figa}
and \ref{figc} above. 
Any fixed value of $\lambda$ and $\alpha$ in the neighbourhood of a line 
$\lambda=q$ within which the $2\times 2$ truncations are valid lies inside
only finitely many of the parabolas centred on that line, and thus sees only 
finitely many complex
eigenvalues.

\smallskip

\[
\begin{array}{c}
\!\!\!\!\!\!
\!\!\!\!\!\!\includegraphics[width=0.6\linewidth]{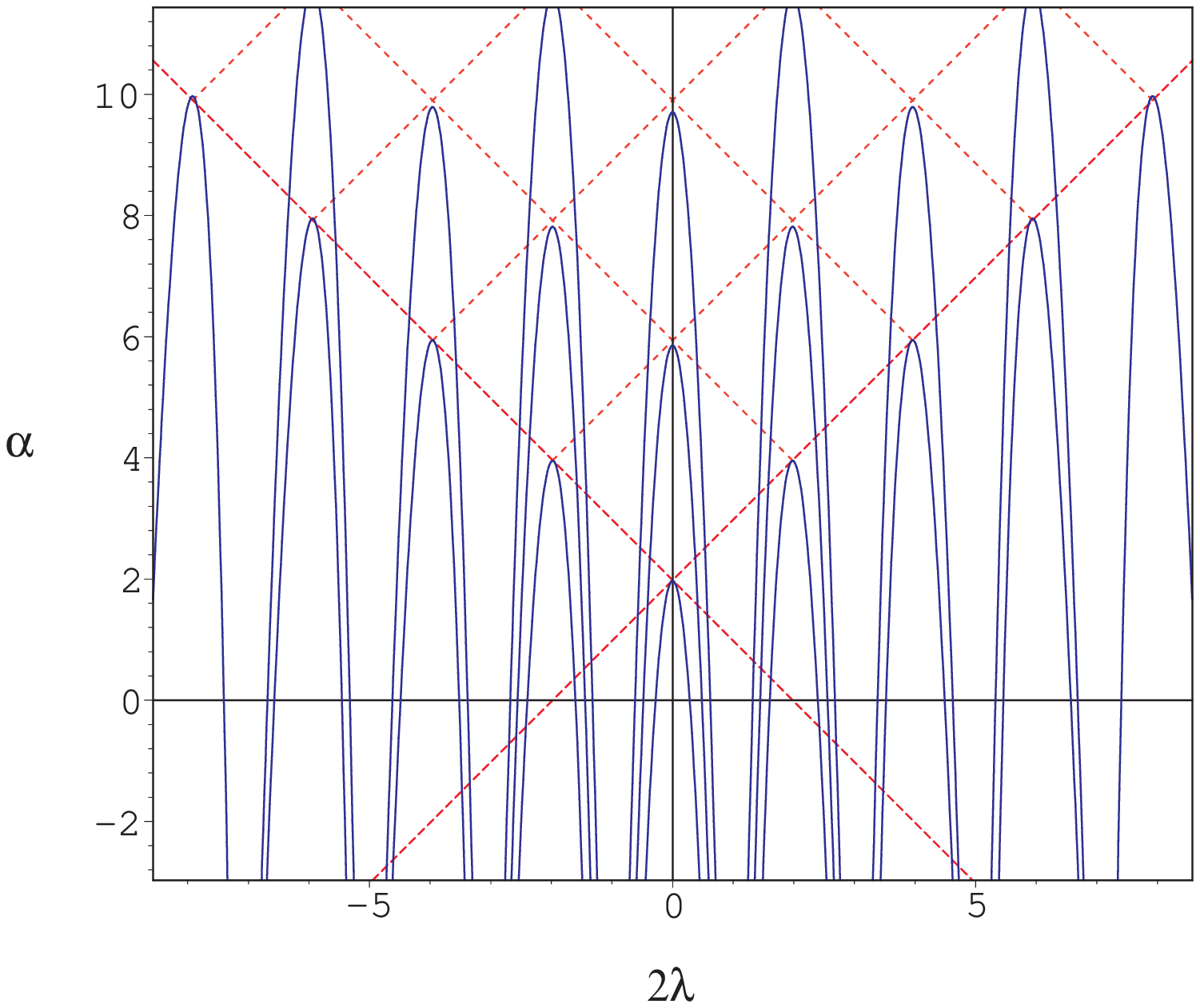}
\\[11pt]
\parbox{0.65\linewidth}{
{\small Figure \ref{figd}: Perturbative lines of exceptional
points for $M=0.98$, with only a subset of the lines shown.
}}
\end{array}
\]
\refstepcounter{figure}
\label{figd}

By contrast, for $\epsilon<0$ the parabolas are
oppositely-oriented, as in figure \ref{figd}.
Any given point
$(2\lambda,\alpha)$ near to a line $\lambda=q$
now lies inside infinitely many of the parabolas centred on that line, and 
outside only a finite number of them.
Thus truncation predicts that infinitely-many eigenvalues
will be complex, with only
finitely many remaining real, these real levels lying at the bottom of the
spectrum. This is exactly as is observed in the full problem.
The transition to infinitely-many complex eigenvalues was
first noted by Bender and Boettcher \cite{BB} for
$\lambda=1/2$, $\alpha=0$. It was subsequently treated analytically,
for general $\lambda$ though still with $\alpha=0$,
in \cite{Dorey:2004fk}, using
a non-linear integral equation for the eigenvalues found via the Bethe
Ansatz approach to the problem. While the latter approach is 
more systematic, the perturbative
understanding of the phase transition just given is particularly
transparent, and gives a more immediate understanding of the regions
in the $(2\lambda,\alpha)$ plane where complex levels are first to be
found.

A check on the truncation method can be made using the asymptotic
obtained in  \cite{Dorey:2004fk} for the value of
$M=M_{\rm crit}<1$ at which high-lying eigenvalues $E$ merge. With
$M_{\rm crit}=1+\ep_{\rm crit}$ and $\lambda\neq 1/2$, this
is\footnote{When comparing with
eq.(5.37) of \cite{Dorey:2004fk}, note that the
$\ep$ used there is equal to $2M-2$, and not $M-1$.}:
\eq
\ep_{\rm crit}\sim \frac{4\ln|\cos(\pi\lambda)|}{\pi^2E}\,.
\label{oldresult}
\en
For $\lambda=q+\eta$ and $\eta$ small, this implies
$\ep_{\rm crit}\sim -2\eta^2/E$.
This is easily seen to match the result just obtained, since
(\ref{Eapprox}) places the exceptional points at
$\ep=\eta^2/\kappa$, and for $E$ large, $E\sim-2\kappa$.

\medskip

In table \ref{comptable} the various approximations used in
this section are compared with numerical data obtained from a direct
solution of the ordinary differential equation.
The numerical eigenvalues found by solving the full problem are denoted
by $E_{exact}$; their numerical errors are smaller than the last quoted digit.
The result using the $2\times 2$ truncation and the
exact matrix elements is $E_{trunc}$, the initial approximated truncation
(including the terms proportional to
$\ep$ and $\ep^2/\eta$)  is $E_{\pm}$, and the final
approximation (retaining only terms proportional to $\eta$ and
$\ep/\eta$ in the matrix elements) is 
$E_{approx}$. The table shows the comparison for sample values of
$\ep$, $\alpha$ and $\eta$, for
for $p=q=0$ (i.e.\ $\lambda=\eta$) and $p=0$,
$q=1$ (i.e.\ $\lambda=1+\eta$).

\medskip

\begin{table}[htbp]
\begin{center}
\begin{tabular}{|c|c|c|c|c|}
\hline
\multicolumn{5}{|c|}{} \\[-10pt]
\multicolumn{5}{|c|}{$p=q=0$} \\[2pt]
\hline
&\multicolumn{2}{|c|}{}&\multicolumn{2}{|c|}{} \\[-11pt]
&\multicolumn{2}{|c|}{$\ep=0.001$, $\al=0.9$, $\eta=0.01$} &
\multicolumn{2}{|c|}{$\ep=0.001$, $\al=0.9$, $\eta=0.25$} \\[1pt]
\hline
$E_{exact}$ &   1.05069482 & 1.15266823 &  0.599733995 & 1.60332810\\
$E_{trunc}$ &   1.05069431 & 1.15266441 &  0.599733083 & 1.60332370 \\
$E_{\pm}$ &     1.05067066 & 1.15269439 &  0.599485149 & 1.60387991\\
$E_{approx}$ &  1.04900980 & 1.15099019 &  0.597804819 & 1.60219518\\
\hline
\multicolumn{5}{|c|}{} \\[-10pt]
\multicolumn{5}{|c|}{$p=0$, $q=1$} \\[2pt]
\hline
&\multicolumn{2}{|c|}{}&\multicolumn{2}{|c|}{} \\[-11pt]
&\multicolumn{2}{|c|}{$\ep=0.01$, $\al=3.9$, $\eta=0.01$} &
\multicolumn{2}{|c|}{$\ep=0.01$, $\al=3.9$, $\eta=0.25$} \\[1pt]
\hline
$E_{exact}$ & 0.07899348 & 0.18089945  & -0.36215520 & 0.62170890 \\
$E_{trunc}$ & 0.07897778 & 0.18034086  & -0.36225580 & 0.62111404 \\
$E_{\pm}$ & 0.07913480 & 0.18078797    & -0.36280969 & 0.62273248 \\
$E_{approx}$ & 0.05101020 & 0.14898979 & -0.40199601 & 0.60199601 \\
\hline 
\end{tabular}
\caption{\small Comparison of the various approximation
methods used for $M\approx 1$.}
\label{comptable}
\end{center}
\end{table}

The treatment so far has concerned the limiting region 
$|\ep|\ll\eta\ll 1$, which suffices to capture the behaviour of the
exceptional lines as $\eta\to 0$. 
Other limits are also interesting, and in closing this section
we remark that other
presentations of the Hamiltonian may then be useful. As an example,
we return to the toy model (\ref{toy1}), (\ref{toy2}), at $\alpha=1$,
and consider taking $\eta\to 0$ {\em before}\/ $\ep\to 0$. As
in \cite{GRS}, one can introduce a pair of matrices
\eq
P=\frac{1}{\sqrt{2}}
\left(
\begin{matrix}
1&i\\[3pt]
i&1
\end{matrix}
\right)\,,\quad
R=
\left(
\begin{matrix}
q&0\\[3pt]
0&1/q
\end{matrix}
\right)
\en
where $q^2=2i\epsilon/\eta$. Then
$H_{1+\epsilon}=H_1+V_{\ep}$ is similar to
\eq
\widehat H_{1+\epsilon}
= R^{-1}P^{-1} H_{1+\epsilon} PR
=
\left(
\begin{matrix}
0&1+\eta^2/\epsilon\\[3pt]
4\epsilon&0
\end{matrix}
\right).
\en
It is now possible to set $\eta=0$, showing that the Jordon block is
indeed recovered as the limit is taken.

\resection{Perturbation theory about $M= \infty$}
\label{Minfsection}
In this section we complete our analysis with a perturbative study about  
the  model at $M=\infty$, which is shown in appendix 
\ref{csw} to be exactly solvable. 
For large $M$, the second duality of appendix \ref{dualapp}  maps the
original eigenproblem (\ref{PTg}) into the Schr\"odinger equation
\eq
H_\ep \phi(z)=-\frac{d^2}{dz^2}\phi(z)+\left[z^2
+\frac{\tilde{\lambda}^2-\frac{1}{4}}{z^2}-\tilde{\al}\right]\phi(z)=
-\frac{1}{z^2}\tilde{E}(-iz)^{2\ep}\phi(z)
\label{hamm}
\en
where
\eq
\tilde M= -1+\frac{2}{M+1}={-1+\ep}~~,~
\tilde E=\left(\frac{2}{M{+}1}\right)^{\frac{2M}{M+1}}E~,~~~
\tilde\lambda=\frac{2}{M{+}1}\,\lambda~,~~
\tilde\alpha=\frac{2}{M{+}1}\,\alpha
\en
and we have set $\ep=2/(M+1)$.   Under the duality transformation, the 
contour ${\cal C}$ transforms into a curve equivalent to an
$M$-independent straight
line running just below  the real axis.  

The inhomogeneous complex square well of appendix \ref{csw} 
appears from (\ref{hamm}) in the large-$M$ (small $\ep$)
limit, when
the right-hand side 
reduces to an  additional
 angular momentum  term so that (\ref{hamm})  becomes 
 the ($\PT$-symmetric)
simple harmonic oscillator, when viewed as an eigenproblem for  
$\tilde \alpha$.  The (unnormalised) eigenfunctions 
\eq
\phi_n^{\pm}(z)=z^{\frac{1}{2} \pm \Lambda}
e^{-\frac{z^2}{2}}L_n^{\pm \Lambda}(z^2)
\quad,\quad \Lambda=\sqrt{\tilde{\lambda}^2+\tilde{E_n}}
\label{un}
\en
correspond to the $\tilde\al$ eigenvalues
\eq
\tilde{\al}_n^{\pm}=4n+2\pm 2\sqrt{\tilde{\lambda}^2+\tilde{E_n}}\,.
\label{aspect}
\en
Alternatively the problem at $M=\infty$
can be considered at fixed $\tilde\al$ as a
generalised eigenproblem for $\tilde E$, with the (entirely real)
spectrum following on rearranging (\ref{aspect}):
\eq
\tilde {E}_n = (2n+1-\hf \at)^2 -\lamt^2 + \quad n=0,1,\dots~.
\en
The pair of levels 
 $\tilde E_n$ and $\tilde E_m$, $n\neq m$, will be degenerate whenever 
$\tilde \alpha=2(n+m+1)$.
Thus degeneracies occur in the spectrum
on the horizontal lines $\tilde \alpha =4$, $6$, $8$, \dots~in 
the $(2\tilde \lambda,\tilde \alpha)$ plane, and a
perturbative treatment will be reliable close to these lines.

The eigenvalue problem at large but finite $M$ can be explored
by taking $\ep$ small and truncating 
the full Hamiltonian $H_\ep$ to the $2 \times 2$ subspace 
spanned by the eigenfunctions $\phi^{\pm}$ associated with the 
 levels 
\eq
\tilde E^{+} = (q-2p+\eta/2)^2 -\lamt^2\quad,\quad 
\tilde E^{-} = (q-2p-\eta/2)^2 -\lamt^2\quad ,\quad q \in \ZZ^+ ~.
\en
This pair of eigenvalues will be almost-degenerate when 
$\at=2(q+1)+\eta$ and $p=0,1,\dots 
[(q-1)/2]$ provided 
$\eta$ is small. 
When $\eta$ is zero the eigenvalues merge  to the single eigenvalue
$E_0:=E_p^{+}=E_{q-p}^{-}$.    
Since the  eigenfunctions (\ref{un}) satisfy the nonstandard
eigenproblem 
\eq
H_0 \phi^{\pm} = -\frac{1}{z^2} \tilde E^{\pm} \phi^{\pm}~, 
\label{hpm}
\en
the usual inner product must be weighted by a factor of
$z^{-2}$, and so we define 
\eq
( \phi_n | \phi_m ) = \int_{\RR-i\eps} 
\phi_n(z) \phi_m(z) 
z^{-2} dz
\en
with a  small positive $\eps$ to avoid any singularities at $z=0$. 
Using the integral (\ref{lagint}) and analytic continuation as necessary,
the orthonormal  eigenfunctions  are  
\eq
\phi^{+}(z)=\frac{\sqrt{2p!(q-2p+\eta/2)}}{\sqrt{(1-e^{\pi
 i\eta})\Gamma(q-p+\eta/2+1)}}   \,
 z^{1/2+q-2p+\eta/2}e^{-z^2/2}L_p^{q-2p+\eta/2}(z^2) 
\en
and 
\eq 
\phi^{-}(z)=\phi^{+}(z)|_{p\to
  q-p}=\frac{\sqrt{2(q-p)!(2p-q+\eta/2)}}{\sqrt{(1-e^{\pi 
 i\eta})\Gamma(p+\eta/2+1)}}  
 \, z^{1/2+2p-q+\eta/2} e^{-z^2/2}L_{q-p}^{2p-q+\eta/2}(z^2)~.
\en

In the truncated basis any eigenfunction $\phi$ can be approximated  
as $\phi=\mu \phi^{+} + \nu \phi^{-}$ for some constants $\mu$ and
$\nu$.   Applying  $H_{\ep} $ to $\phi $, the corresponding
approximate eigenvalue $\tilde E$ 
must satisfy  
\eq
\tilde E^{+} \frac{\phi^{+} }{z^2} + \nu \tilde E^{-}
\frac{\phi^{-} }{z^2}  = \frac{1}{z^2} (-iz)^{2\ep} \tilde E (\mu
\phi^{+} + \nu \phi^{-}),  
\label{eeq}
\en
given  that $\phi^{\pm}$  are eigenfunctions of the
unperturbed Hamiltonian (\ref{hpm}). Thus taking
inner product of  (\ref{eeq})  with
$\phi^\pm$ in turn, we  obtain   
\eq
\left(\begin{matrix}
\tilde E^{+} & 0 \\ 
0 & \tilde E^{-} 
\end{matrix}\right) 
\lf(\begin{matrix}
\mu \\ \nu \end{matrix} \ri)
=  \tilde E 
\left(\begin{matrix}
(  \phi^{+}| (-iz)^{2\ep} | \phi^{+} ) & ( \phi^{+}|
(-iz)^{2\ep} | \phi^{-} )  \\ 
  ( \phi^{-}| (-iz)^{2\ep} | \phi^{+} ) & ( \phi^{-}|
(-iz)^{2\ep} | \phi^{-} )  
\end{matrix}\right)  
\lf(\begin{matrix}
\mu \\ \nu \end{matrix} \ri)~. 
\label{hame}
\en

We use the  integral (\ref{lagint}) in
appendix~\ref{matrixelements} to calculate the required exact matrix elements.
To leading order in $\ep$ and $\eta$, 
the matrix elements $H_{ab} = (\phi^a | (-iz)^{2\ep} | \phi^b )$  are 
\bea  
H_{++}&\approx& 1+\frac{2\ep}{\eta} +\Bigl (\psi(q{-}2p) +
\psi(q{-}2p{+}1)
- \psi(q{-}p{+}1)\Bigr)\ep \nn \\
&&~+2\Bigl (\psi(q{-}2p) + \psi(q{-}2p{+}1) -
\psi(q{-}p{+}1)\Bigr)\frac{\ep^2}{\eta}~; \\[4pt] 
H_{--}&\approx & 1-\frac{2\ep}{\eta}+\Bigl ( \psi(q{-}2p)  +
\psi(q{-}2p{+}1)-\psi(p{+}1)\Bigr )\ep 
\nn \\ &&~-2\Bigl (\psi(q{-}2p) + \psi(q{-}2p{+}1) -
\psi(q{-}p{+}1)\Bigr)\frac{\ep^2}{\eta}~; \\[4pt]  
H_{+-}&\approx & i (-1)^q \Bigl [ -\frac{2\ep}{\eta} +\frac{1}{2}\Bigl
(\psi(q{-}p{+}1)-\psi(p{+}1) 
\Bigr )\ep  \nn \\
&&~+2\Bigl(\psi(q{-}p{+}1)-\psi(q{-}2p)-\psi(q{-}2p{+}1) 
\Bigr )\frac{\ep^2}{\eta}  \Bigr ]~.
\eea

Diagonalising the RHS of (\ref{hame}),  the approximate eigenvalues at
$\ep=2/(M+1)$ and $\at=2(q+1)+\eta$ are 
\bea
 \tilde E_{\pm}&=& E_0
+\biggl( \frac{E_0}{2}\lf (\psi(q{-}p{+}1)+\psi(p{+}1)-4\psi(q{-}2p{+}1)\ri)
-\frac{E_0}{q{-}2p} -2(q{-}2p)  \biggr)\ep \nn \\ 
&& \pm \Biggl [ 
-4(q{-}2p)E_0\ep+(q{-}2p)^2\eta^2+
(q{-}2p)
(\psi(q{-}p{+}1)-\psi(p{+}1))E_0\ep \eta  \nn \\ 
 &&  + \biggl ( 2(q{-}2p)\lf(4\psi(q{-}2p{+}1)-3\psi(p{+}1)
-\psi(q{-}p{+}1)+4\psi(q{-}2p)\ri)E_0+4(q{-}2p)^2 \biggr)\ep^2   
\Biggr ]^{1/2}\nn  \\
\label{sqm} 
\eea
where $E_0=(q-2p)^2-\lamt^2$.    Just as for $M\approx 1$,
the exceptional points can be located by finding where
the argument of the square root in (\ref{sqm}) vanishes. 
Figure \ref{fig250p0} shows the resulting curves of exceptional
points in the $(2\tilde \lambda , \at)$ plane 
for $M=250$, taking
$q=1\dots 5$
and $p=0\dots[(q-1)/2]$\,. The match with the results from
a numerical solution to the full problem is excellent, and indeed even
at $M=30$ the truncation method gives a plot essentially indistinguishable 
from that shown earlier in figure \ref{fig30p0}.

\medskip

\[ 
\begin{array}{c}
\!\!\!\!\!\!\includegraphics[width=0.6\linewidth]{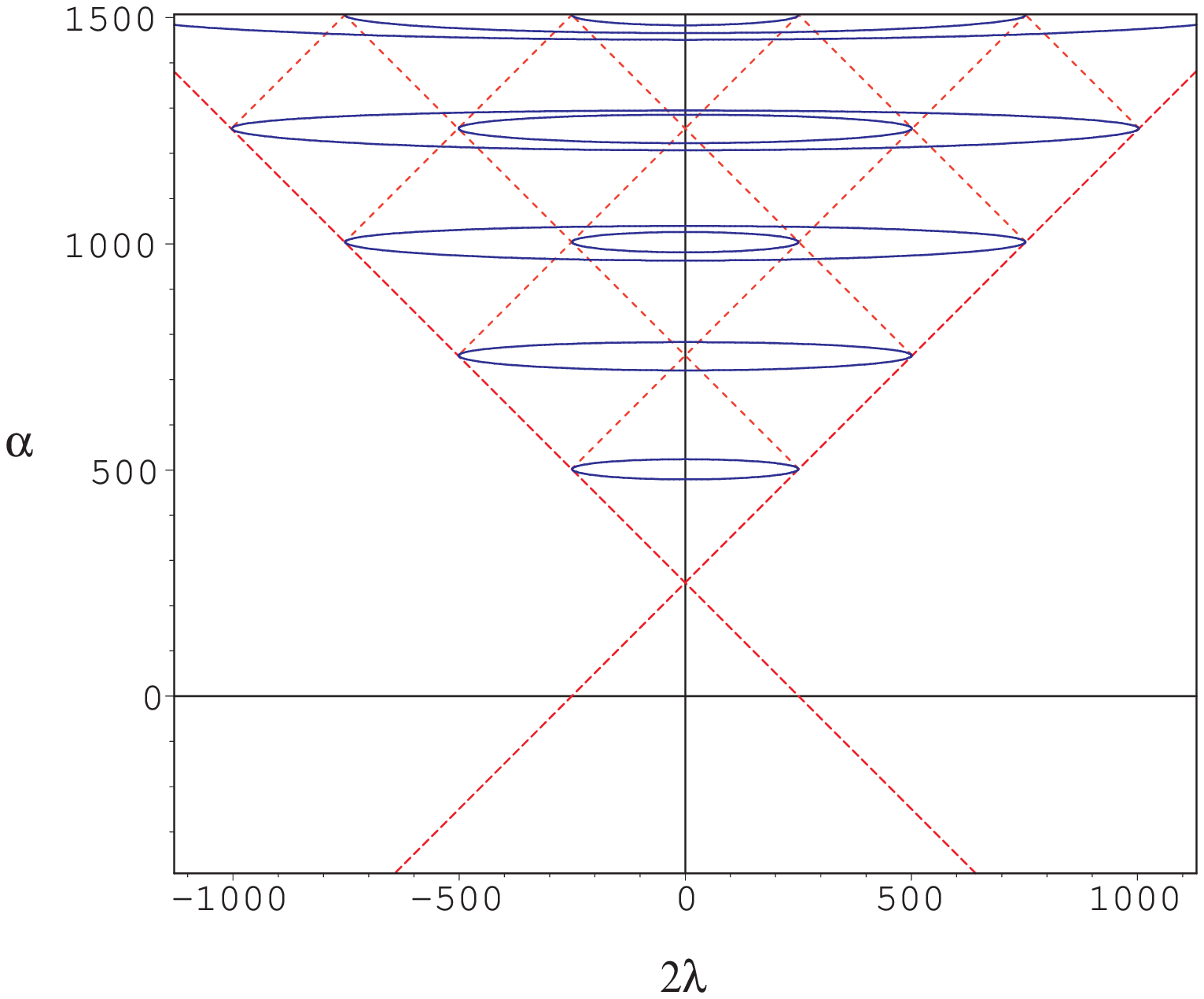}
\\[11pt]
\parbox{0.71\linewidth}{ 
{\small Figure \ref{fig250p0}: Perturbative predictions for the
  exceptional lines for $M=250$.
}}
\end{array}
\]
\refstepcounter{figure}
\label{fig250p0}   

The main features of the transitions are most clearly understood
 if only the terms proportional to $\eta$ and
$\ep/\eta$ are kept in the matrix elements $H_{ab}$.  Rediagonalising
(\ref{hame}),
the  approximate eigenvalues are 
\eq
 \tilde E_{\rm approx}= E_0  \pm  \sqrt{  
(q-2p)^2\eta^2 -4E_0(q-2p)\ep}~.
\en
Demanding once again that the argument of the square root vanishes leads
to the prediction that the 
exceptional points lie on the ellipses  
\eq
4\lf (\frac{\alpha}{M+1}-q-1\ri)^2(q-2p)-4\ep\lf((q-2p)^2-4
\frac{\lambda^2}{(M+1)^2}\ri )=0 
\en
in the $(2\lambda, \alpha)$ plane.  Thus as $M$ decreases from
infinity isolated ellipses of unreality appear, starting from segments
of the degenerate lines $\at =4$, $6$, $8$, \dots~at 
$M=\infty$ and acquiring
exactly the `nested' structure seen in 
figures \ref{fig30p0} and \ref{fig250p0}.

Table \ref{comptableinf} compares the various  levels of
approximation used in
this section  with numerical data obtained from a direct
treatment of the ordinary differential equation, in the same
notation as  table \ref{comptable}.   
 The table shows the comparison for sample values of 
$\ep$, $\tilde \lambda$ and $\eta$, for
$p=0,\ q=1$ (i.e.\ $\tilde \alpha=4+\eta$) and for $p=0$,
$q=2$ (i.e.\ $\at=6+\eta$). 

\medskip

\begin{table}[htbp]
\begin{center}
\begin{tabular}{|c|c|c|c|c|}
\hline
\multicolumn{5}{|c|}{} \\[-10pt]
\multicolumn{5}{|c|}{$p=0$, $q=1$} \\[2pt]
\hline
&\multicolumn{2}{|c|}{}&\multicolumn{2}{|c|}{} \\[-11pt]
&\multicolumn{2}{|c|}{$\ep=0.001$, $\tilde \lambda=-1.2$, $\eta=0.01$}
& \multicolumn{2}{|c|}{$\ep=0.001$, $\lambda=-1.2$, $\eta=0.25$}
\\[1pt] \hline
$E_{exact}$ &  -0.48512051 &   -0.3988914 &  -0.679527538 &-0.17313365\\
$E_{trunc}$ &  -0.48511885 & -0.3988926 &  -0.679526948 &-0.17313370 \\
$E_{\pm}$ &  -0.48514998 & -0.3989179 &    -0.695319170 &-0.18874878 \\
$E_{approx}$&-0.48312771 & -0.3968723 &    -0.693495562 &-0.18650444
\\
\hline \multicolumn{5}{|c|}{} \\[-10pt] \multicolumn{5}{|c|}{$p=0$,
  $q=2$} \\[2pt] \hline &\multicolumn{2}{|c|}{}&\multicolumn{2}{|c|}{}
\\[-11pt] &\multicolumn{2}{|c|}{$\ep=0.01$, $\tilde \lambda=-3$,
  $\eta=0.01$} & \multicolumn{2}{|c|}{$\ep=0.01$, $\tilde \lambda=-3$,
  $\eta=0.25$} \\[1pt] \hline
$E_{exact}$ &-5.59855357 &-4.36272053  &-5.72941123 &-4.20419136 \\
$E_{trunc}$ &-5.59525382 &-4.36529048  &-5.72605257 &-4.20649571 \\
$E_{\pm}$ &  -5.60435172 &-4.35836985  &-5.75706543 &-4.20565613 \\ 
$E_{approx}$&-5.63277168 &-4.36722832  &-5.80622578 &-4.19377423   \\
\hline
\end{tabular}
\caption{\small Comparison of the various
approximation methods used in section \ref{Minfsection}.
\label{comptableinf}}
\end{center} 
\end{table}

\resection{Conclusions}
In this paper we have continued the project initiated in
\cite{DDT3,Dorey:2001hi}, and mapped out the phase diagram of a
three-parameter family of $\PT$-symmetric eigenvalue problems
related to the
Perk-Schultz models. Special features have enabled us to make precise
the Jordan block structures at a subset of the exceptional points,
going beyond the finite-dimensional examples which were the subject of
most previous work. We have also uncovered some novel properties of
the Bender-Dunne polynomials. The resulting phase diagrams at
fixed $M$, consisting of lines of quadratic exceptional points 
punctuated by triply-exceptional (cubic) cusps, 
generalise the previously-observed story at
$M=3$ in an appealing way, and the perturbative treatment about $M=1$
has allowed us to understand the 
transition to infinitely-many complex eigenvalues which occurs as
$M$ decreases below $1$ from a new perspective. The dualities that
we have used were crucial in making a reliable numerical treatment of
the problem, and may be of independent theoretical
interest, especially given the
roles that this set of models plays as possibly the simplest example
of an ODE/IM correspondence.

%
%

\bigskip
\bigskip

\noindent{\bf Acknowledgements --}
We would like to thank Adam Millican-Slater for previous
collaboration, and Carl Bender,
Uwe Gunther, Deiter Heiss, Joey Oliver,
Mark Sorrell and Farid
Tari for useful conversations and help.
PED, TCD and AL thank Torino University, 
and PED, TCD and RT thank APCTP, Pohang and the Galileo
Galilei Institute, Florence, for hospitality at various
stages of this project.
PED was partially supported by the International
Molecule program `Aspects of Quantum Integrability', and thanks 
the Yukawa Institute for Theoretical Physics for its hospitality
during this period. 
The project was also partially supported by INFN grants TO12 and PI11, NATO
grant number PST.CLG.980424, STFC rolling grant ST/G000433/1,
a Nuffield Foundation grant
number NAL/32601, and a grant from the Leverhulme Trust.

\bigskip

\appendix
\resection{Basis for an $n\times n$ Jordan block} \label{jordan}
 (See \cite{Sokolov:2006vj} for a discussion of the $n=2$
case.)
To illustrate a method we can use to construct the basis of an $n\times n$
Jordan block, which arises when $n$ eigenstates merge, we will work with a toy
model. Take an $n\times n$ matrix $L$, depending on one parameter $\ep$:
\eq
L(\ep)=
\left(\begin{array}{cccc}
0 & 1 & 0 & \ldots \\
\vdots & \ddots & \ddots & 0 \\
0 & \ldots & 0 & 1\\
\ep & 0 & \ldots & 0 \\
\end{array}\right).
\en
This has $n$ independent eigenvectors: 
\eq
\psi_j =
\left(\begin{array}{c}
1 \\
e^{(2\pi ij/n)} \ep^{1/n} \\
e^{(4\pi ij/n)} \ep^{2/n} \\
\vdots \\
e^{2\pi(n-1)ij/n} \ep^{(n-1)/n}\\
\end{array} \right)\,,~~~j=1\ldots n~.
\en
When $\ep=0$, $L(\ep)$ has a Jordan block form, but at this point all $n$
eigenvectors $\psi_j$ become equal and so no longer form a basis.
We therefore need to construct a new basis consisting of the vectors
$\phi^{(k)}$, $k=0\ldots n-1$ which satisfy a Jordan chain
\bea
\left.L(\ep)\phi^{(0)}\right|_{\ep=0} &=& 0  \label{L0}\\
\left.L(\ep)\phi^{(k)}\right|_{\ep=0} &=& \left.\phi^{(k-1)}\right|_{\ep=0}
\,,~~~ k=1\ldots n-1~. \label{Lk}
\eea
For simplicity we begin with the eigenvector $\psi_n$
where
\eq \label{0}
L(\ep)\psi_n(\ep) = \ep^{1/n}\psi_n(\ep)~.
\en
Clearly $\phi^{(0)}=\psi_n(\ep)$ satisfies the 
condition (\ref{L0}) when $\ep=0$.  We could choose  $\phi^{(0)}$ to
be any  of the   $\psi_j$ here; each one would lead to a 
 different normalisation for the $\phi^{(k)}$ below.

Before we construct the other basis vectors, we introduce some notation.
Let
\eq
D\equiv n\ep^{\frac{n-1}{n}}\frac{d}{d\ep}
\en
and
\eq
\tilde{L}\equiv \frac{dL}{d\ep}.
\en
Note that $L$ is linear in $\ep$ so
$\frac{d\tilde{L}}{d\ep}=0$. We now have the following commutation relations
\bea
\lbrack D,\ep^{k/n}\rbrack &=& k\ep^{(k-1)/n} \\
\lbrack D,L\rbrack &=& n\ep^{(n-1)/n}\tilde{L} \\
\lbrack D,\tilde{L}\rbrack &=& 0~.
\eea
Finally, define
\eq \label{psik}
\phi^{(k+1)}\equiv \frac{1}{k+1}D\phi^{(k)}.
\en

By induction, it is easy to show that acting with $D$ on (\ref{0}) $k$
times for $1\leq k\leq n-1$ gives 
\eq \label{k}
\sum_{j=0}^{k-1}\prod_{i=0}^{j}\frac{(n-i)}{(j+1)!}\ep^{\frac{n-j-1}{n}}
\tilde{L}\phi^{(k-j-1)}+L\phi^{(k)}=\phi^{(k-1)}+\ep^{\frac{1}{n}}\phi^{(k)}.
\en
When $\ep=0$ this satisfies (\ref{Lk}), so an appropriate basis is
\eq
\phi^{(0)} = \psi_n
\en
and
\eq
\phi^{(k)} = \frac{1}{k}D\phi^{(k-1)}
\en
for $k=1\ldots n-1$.

\resection{Two dualities}
\label{dualapp}
As noted in
\cite{Bazhanov:1998wj}, useful relations between spectral
problems which arise in the ODE/IM correspondence
can often be found by simple variable changes.
Here, starting from (\ref{PTg}) and setting $z=ix$ as in
(\ref{PT3}) to obtain
\eq
-\frac{d^2}{dz^2}\,\psi(z) + \left[\, z^{2M}+\alpha z^{M-1} +
\frac{\lambda^2-\frac{1}{4}}{z^2}+E\,\right]\psi(z) = 0
\label{PTz}
\en
we exploit the fact that, for arbitrary $\beta$,
the combined substitutions $z=y^{\beta}$,
$\psi(z)=y^{(\beta-1)/2}\phi(y)$, transform
$d^2\psi/dz^2$ without introducing a first derivative term:
\eq
\frac{d^2}{dz^2}\psi(z)=
\frac{y^{3/2-3\beta/2}}{\beta^2}\left[
\frac{d^2}{dy^2}-\frac{\beta^2-1}{4y^2}\right]\phi(y)
\en
so that the equation becomes
\eq
-\frac{d^2}{dy^2}\,\phi(y) + \beta^2\!\left[\, y^{2(M{+}1)\beta-2}+
\alpha y^{(M{+}1)\beta-2} 
+ \frac{\beta^2\lambda^2-\frac{1}{4}}{\beta^2\,y^2}
+ Ey^{2\beta-2}\,
\right]\phi(y) = 0~.
\label{PTw}
\en
Two important special cases are
$\beta=1/(M{+}1)$ and $\beta=2/(M{+}1)$.

\medskip

\noindent {\bf 1)} 
$\beta=1/(M{+}1)$\,: 
setting $y=\kappa w$ with
$\kappa=((M{+}1)/\sqrt{-E})^{M+1}$
leads to
\eq
-\frac{d^2}{dw^2}\,\phi(w) + \left[\, -w^{2\tilde
M}+\tilde\alpha\sqrt{\tilde E}\, w^{-1} +
\frac{\tilde\lambda^2-\frac{1}{4}}{w^2}+\tilde E\,\right]\phi(w) = 0
\en
where
\eq
\tilde M=-\frac{M}{M{+}1}~,~~~
\tilde E=\frac{(M{+}1)^{2M}}{(-E)^{M+1}}~,~~~
\tilde\lambda=\frac{1}{M{+}1}\,\lambda~,~~
\tilde\alpha=\frac{1}{M{+}1}\,\alpha\,.
\label{dual1}
\en
This generalises the duality used in
\cite{Bazhanov:1998wj} to inhomogeneous potentials\footnote{It is
interesting that, while \cite{Bazhanov:1998wj} is indeed the first
time that this duality was applied in the context of integrable
quantum field theory,
the homogeneous case can be traced back to
(Isaac) Newton: see \cite{Quigg:1997bx,Grant:1993rh}.}.

\medskip

\noindent {\bf 2)} 
$\beta=2/(M{+}1)$\,: 
setting $y=\kappa w$ with $\kappa=\sqrt{(M{+}1)/2}$ yields
\eq
-\frac{d^2}{dw^2}\,\phi(w) + \left[\, w^2
+\tilde E \,w^{2\tilde M}+
\frac{\tilde\lambda^2-\frac{1}{4}}{w^2}+\tilde \alpha\,\right]\phi(w) = 0
\label{dual2}
\en
where
\eq
\tilde M= {-1+\frac{2}{M{+}1}} ~,~~
\tilde E=\left(\frac{2}{M{+}1}\right)^{\frac{2M}{M+1}}E~,~~~
\tilde\lambda=\frac{2}{M{+}1}\,\lambda~,~~
\tilde\alpha=\frac{2}{M{+}1}\,\alpha\,.
\en

\medskip

To obtain an equivalence between eigenvalue problems,
the transformation of the boundary conditions
under the mappings must be tracked.
The boundary conditions from section \ref{intro} translate into
the requirement that eigenfunctions of the
initial problem (\ref{PTz}) should decay in $i\CS_{-1}$
and $i\CS_{1}$, where the sectors $\CS_k$ were defined in
(\ref{stokesdef}).
After the transformation the simultaneous decay should instead be
in $i\,\widetilde\CS_{-1}$ and $i\,\widetilde\CS_1$, where
for case {\bf 1},
Newton's duality,
\eq
\widetilde\CS_k=\left\{ x\in{\mathbb C}\,:\,
\left|\arg(ix)-\pi k\,\right|<\pi/2\,\right\},
\en
while for case {\bf 2} ($\beta=2/(M{+}1)\,$),
\eq
\widetilde\CS_k=\left\{ x\in{\mathbb C}\,:\,
\left|\arg(ix)-\pi k/2\,\right|<\pi/4\,\right\}.
\en
In both cases the transformed
sectors are independent of $M$, reflecting the fact that the leading
terms in (\ref{dual1}) and (\ref{dual2}) at large $|w|$, $\tilde E$
and $w^2$ respectively, are themselves independent of $M$. For the
first duality it might appear that the sectors $i\,\widetilde\CS_{\pm
1}$ coincide, but this is not so -- the branch cut in the
original problem (\ref{PTg}) becomes a cut along the negative real
axis of the $w$ plane, and so the two sectors lie on top of each other
on the full Riemann surface of the problem. For the second duality
the sectors are those of the simple harmonic oscillator
and this makes (\ref{dual2}) particularly useful for numerical work:
eigenvalues can be found by solving the ODE on a straight,
$M$-independent contour,
running vertically (parallel to the imaginary axis)
in the right half of the
complex $w$ plane.
An efficient approach uses
WKB asymptotics at large $|w|$ as initial conditions for a pair of
numerical
solutions, $\phi_{-1}$ and $\phi_1$, decaying as $\Im m\,w\to\pm\infty$,
and then locates the eigenvalues by looking for zeros of the
Wronskian $W[\phi_{-1},\phi_1]$, evaluated in the neighbourhood of the
origin where both numerical solutions are reliable.
This method was used to produce many of the figures
in this paper.

Replacing $w$ by $w/i$ trivially rotates the
dual problems back to a more usual `$\PT$-symmetric' form.
The mappings can also be used to give equivalences for
spectral problems initially specified by the simultaneous decay of
eigenfunctions on more widely-separated pairs of Stokes sectors than
$\CS_{-1}$ and $\CS_1$. The homogeneous cases of these
problems were discussed in \cite{Bender:1998gh}, and related to fused
transfer matrices in integrable models in \cite{Dorey:1999uk}.

\resection{Useful formulae}
\label{matrixelements}
This appendix records a number of formulae used in the main text.
All can be inferred from the following basic integral, involving a
pair of Laguerre polynomials: 
\bea
 &&
\!\!\!\!\!\!\!\!
\!\!\!\!\!\!\!\!
\!\!\!\!\!\!\!\!
\!\!\!\!\!\!\!\!
\int_{0}^\infty t^{\alpha}t^{(\gamma+\rho)/2} e^{-t} 
L_m^{\rho}(t) 
L_n^{\gamma}(t)
\,dt \nn\\[3pt]
& = &
\frac{(\half(\gamma{-}\rho){-}\alpha)_n(\rho{+}1)_m}{n!\,m!}\,
\Gamma(\half(\gamma{+}\rho){+}1{+}\alpha)
\,\times ~
\nn\\ &&\qquad
{}_3F_2
(-m,\half(\rho{+}\gamma){+}1{+}\alpha,\half(\rho{-}\gamma){+}1{+}\alpha;
\rho{+}1,\half(\rho{-}\gamma){+}1{+}\alpha{-}n;1) \nn\\[3pt]
& = &
\frac{\Gamma(\half(\gamma{+}\rho){+}1{+}\alpha)}{m!\,n!}
\,\times ~
\nn\\ &&\qquad
\sum_{k=0}^m\binom{m}{k}
(\rho{+}1{+}k)_{m-k}(\half(\rho{+}\gamma){+}1{+}\alpha)_k
(\half(\rho{-}\gamma){+}1{+}\alpha)_k
(\half(\gamma{-}\rho){-}\alpha)_{n-k}
\label{lagint}
\eea
where $(a)_n=a(a{+}1)\dots (a{+}n{-}1)$ is the Pochhammer symbol and
${}_3F_2$ is a generalised hypergeometric function. The first version
of this result can be 
found in \cite{Adam}; it generalises a formula 
for the case $\gamma=\rho$ that was
given in \cite{hallsaad}. 
The symmetry of the final expressions under
the simultaneous exchanges
$m\leftrightarrow n$,
$\rho\leftrightarrow \gamma$ is not obvious, though it can be checked.

In section \ref{pertsec} the matrix elements
$\langle\phi_n^{\pm}(x)|(ix)^{2M}|\phi_m^{\pm}(x)\rangle$ and
$\langle\phi_n^{\pm}(x)|(ix)^{2M}|\phi_m^{\mp}(x)\rangle$ were needed
for general $M$, where
$\phi_n^+(x)$ and $\phi_n^-(x)$ are the normalised wavefunctions
given by (\ref{efns}).
The relevant
calculations were also carried out by Millican-Slater in \cite{Adam}, and
we reproduce his final results here.
The matrix element
$\langle{\phi}_n^{+}|(ix)^{2M}|{\phi}_m^{+}\rangle$ is
\eq \label{Phipp}
\begin{split}
&
\!\!\!\!  \!\!\!\!  \!\!\!\!  \!\!\!\!  \!\!\!\!  \!\!\!\!
\langle{\phi}_n^+|(ix)^{2M}|{\phi}_m^+(x)\rangle =\\
&\left(\cos(M\pi)+\sin(M\pi)\cot(\lambda\pi)\right)\frac{(-M)_n(\lambda+1)_m}
{\sqrt{n!m!\Gamma(\lambda+m+1)\Gamma(\lambda+n+1)}}\,\times \\
&\qquad\Gamma(\lambda{+}M{+}1)\,{}_3
F_2(-m,\lambda{+}M{+}1,1{+}M\,;\,\lambda{+}1,1{+}M{-}n\,;\,1)\,.
\end{split}
\en

For $M=1$ (one of the cases needed)
there is a negative integer in one of the second group of
entries of the hypergeometric function in (\ref{Phipp}), and so for
certain values of $n$ and $m$ these functions may be undefined. This
is the case when $n-2<m$. However, for $|n-m|\geq 2$
the symmetry of the inner products in $n$ and $m$ can be used to avoid
the problem. In these cases, when $M=1$, the $(-M)_n$
in the expressions above become $(-1)_n=0$ so the inner products
are zero. For $n=m$ and $n=m\pm 1$,
by taking the limit $M\rightarrow 1$ in (\ref{Phipp})
it can be shown \cite{Adam} that the only
non-zero inner products are
\bea
\langle{\phi}_n^+(x)|x^2|{\phi}_n^+(x)\rangle &=&
1+\lambda+2n \label{Phipp1}\\
\langle{\phi}_{n+1}^+(x)|x^2|{\phi}_n^+(x)\rangle &=&
\sqrt{\frac{n+1}{n+\lambda +1}}
\left(\lambda-n\right)\label{Phipp2}~.
\eea
The matrix elements corresponding to (\ref{Phipp}),
(\ref{Phipp1}) and (\ref{Phipp2}) for ${\phi}_n^-$ can be found by
sending $\lambda\rightarrow -\lambda$.

The matrix element
$\langle{\phi}_n^{+}(x)|(ix)^{2M}|{\phi}_m^{-}(x)\rangle$ is
given by
\eq \label{Phipm}
\begin{split}
\langle{\phi}_n^+(x)|(ix)^{2M}|{\phi}_m^-(x)\rangle =
i\frac{\sin(M\pi)(1-\lambda)_m(\lambda -M)_n
\Gamma(M+1)}{|\sin(\pi \lambda)|
\sqrt{\Gamma(1-\lambda +m)\Gamma(1+\lambda +n)m!n!}} \\
\times\,{}_3F_2(-m,1+M,M+1-\lambda\,;\,1-\lambda,M+1-\lambda -n\,;\,1)~,
\end{split}
\en
which, unlike (\ref{Phipp}), is always well defined at $M=1$.

\resection{The inhomogeneous complex square well}
\label{csw}
In the main text, the large-$M$ limit of the spectrum of
\eq
\Bigl[-\frac{d^2}{dx^2}-(i x)^{2M}
-\alpha (i x)^{M-1}+ \frac{\lambda^2-\frac{1}{4}}{x^2}
\Bigr]\psi(x)=E\,\psi(x)\, ,
{}~~~\psi(x) \in L^2(\CaC)\,,
\label{PTga}
\en
was needed. The $\alpha=0$, $\lambda=1/2$ case was investigated in
\cite{Bender:1999ds}, where it was dubbed the `complex square well'.
To treat the more general case, we start with the same variable change
as in \cite{Bender:1999ds}, and set
\eq
x=\left(-i+\frac{z\pi}{2M}\right)E^{\frac{1}{2M}}.
\en
Taking the limit $M\to\infty$, using the identity
$\lim_{M\to\infty}(1+x/M)^M=e^x$ and dropping all subleading terms,
(\ref{PTga}) becomes
\eq
\left[\,\frac{d^2}{dz^2} +
\frac{\pi^2}{16}\,\tilde E(1+e^{i\pi z})+
\frac{\pi^2}{16}\sqrt{E}\tilde\alpha\,e^{i\pi z/2}+
\frac{\pi^2}{16}\,\tilde\lambda^2\,\right]\psi(z)=0
\label{Minf}
\en
where
\eq
\tilde E=\left(\frac{2}{M{+}1}\right)^2\,E
\en
and the scaled parameters
\eq
\tilde\lambda=\frac{2\lambda}{M+1}~,\quad
\tilde\alpha=\frac{2\alpha}{M+1}
\en
were used to ensure the survival of the inhomogeneous and
angular-momentum terms in the limit. Notice that in terms
of $\tilde\lambda$ and $\tilde\alpha$, the parameters $\alpha_{\pm}$ of
(\ref{apm}) are simply
\eq
\alpha_{\pm}=\frac{1}{4}\,(\,\tilde\alpha-1\pm 2\tilde\lambda\,)\,.
\en
The special feature of this limit is that the resulting ODE
(\ref{Minf}) is exactly solvable. Here we highlight the link with the
simple harmonic oscillator by making a further variable change to
$w=\tilde E^{1/4}e^{i\pi z/4}$ and trading $\psi(w)$ for
$\phi(w)=\sqrt{w}\,\psi(w)$\,. Substituting in, $\phi(w)$ satisfies
\eq
-\frac{d^2\phi}{dw^2}+\left[\, w^2+
\frac{\tilde E+\tilde\lambda^2-\frac{1}{4}}{w^2}\,
\right]\phi
= -\tilde\alpha\,\phi\,.
\label{Minfb}
\en
Boundary conditions should be imposed on the asymptotic Stokes lines
$z=\pm 2-iy$, $y\to\infty$
\cite{Bender:1999ds}, which translate into the positive and negative
imaginary axes in the complex $w$ plane. That said, the spectrum of
(\ref{Minfb}) can be recognised as that of the $\PT$-symmetric simple
harmonic oscillator \cite{Znojil:1999qt,Dorey:1999uk}, with `energy'
$\tilde\alpha$ and `angular momentum'
$-1/2\pm \sqrt{\tilde E+\tilde\lambda^2}$ (the reversed sign of the energy
is a result of the rotated quantisation contour for (\ref{Minfb})
compared to that used in \cite{Dorey:1999uk}\,). Hence, from
\cite{Dorey:1999uk}, (\ref{Minfb}) has a wavefunction normalisable on the
quantisation contour if and only if
\eq
\tilde\alpha=4n+2\pm 2\sqrt{\tilde E+\tilde\lambda^2}~,\quad n=0,1,\dots
\en
which translates into our main result for the exact spectrum of
(\ref{PTga}) in the $M\to\infty$ limit:
\eq
\tilde E_n=
(2n+1-\fract{1}{2}\tilde\alpha)^2-
\tilde\lambda^2\,, \quad n=0,1,\dots~.
\en
Via $\tilde E_n=4E_n/(M{+}1)^2$, this result also gives the leading
behaviour of the original levels $E_n$ as the linit is taken.
For $\tilde\lambda=\tilde\alpha=0$, this reproduces the result
of \cite{Bender:1999ds}. Notice that the spectrum is entirely real for
all values of $\tilde\lambda$ and $\tilde\alpha$, matching the
situation at $M=1$, the other exactly-solvable point.

%

\end{document}